\documentclass{aa} 
\usepackage{graphicx}
\usepackage{txfonts}
\usepackage{natbib}
\bibpunct{(}{)}{;}{a}{}{,} 
\usepackage{color}
\usepackage{float}
\usepackage{subfigure}
\usepackage{hyperref}
\usepackage{longtable}
\usepackage{rotating}
\usepackage{lscape}
\usepackage{soul}
\newcommand{\mosaic}{\textsf{mosaic}}

\newcommand{\flcgs}{\mbox{erg s$^{-1}$ cm$^{-2}$}}
\newcommand{\sbl}{\mbox{erg s$^{-1}$ cm$^{-2}$ arcsec$^{-2}$}}

\def\HI{\mbox{H\,{\sc i}}}
\newcommand{\OII}{[O\,{\sc ii}]}

\newcommand{\MgII}{Mg\,{\sc ii}}
\newcommand{\FeII}{Fe\,{\sc ii}*}
\newcommand{\CII}{[C\,{\sc ii}]}
\newcommand{\lya}{\mbox{Ly$\alpha$}}
\begin{document}

\title{The MUSE eXtremely Deep Field: Individual detections of $\lya$ haloes around rest-frame UV-selected galaxies at $z\simeq2.9$--$4.4$ \thanks{Based on observations made with ESO telescope at the La Silla
Paranal Observatory under the large program 1101.A-0127}}
\titlerunning{$\lya$ haloes around individual high-$z$ UV-selected galaxies}

\author{Haruka Kusakabe\inst{\ref{inst1}\thanks{e-mail: haruka.kusakabe@unige.ch}}
\and Anne Verhamme\inst{\ref{inst1},\ref{inst2}}
\and J{\'e}r{\'e}my Blaizot\inst{\ref{inst2}}           
\and Thibault Garel\inst{\ref{inst1},\ref{inst2}}
\and Lutz Wisotzki\inst{\ref{inst3}}
\and Floriane Leclercq \inst{\ref{inst1}}
\and Roland Bacon\inst{\ref{inst2}}
\and Joop Schaye\inst{\ref{inst4}}
\and Sofia G. Gallego\inst{\ref{inst5}}
\and Josephine Kerutt\inst{\ref{inst1}}
\and Jorryt Matthee\inst{\ref{inst5}}
\and Michael Maseda\inst{\ref{inst6}}
\and Themiya Nanayakkara\inst{\ref{inst7}}
\and Roser Pell\'{o}\inst{\ref{inst8}}
\and Johan Richard\inst{\ref{inst2}}
\and Laurence Tresse\inst{\ref{inst8}}
\and Tanya Urrutia\inst{\ref{inst3}}
\and Elo\"{i}se Vitte\inst{\ref{inst1},\ref{inst9}}
}
\institute{Observatoire de Gen\`{e}ve, Universit\'e de Gen\`{e}ve, 51 Chemin de P\'egase, 1290 Versoix, Switzerland\label{inst1}
\and
Univ. Lyon, Univ. Lyon1, Ens de Lyon, CNRS, Centre de Recherche Astrophysique de Lyon UMR5574, F-69230, Saint-Genis-Laval, France\label{inst2}
\and
Leibniz-Institut f\"{u}r Astrophysik Potsdam (AIP), An der Sternwarte 16 14482 Potsdam, Germany\label{inst3}
\and
Leiden Observatory, Leiden University, PO Box 9513, NL-2300 RA Leiden, The Netherlands\label{inst4}
\and
Department of Physics, ETH Z\"urich, Wolfgang-Pauli-Strasse 27, 8093 Z\"urich, Switzerland\label{inst5}
\and
Department of Astronomy, University of Wisconsin, 475 N. Charter Street, Madison, WI 53706, USA\label{inst6}
\and
Centre for Astrophysics and Supercomputing, Swinburne University of Technology, Hawthorn, Victoria 3122, Australia\label{inst7}
\and
Aix Marseille Univ, CNRS, CNES, LAM, 38 rue Fr\'ed\'eric Joliot-Curie, 13013 Marseille, France\label{inst8}
\and
ESO Vitacura, Alonso de C{\'o}rdova 3107,Vitacura, Casilla 19001, Santiago de Chile, Chile \label{inst9}
}
\date{Received September 27th, 2021; accepted January 14th, 2022}

 
  \abstract
  {Hydrogen $\lya$ haloes (LAHs) are commonly used as a tracer of the circumgalactic medium (CGM) at high redshifts. In this work, we aim to explore the existence of $\lya$ haloes around individual UV-selected galaxies, rather than around $\lya$ emitters (LAEs), at high redshifts. Our sample was continuum-selected with $F775W\leq27.5$, and spectroscopic redshifts were assigned or constrained for all the sources thanks to the deepest (100- to 140-hour) existing Very Large Telescope (VLT)/Multi-Unit Spectroscopic Explorer (MUSE) data with adaptive optics. The final sample includes 21 galaxies that are purely $F775W$-magnitude selected within the redshift range $z\approx2.9$-$4.4$ and within a UV magnitude range $-20\leq M_{1500}\leq-18$, thus avoiding any bias toward LAEs. We tested whether galaxy's $\lya$ emission is significantly more extended than the MUSE PSF-convolved continuum component. We find 17 LAHs and four non-LAHs. We report the first individual detections of extended $\lya$ emission around non-LAEs. The $\lya$ halo fraction is thus as high as $81.0^{+10.3}_{-11.2}$\%, which is close to that for LAEs at $z=3$--$6$ in the literature. This implies that UV-selected galaxies generally have a large amount of hydrogen in their CGM. We derived the mean surface brightness (SB) profile for our LAHs with cosmic dimming corrections and find that $\lya$ emission extends to 5.4 arcsec ($\simeq40$ physical kpc at the midpoint redshift $z=3.6$) above the typical $1\sigma$ SB limit. The incidence rate of surrounding gas detected in $\lya$ per one-dimensional line of sight per unit redshift, $dn/dz$, is estimated to be $0.76^{+0.09}_{-0.09}$ for galaxies with $M_{1500}\leq-18$ mag at $z\simeq3.7$. Assuming that $\lya$ emission and absorption arise in the same gas, this suggests, based on abundance matching, that LAHs trace the same gas as damped $\lya$ systems (DLAs) and sub-DLAs. 
 }


   \maketitle
%

\section{Introduction}\label{sec:intro}

The circumgalactic medium (CGM) is the gas surrounding galaxies and corresponds to the reservoir of material fuelling galaxy formation. It serves as the interface between the interstellar medium (ISM) and the intergalactic medium (IGM). The boundary of the CGM has not been well-defined yet, but is commonly considered to be outside the ISM and inside the virial radius \citep[e.g.,][]{Tumlinson2017}. Gas is exchanged between the CGM and the ISM in galaxies via inflows and outflows. Outflows enrich the CGM with metals, and part of the outflowing gas is expected to be recycled through a halo fountain \citep[e.g.,][]{Oppenheimer2008}. Local observations suggest that the CGM is a multiphase medium in terms of its density, temperature, ionization state, kinematics, and metallicity \citep[e.g.,][]{Lanzetta1995a,Werk2014,Werk2016}. State-of-the-art cosmological simulations predict the time evolution of complicated multiphase structures in the medium, such as the Evolution and Assembly of Galaxies and their Environment (EAGLE) simulations \citep[e.g.,][]{Schaye2015, Rahmati2015,Oppenheimer2016} and the Illustris TNG50 simulations \citep[e.g.,][]{Nelson2019,Pillepich2019,Nelson2021arXiv2}, but physical mechanisms of gas exchanges, heating by feedback, and metal pollution are still poorly constrained. As a result, the typical mass distribution and kinematics of the various phases at play in the CGM are not well understood yet \citep[e.g.,][]{JDavies2020}.  

Observations of the CGM are traditionally based on transverse absorption-line studies (or tomographic mapping), providing the \HI\ column densities along the line of sight to bright background sources such as quasars and bright galaxies \citep[e.g.,][]{Wolfe1986,Tumlinson2017,Peroux2020}. This technique is sensitive to low \HI\ column densities, $N(\HI)$, and allows us to study a wide range of $N(\HI)$. Neutral gas clouds with $N(\HI)>2\times10^{20}$ cm$^{-2}$ are often referred to as damped $\lya$ systems \citep[DLAs, e.g.,][]{Wolfe2005}, while partially ionized gas regions with $10^{19}\leq N(\HI)\leq2\times10^{20}$ cm$^{-2}$ are identified as sub-DLAs \citep[e.g.,][]{Peroux2003a,Dessauges-Zavadsky2003}. Lyman-limit systems (LLSs), whose lower boundary corresponds to unit optical depth at the Lyman limit, have $1.6\times10^{17}\leq N(\HI)\leq10^{19}$ cm$^{-2}$ \citep[e.g.,][]{Tytler1982}. This method is sensitive to multiple ionization states of the gas, can provide information on metallicity and kinematics information, and is commonly used to study gas reservoirs from low to high redshifts, $z$, \citep[e.g.,][]{Steidel2002,Noterdaeme2012,Werk2014,Schroetter2016, Krogager2017,Zabl2019,Ho2020}. However, it is limited to the lines of sight to rare bright sources, which cannot probe the spatially resolved distribution of the CGM except for sources within the local Universe \citep[e.g.,][]{Tumlinson2017}.  

The CGM can be also observed in emission, which enables us to directly ``take a picture'' of gas in and around galaxies. While hot gas is routinely detected around very massive objects through its X-ray emission \citep[e.g.,][]{Spitzer1956,Li2013}, detecting emission from cooler gas is extremely challenging. Observations of the \HI\ CGM with 21cm emission are only possible in the local Universe. Even with forthcoming telescopes such as the Square Kilometre Array (SKA), 21cm direct mapping of the CGM will not be possible at $z\gtrsim2$. Popular probes of the CGM at $z\gtrsim1$ are spatially extended emission (i.e., haloes) in Ly$\alpha$ $\lambda$1216 \AA\ \citep[e.g.,][]{Momose2014,Wisotzki2016, Leclercq2017}, \MgII\ $\lambda\lambda$2796, 2803 \AA\ \citep[e.g.,][]{Rubin2011,Erb2012,Martin2013,Burchett2021,Zabl2021arXiv,Leclercq2022arXiv}, \OII$\lambda\lambda3726, 3729$ \citep[e.g.,][]{Yuma2013,Yuma2017,Epinat2018,Johnson2018}, \FeII$\lambda2365$, $\lambda2396$, $\lambda2612$, and $\lambda2626$ \citep[e.g.,][]{Finley2017a}, and \CII\ $\lambda$ 158 $\mu$m \citep[e.g.,][]{Fujimoto2019a,Ginolfi2020}. Among them Ly$\alpha$ haloes have several advantages. $\lya$ is intrinsically the brightest nebular recombination line of hydrogen atoms and often the strongest feature in the rest-frame UV spectra of galaxies. The emissivity of $\lya$ does not depend on metallicity, unlike the other emission lines. At $z\gtrsim2$, $\lya$ can be observable with ground-based telescopes. Moreover, $\lya$ haloes have been explored intensively in both theoretical and observational studies over wide redshift and mass ranges.

One of the popular methods of detecting $\lya$ haloes has been narrow-band (NB) stacking analyses, in particular at $z\gtrsim2$, due to the faintness of $\lya$ haloes and the cosmic dimming effect \citep[e.g.,][see also \citealt{Kakuma2021} and \citealt{Kikuchihara2021arXiv} for the NB intensity mappings]{Hayashino2004,Steidel2011, Matsuda2012,Momose2014, Matthee2016}. Until recently, individual detections have been limited to local galaxies, active galactic nuclei (AGNs), quasi-stellar objects (QSOs), and high-$z$ gravitationally lensed galaxies \citep[e.g.,][see also \citealt{Rauch2008} for 92-hour long-slit spectroscopy]{Keel1999,Kunth2003,Swinbank2007, Ostlin2009,Matsuda2011, Hayes2013}. Large samples of individual $\lya$ haloes around high-$z$ star-forming galaxies have become available thanks to wide-field optical integral field units (IFUs) like the Very Large Telescope (VLT)/ Multi-Unit Spectroscopic Explorer \citep[MUSE; e.g.,][]{Bacon2010,Wisotzki2016,Bacon2017,Herenz2017,Inami2017,Leclercq2017,Urrutia2019} and the Keck II telescope/ Keck Cosmic Web Imager \citep[KCWI;  e.g.,][]{Morrissey2018,Chen2021}. For instance, \citet{Leclercq2017} found 145 $\lya$ haloes around MUSE Ly$\alpha$ emitters (LAEs) at $z\approx3$--$6$. While the physical origins of $\lya$ haloes are still unclear, various scenarios have been suggested: CGM scattering for $\lya$ from star-forming regions \citep[e.g.,][]{Laursen2007,Zheng2011}, gravitational cooling radiation \citep[cold streams; e.g.,][]{Haiman2000,Fardal2001, Rosdahl2012}, star formation in satellite galaxies \citep[one-halo term; e.g.,][]{Zheng2011,Mas-Ribas2017a}, fluorescence \citep[photo-ionization; e.g., ][]{Furlanetto2005, Cantalupo2005,Kollmeier2010, Mas-Ribas2016}, shock heating by gas outflows \citep[e.g.,][]{Taniguchi2000}, major mergers \citep[e.g.,][]{Yajima2013}, and combination of the aforementioned \citep[e.g.,][see also \citealt{Ouchi2020} and reference therein]{Furlanetto2005,Lake2015,Smith2019,Byrohl2021,Garel2021,Mitchell2021}. Despite intensive observations of $\lya$ haloes and studies on their origins, no clear correlations between $\lya$ halo properties and their galaxy hosts' properties have been found at high redshifts \citep[e.g.,][]{Wisotzki2016,Leclercq2017}. So far, the most remarkable correlation is that between the $\lya$ peak velocity shift and the width of $\lya$ lines both at ISM and CGM scales, which supports a $\lya$ halo scenario of resonant scattering in the outflowing medium \citep[][see also \citealt{Verhamme2018} and \citealt{Chen2021}]{Claeyssens2019,Leclercq2020}. 

Although simulations have predicted the presence of the CGM around high-$z$ galaxies, most of the observational $\lya$ halo studies so far have focused on LAEs \citep[e.g.,][]{Leclercq2017,Wisotzki2016}. Only $\simeq$ 10 to 30\% of star-forming galaxies are LAEs at $z\simeq3$--$6$ \citep[e.g.,][]{Kusakabe2020}. $\lya$ haloes around UV-selected galaxies (Lyman break galaxies, LBGs) have previously been studied with NB stacks, which are biased s overdense regions \citep[e.g.,][]{Steidel2011,Xue2017}, and stacking analyses cannot provide information on the individual presence of $\lya$ haloes. The sample in KBSS-KCWI (Keck Baryonic Structure Survey using the KCWI) includes continuum-selected galaxies as well as LAEs, and their sample construction is not easy to characterize \citep{Chen2021}. Therefore, it is still unknown whether star-forming galaxies such as UV-selected galaxies generally have a $\lya$ halo. To minimize biases s LAEs and LBGs in overdense regions, a spectroscopically complete sample without preselection on the $\lya$ emission basis is required. Moreover, only the large field of view of MUSE with a long integration time enables us to do a volume-limited search for diffuse CGM emission at high redshifts. 

In this study, we construct a sample of UV-selected galaxies with spectroscopic redshifts, rather than $\lya$ emission selected-galaxies, in the MUSE eXtremely Deep Field \citep[MXDF;][Bacon et al. in prep.]{Bacon2021}, making use of more than 100-hour integration time of MUSE adaptive optics (AO) data. The high spatial resolution and the unprecedented depth are key to obtain spectroscopic redshifts for galaxies and spatial profiles of diffuse emission. Using the MUSE data, we investigate the existence of $\lya$ haloes around the galaxies and for the first time derive the $\lya$ halo fraction for UV-selected galaxies. We can resolve compact and faint haloes and assess the presence of a $\lya$ halo with the highest accuracy to date. This allows us to connect the separate views of \HI\ gas observed through $\lya$ emission and $\lya$ absorption, by extending the work by \citet{Rauch2008} and \citet{Wisotzki2018}.

The paper is organized as follows. In Section \ref{sec:data_sample}, we describe the data and the sample construction. Section \ref{sec:uvlah} presents methods and results of halo tests with individual $\lya$ surface brightness (SB) profiles, $\lya$ halo fractions, and completeness simulations. In Section \ref{sec:discussion}, we discuss implications from our $\lya$ fractions and incidence rates of $\lya$ emission compared with those of $\lya$ absorbers, as well as implications from non-$\lya$ haloes. Finally, the summary and conclusions are given in Section \ref{sec:conclusions}. Throughout this paper, we assume the Planck 2018 cosmological model \citep{PlanckCollaboration2020_2018VI} with a matter density of $\Omega_{\rm m} = 0.315$, a dark energy density of $\Omega_{\Lambda} = 0.685$, and a Hubble constant of $H_0 = 67.4$ km s$^{-1}$ Mpc$^{-1}$ ($h_{100} = 0.67$). Magnitudes are given in the AB system \citep{Oke1983}. All distances are in physical units (kpc), unless otherwise stated.

\section{Data and sample}\label{sec:data_sample}

We constructed a sample of UV-selected galaxies with spectroscopic redshifts to search for and investigate $\lya$ haloes, using a 3D data cube of the deepest MUSE observations with AO, the MXDF. The details of the MXDF data set are given in Section \ref{subsec:data}, the MXDF catalog and our sample selection are explained in Section \ref{subsec:sample}, and continuum subtractions for the MXDF data cube are described in Section \ref{subsec:consub}. 

\subsection{Data}\label{subsec:data}
The MXDF data were obtained as a part of the MUSE guaranteed time observations (GTO) program (PI: R. Bacon). The survey design of MXDF is presented in Bacon et al. in prep. \citep[see also][]{Bacon2021}. The field has a circular shape, centered at R.A.=53.$^{\circ}$16467 and DEC.=-27.$^{\circ}$78537 (J2000 FK5) with a radius of $r=44$ arcsec (1.7 arcmin$^2$), and is located inside the {\it Hubble} eXtreme Deep Field \citep[XDF;][]{Illingworth2013} with deep {\it Hubble} Space Telescope (HST) data. It is also covered by the MUSE-{\it Hubble} Ultra Deep Field (HUDF) with 10- to 30-hour MUSE integration in a 9 arcmin$^2$ area \citep[][]{Bacon2017}. The MXDF is the deepest MUSE survey with AO, reaching up to 140 hours at $r\lesssim29$ arcsec and 100 hours out to $r=31$ arcsec, while the outer edge has a 10-hour integration ($r\lesssim41$ arcsec). In this paper we use the very deep area with 100-to 140-hour integration, located inside a radius of 31 arcsec from the center of the field  \citep[0.84 arcmin$^2$, see Figure 1 in][]{Bacon2021}, which we used in this paper. The corresponding survey volume in this work is $4.0\times10^3$ cMpc$^3$ ($z=2.86$--$4.44$, excluding an AO gap, see below for more details).

\begin{figure}
   \centering
   \includegraphics[width=\hsize]{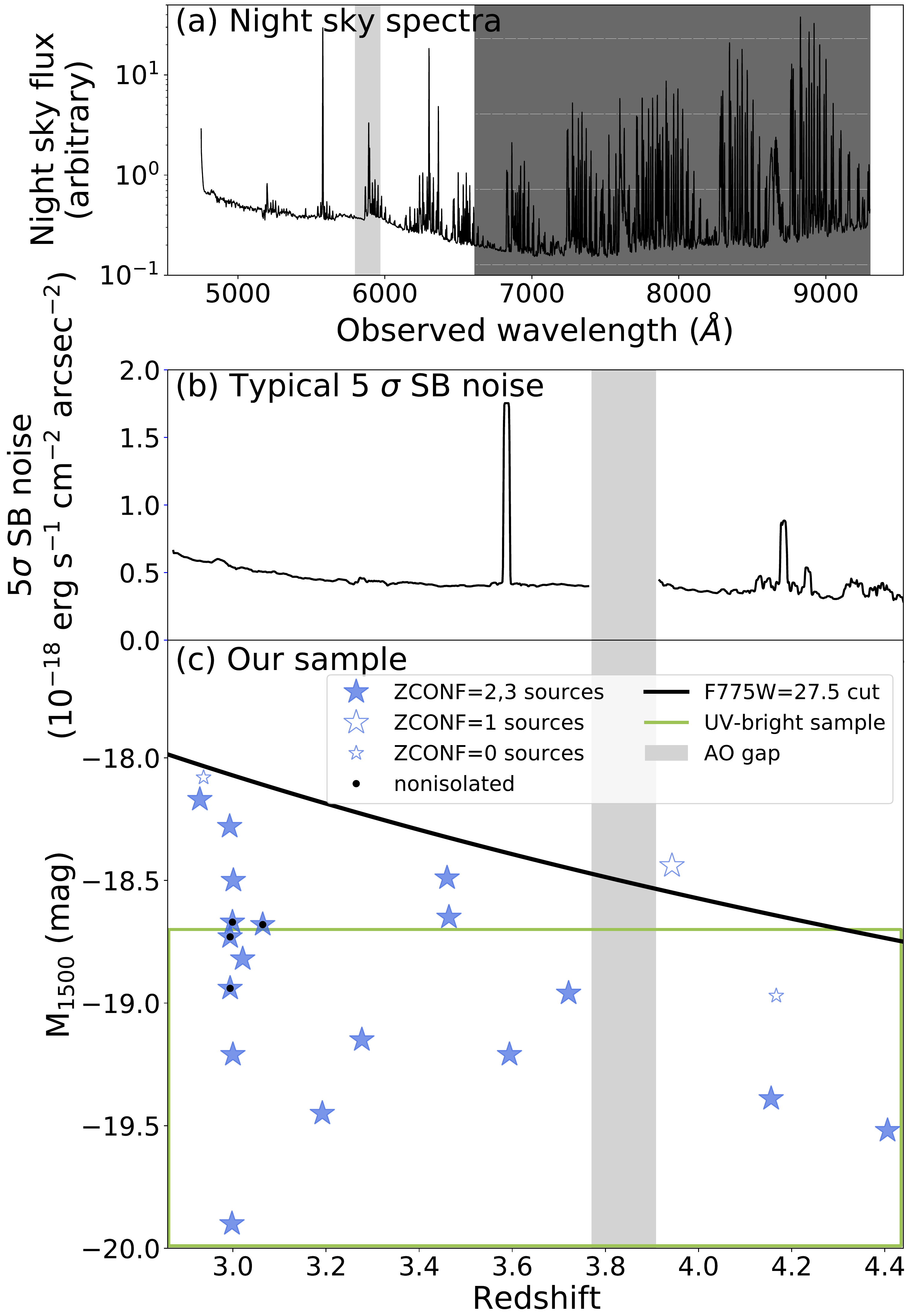}
      \caption{  {\it (a)} Night sky spectrum as a function of observed wavelength. The black line shows the night sky spectrum with an arbitrary normalization. The light gray and dark shaded areas indicate the wavelength range of the AO gap and the wavelength range that is not used in this work, respectively. 
                 {\it (b)} 5$\sigma$ noise surface brightness (SB) for the typical parameters for optimized NBs (see Section \ref{subsec:3dcog}) as a function of redshift (black line). The light gray shaded area indicates the redshift range of the AO gap.
                 {\it (c)} $M_{1500}$ as a function of $z$ for the sample. The large filled blue stars, large open blue stars, and small open blue stars indicate our sources with ZCONF=2,3, ZCONF=1, and ZCONF=0, respectively. The black dots show nonisolated galaxies (see Section \ref{subsec:sample}). The black and green lines represent the $F775W$ apparent magnitude cut, and the criteria for the UV-bright sample, respectively. 
              }
         \label{fig_sky_lambda_MUV_zs}
\end{figure}

The MUSE data cover the optical wavelength range from 4700 \AA\ to 9350 \AA, which is 50 \AA\ longer at the blue edge of the spectrum than that of the MUSE-HUDF Survey. It has an AO gap from 5800 \AA\ to 5966.25 \AA, and $\lya$ lines at $z=2.86$--$3.77$ and $3.91$--$6.65$ are observable. The spectral resolving power of MUSE varies from $R=1610$ to 3750 at 4700 \AA\ to 9350 \AA, respectively, and the median value for $\lambda\simeq4700$--$7000$ \AA, which is used for $\lya$ in this paper, is $R\simeq2200$. The FWHM of the Moffat point spread function \citep[Moffat PSF,][]{Moffat1969} is $0\farcs6$ at 4700 \AA\ and $0\farcs4$ at 9350 \AA\ (PSF calibrated for DR2 v0.8 catalog; see Bacon et al. in prep. for more details). The average $5\sigma$ surface brightness limit in the region with more than 100-hour depth is $1.3\times10^{-19}$ \sbl\ at 7000 \AA, which is not affected by OH sky emission, for an unresolved emission line with a line width of 3.75 \AA\, (3 spectral slices) and 1 arcsec$^2$ ($5\times5$ spaxels). The corresponding $5\sigma$ limiting flux for a point source with the same line width is $2.3\times10^{-19}$ \flcgs\ (see Bacon et al. in prep. for more details). Here we derived the variance in the same manner as that used for the MUSE-HUDF data, while the MUSE pipeline formally underestimates the noise in standard deviation by a factor of 1.6  \citep[2.7dev development version; see Section 4.6 in][for more details]{Weilbacher2020}. Figure \ref{fig_sky_lambda_MUV_zs}a shows a spectrum of the night sky emission as a function of wavelength. The background sky is relatively stable at $5000$--$7000$ \AA, compared to that at longer wavelengths.

\subsection{Catalog and sample selection}\label{subsec:sample}
\begin{table*}
\caption{Overview of the galaxy sample}\label{table:sample}     
\begin{tabular}{rrllllllllll}
\hline 
   RID &   MID &  line for $z_{\rm s}$ & $z_{\rm s}$ &   $z_{\rm p}$ (95\%)&  F775W & $M_{1500}$($1\sigma$)& ZCONF &Method &L17 & Isolated & UV-bright\\
   & & & &  & (mag) & (mag)& & & &\\
\hline 
  4587 &  8465 & Ly$\alpha$  &   3.001 &  $2.976^{+0.194}_{-0.196}$ &  27.05 &   $-18.5^{+0.05}_{-0.05}$ &     2 &  ODHIN & No & Yes & No \\
  4764 &  8339$^{\diamondsuit}$ & Ly$\alpha$$^{\diamondsuit,+}$  &     3.192$^{\diamondsuit}$ &  $3.291^{+0.189}_{-0.201}$ &  26.07 &  $-19.45^{+0.02}_{-0.02}$ &     3 &    ORIGIN$^{\diamondsuit}$& No & Yes & Yes \\
  4838 &  8469 & Abs. &      3.064 &  $3.092^{+0.178}_{-0.192}$ &   26.7 &  $-18.68^{+0.04}_{-0.04}$ &     2 &  ODHIN& No & No & No \\
  5479 &  7089 & Ly$\alpha$$^\star$  &     4.156 &  $3.893^{+0.277}_{-0.303}$ &  26.75 &  $-19.39^{+0.03}_{-0.02}$ &     3 &    ORIGIN& No & Yes & Yes \\
  6693 &      &   &    4.167 &     $4.21^{+0.25}_{-0.27}$ &  27.17 &  $-18.97^{+0.02}_{-0.02}$ &     0 &          & No & Yes & Yes \\  
  7067 &  7091 & Ly$\alpha$$^+$  &     4.406 &  $4.302^{+0.238}_{-0.242}$ &  26.76 &  $-19.52^{+0.01}_{-0.01}$ &     3 &    ORIGIN& Yes & Yes & Yes \\
  7847 &  8332 &  Ly$\alpha$$^+$  &    2.999 &  $2.998^{+0.182}_{-0.198}$ &  26.84 &  $-18.67^{+0.04}_{-0.04}$ &     3 &    ORIGIN& No & No & No \\
  7876 &   103$^*$ & Abs.$^\star$ &      2.994 &  $3.218^{+0.182}_{-0.198}$ &  26.85 &  $-18.73^{+0.04}_{-0.03}$ &     3 &    ORIGIN& No & No & Yes \\
  7901 &   180 &  Ly$\alpha$  &    3.460 &  $3.393^{+0.197}_{-0.203}$ &  27.28 &  $-18.49^{+0.03}_{-0.03}$ &     3 &    ORIGIN& Yes & Yes & No \\
  9814 &   149 &  Ly$\alpha$$^\star$  &    3.721 &  $3.602^{+0.228}_{-0.222}$ &   27.0 &  $-18.96^{+0.03}_{-0.03}$ &     3 &    ORIGIN& Yes & Yes & Yes \\
  9863 &   106 &  Ly$\alpha$$^\star$  &    3.277 &      $3.3^{+0.18}_{-0.19}$ &  26.51 &  $-19.15^{+0.03}_{-0.03}$ &     3 &    ORIGIN& Yes & Yes & Yes \\
  9944 & 103$^*$  &  Abs.$^\star$ &    2.994 &  $3.009^{+0.181}_{-0.189}$ &  26.49 &  $-18.94^{+0.03}_{-0.03}$ &  3    &  ORIGIN& No & No & Yes \\
 10018 &  6700 &  Ly$\alpha$$^\star$  &    2.998 &     $3.16^{+0.18}_{-0.19}$ &  25.67 &   $-19.9^{+0.01}_{-0.01}$ &     3 &    ORIGIN& Yes & Yes & Yes \\
 22230 &   163 & Ly$\alpha$  &     3.464 &  $3.337^{+0.193}_{-0.197}$ &  27.16 &  $-18.65^{+0.03}_{-0.03}$ &     2 &    ORIGIN& No & Yes & No \\
 22386 &  8518 &  Ly$\alpha$$^+$  &    2.929 &  $3.075^{+0.195}_{-0.205}$ &  27.25 &  $-18.17^{+0.05}_{-0.04}$ &     3 &  ODHIN& No & Yes & No \\
 22490 &  8377 &  Ly$\alpha$$^+$  &    3.000 &     $3.17^{+0.18}_{-0.19}$ &  26.21 &  $-19.21^{+0.03}_{-0.03}$ &     2 &    ORIGIN& No & Yes & Yes \\
 23124 &  7073 &   Ly$\alpha$  &   3.595 &  $3.628^{+0.202}_{-0.218}$ &  26.62 &  $-19.21^{+0.02}_{-0.02}$ &     3 &  ODHIN&  Yes & Yes & Yes \\
 23135 &  8392 &  Ly$\alpha$  &    3.943 &     $0.76^{+0.11}_{-0.38}$ &   27.5 &  $-18.44^{+0.03}_{-0.03}$ &     1 &    ORIGIN& No & Yes & No \\
 23408 &   174 & Ly$\alpha$$^+$  &     2.993 &  $3.125^{+0.195}_{-0.205}$ &  27.24 &  $-18.28^{+0.05}_{-0.05}$ &     2 &    ORIGIN& No & Yes & No  \\
 23839 &   118 & Ly$\alpha$$^\star$  &     3.021 &  $2.989^{+0.171}_{-0.179}$ &  26.63 &  $-18.82^{+0.02}_{-0.03}$ &     3 &    ORIGIN& No & Yes & Yes \\
54891 &   &   &   2.937 &  $2.306^{+0.184}_{-0.186}$ &  27.22 &  $-18.08^{+0.06}_{-0.05}$ &     0 &  & No & Yes & No \\
\hline 
\end{tabular}
\tablefoot{RID: ID in \citet{Rafelski2015}, MID: MUSE ID in MXDF (Bacon et al. in prep.), line for $z_{\rm s}$: line used to measure spec-$z$ for ZCONF$\geq1$ sources in the v0.8 catalog ($\lya$ and Abs. indicate $\lya$ emission and UV absorption lines, respectively; see Section \ref{subsec:sample} for 2 ZCONF$=0$ sources), $z_{\rm s}$: spec-$z$ in MXDF, $z_{\rm p}$: photo-$z$ with 95\% uncertainties in \citet{Rafelski2015}, $F775W$: F775W magnitude in \citet{Rafelski2015},  $M_{1500}$: absolute UV magnitude with 1$\sigma$ uncertainties, ZCONF: Confidence level for spec-$z$ for MUSE sources in MXDF, Method: the detection method in the MXDF (the two sources without a detection method are not included in the catalog but discussed in Section \ref{subsec:sample}), L17: included in \citet{Leclercq2017} or not, Isolated: isolated source or not (nonisolated), and UV-bright: included in the UV-bright sample or not. $^\diamondsuit$: RID=4764 has an update in a new version of the MXDF catalog (v0.9, Bacon et al. in prep.) as follows: MID=8357, line for $z_{\rm s}$=Abs., $z_{\rm s}$=3.188, and Method=ODHIN, which do not change the result of our halo test. $^*$:RID=7876 and 9944 were assigned for MID=103. $^\star$ ($^+$): Sources that have at least one emission line with S/N$\geq5$ (2$\leq$S/N$<$5) in the v0.9 catalog. We note that the $\lya$ emission line was used to measure the spec-$z$ for most of the sources in the v0.8 catalog, though 13 sources have at least one emission line other than Ly$\alpha$ (see $^\star$ and $^+$). It is because $\lya$ emission typically has the highest-S/N value among all spectral features. We note that it does not mean that our sample is biased s the LAE selection, and that the choices of lines used to measure spec-$z$ are not relevant for our halo test as we took a wide wavelength range when we optimized NBs (Section \ref{subsec:3dcog}). The spec-$z$ values are planned to be improved in an updated MXDF catalog.}
\end{table*}

\begin{figure*}
   \centering
   \includegraphics[width=0.91\hsize]{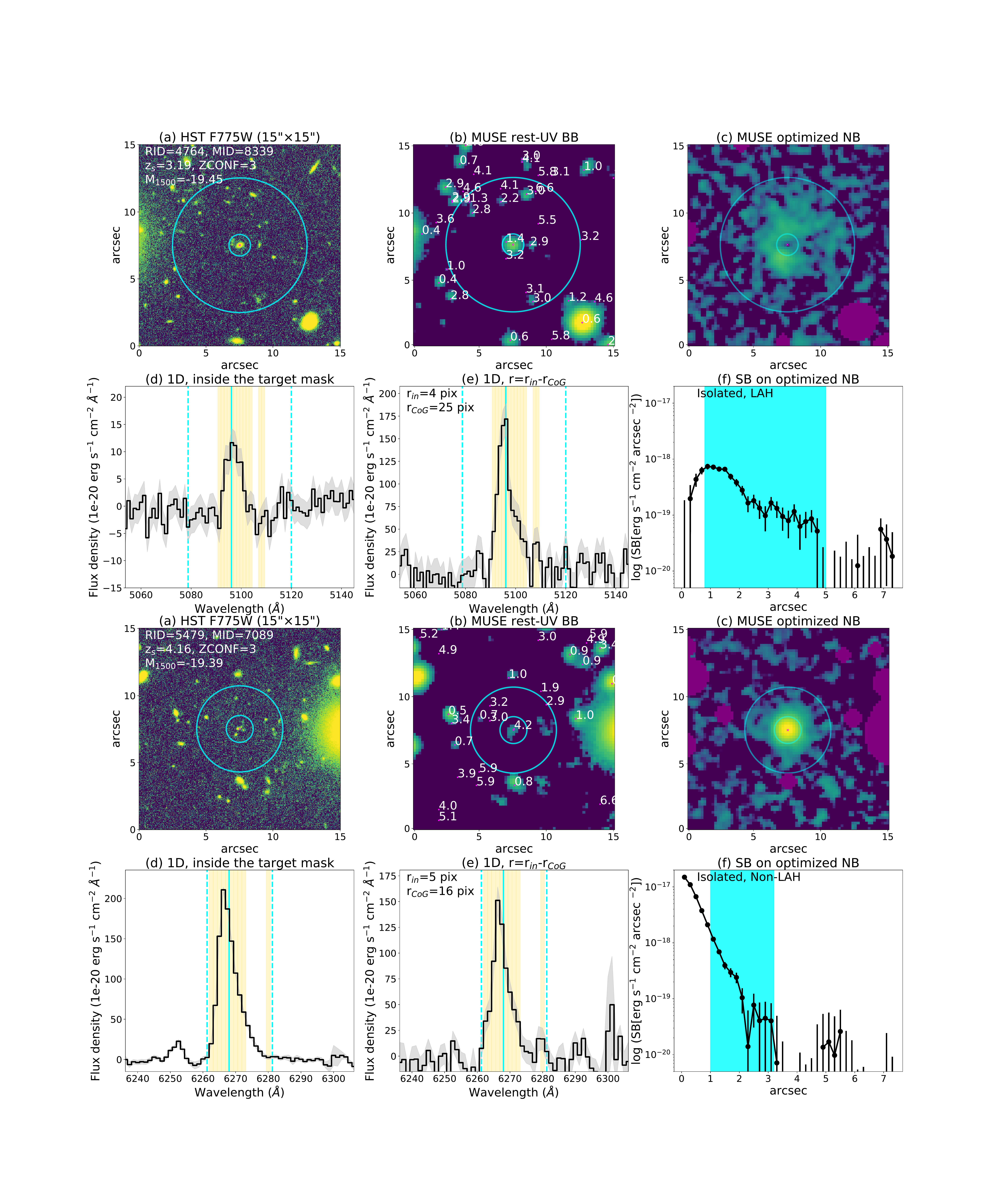}
      \caption{Examples of the overview with two sources (a LAH RID=4764 and a non-LAH RID=5479). The first two rows show RID=4764, while the last two rows show RID=5479. {\it (a):} HST $F775W$ image (rest-frame UV). The Rafelski's ID (RID), the MUSE ID (MID), $z_{\rm s}$, ZCONF, and M$_{1500}$ are indicated. The smaller and larger cyan circles present $r_{\rm in}$ and $r_{\rm CoG}$ (Sections \ref{subsec:mask} and \ref{subsec:3dcog}). {\it (b):} MUSE broad band image summed over the rest-frame UV range 1300--1800 $\AA$. The image is smoothed with a 2D Gaussian filter with a standard deviation of 1 pixel. Spectroscopic redshifts of sources in the MXDF catalog in the minicube are indicated. {\it (c):} Optimized narrow band for $\lya$ with the neighboring object mask (purple areas). The image is smoothed with a 2D Gaussian filter with a standard deviation of 1 pixel. The white contours indicate a SB of $2\times10^{-19}$ \sbl.  {\it (d):} 1D spectrum around the wavelength of $\lya$ extracted inside the target's continuum-component mask. The spectra and the 1$\sigma$ uncertainty are represented by the black line and the gray shaded area, respectively. The spectral width of the flux-maximized NB and the wavelength of $\lya$ converted from $z_{\rm s}$ in the MXDF catalog are shown by the cyan dashed and solid lines, respectively. The spectral slices used to produce the optimized NB are indicated by the yellow shaded areas for sources with high S/N spectral slices at r$=$r$_{\rm in}$--r$_{\rm CoG}$. {\it (e):} 1D spectrum around the wavelength of $\lya$ extracted in r$=$r$_{\rm in}$--r$_{\rm CoG}$. {\it (f):} Radial SB profile of $\lya$ emission in log scale measured on the optimized NB image. The cyan shaded areas represent the radial range, $r=r_{\rm in}$--$r_{\rm CoG}$ (see Section \ref{subsec:testhalo}). The panel shows whether the source is an isolated or nonisolated object, and whether it is a $\lya$ halo (LAH) or non-LAH. All images are 15 arcseconds each on a side. We note that the segmentation maps and the masks as well as zoomed-in HST $F775W$ cutouts (4$''$ $\times$ 4$''$) are shown in Figures \ref{fig_hstmask1}--\ref{fig_hstmask2}.}
         \label{fig_sample_cat1}
\end{figure*}

\begin{figure*}
   \centering
   \includegraphics[width=1.0\hsize]{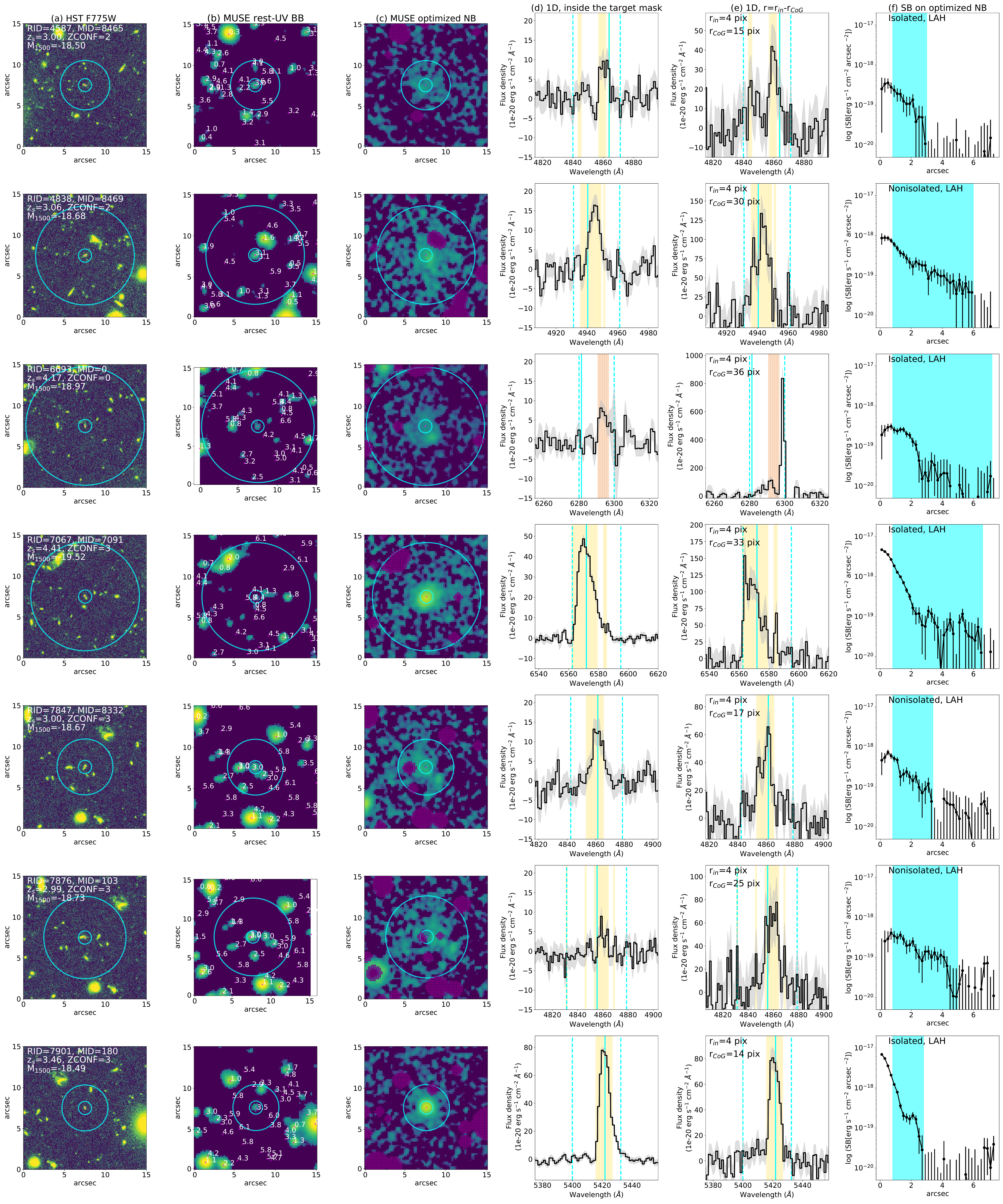}
      \caption{Overview of the first 7 sources in order of RID except for RID=4764 and 5479. Each row shows a different object. Panels (a)--(f) are the same as those of Fig. \ref{fig_sample_cat1}. In panel (a), MID=0 means that no MUSE ID is assigned in the MXDF catalog (i.e., ZCONF=0). In panels (d) and (e), the spectral slices used to produce the optimized NB are indicated by the yellow and orange shaded areas for sources with high S/N spectral slices at r$=$r$_{\rm in}$--r$_{\rm CoG}$ and those with low S/N spectral slices at r$=$r$_{\rm in}$--r$_{\rm CoG}$, respectively (see Section \ref{subsec:3dcog}). We note that the emission ring shown in the NB of RID=7876 is caused by the mask for a source with extended emission. Since the ring is located outside $r_{\rm CoG}$, it does not affect our results. 
      }
         \label{fig_sample_cat2}
\end{figure*}

\begin{figure*}
   \centering
   \includegraphics[width=1.0\hsize]{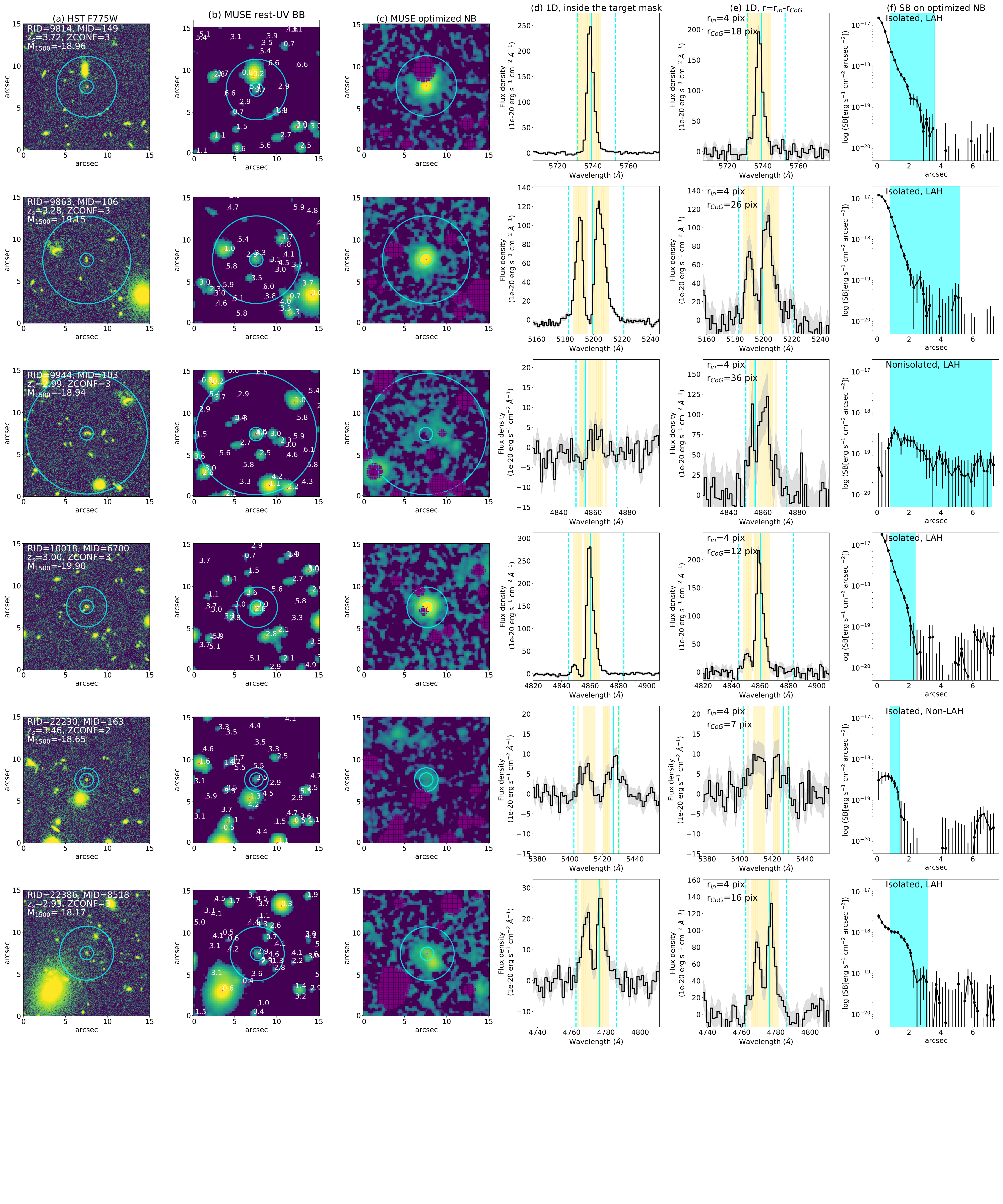}
      \caption{Same as Figure \ref{fig_sample_cat1}, but for the middle 6 sources in order of RID. Each row shows a different object. Panels (a)--(f) are the same as those of Fig. \ref{fig_sample_cat1}. We note that the emission rings shown in the NB of RID=9944 is caused by the mask for a source with extended emission. Although the ring feature of RID=9944 overlaps with r=r$_{\rm in}$--$r_{\rm CoG}$, this object is not included in the isolated sample (as well as RID=7876), and the main conclusion about the $\lya$ halo fraction is not affected by the ring. 
              }
         \label{fig_sample_cat3}
\end{figure*}

\begin{figure*}
   \centering
   \includegraphics[width=1.0\hsize]{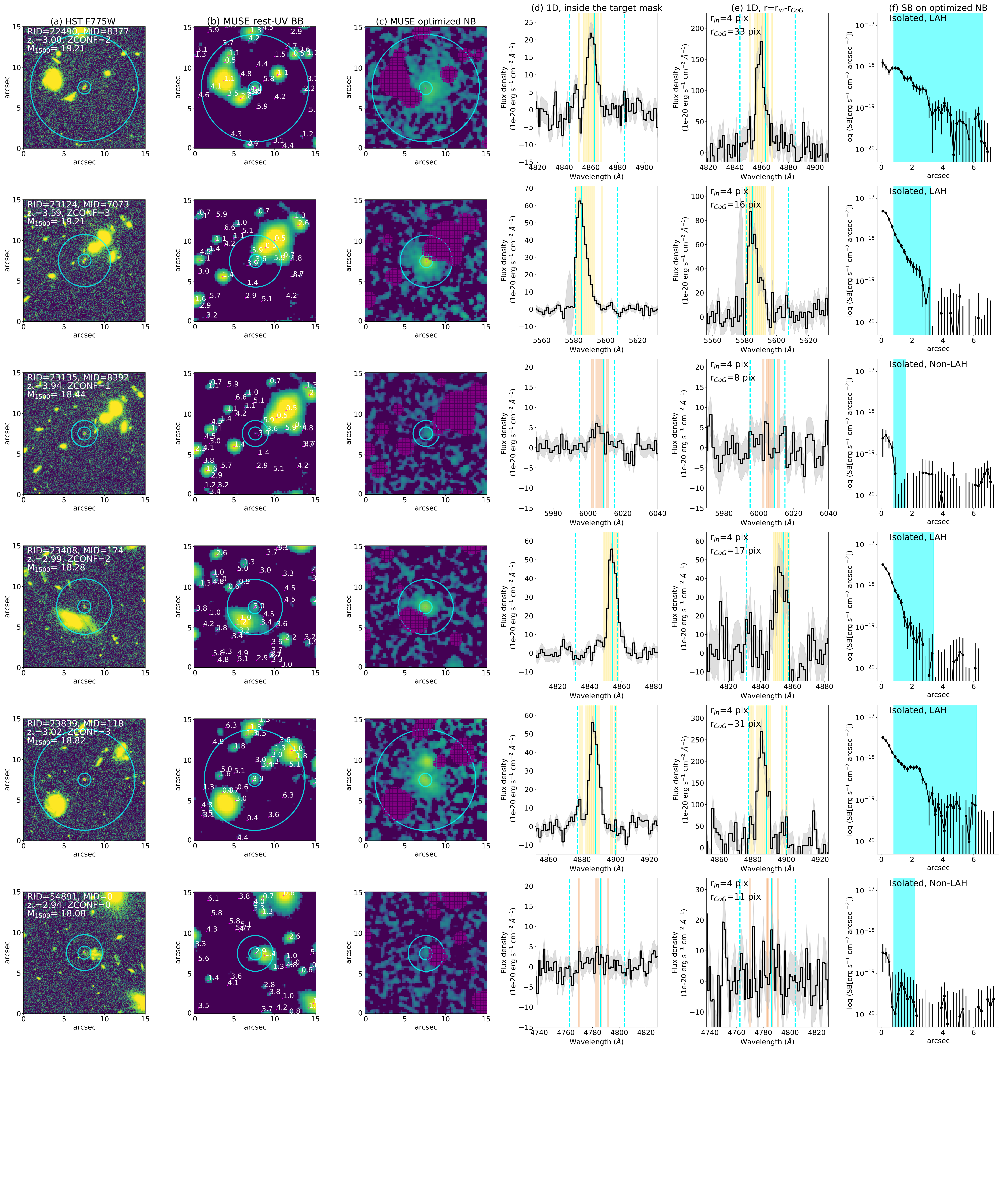}
      \caption{Same as Figure \ref{fig_sample_cat1}, but for the last 6 sources in order of RID. Each row shows a different object. Panels (a)--(f) are the same as those of Fig. \ref{fig_sample_cat1}.
              }
         \label{fig_sample_cat4}
\end{figure*}

The MXDF catalog was constructed in two ways, a blind emission search in the MUSE cube (ORIGIN method) and spectra extraction with prior information (ODHIN method). The two types of information were merged into one catalog (Bacon et al. in prep.). Appendix \ref{ap:catalog} gives a brief summary of the ORIGIN and ODHIN methods, the flow of the visual inspection, assignment of \citet{Rafelski2015}'s ID (RID), and the criteria of confidence levels (ZCONF) of spectroscopic redshifts (spec-$z$ or $z_{\rm s}$). The MUSE sources that are included in the catalog of \citet{Inami2017} keep the same ID (MID) in the MXDF catalog.  Most of the continuum-bright sources in \citet{Rafelski2015} are spectroscopically confirmed in the MXDF catalog. For instance, in the region with more than 100-hour integration ($r<31$ arcsec), 87\% of HST/Advanced Camera for Surveys Wide Field Channel (ACS/WFC) $F775W\leq27.5$ sources have reliable spectroscopic redshifts, ZCONF=2 or 3, and 7\% have ZCONF=1 redshifts. Moreover, 25 sources out of 26 $F775W\leq27.5$ galaxies, whose photometric redshifts (photo-$z$ or $z_{\rm p}$) are within the targeted redshift range below, have ZCONF=2 or 3 (see Figure \ref{fig_zp_zs} in Appendix \ref{ap:zpzs}). The spec-$z$ for the remaining 1 source was assigned as explained below. We note that we used a preliminary version of the MXDF catalog (DR2 v0.8) in this study.

We targeted galaxies at $z=2.86$--$4.44$ (excluding the $\lya$ redshift range in the AO gap) to investigate their $\lya$ haloes. There are two main reasons for this redshift range choice. First, at $z\leq4.7$ for $\lya$ ($4700$--$7000$ \AA), the background noise level is stable compared to those at longer wavelengths (see the black shaded area indicating $z\geq4.44$ in Figure \ref{fig_sky_lambda_MUV_zs}a). The 5$\sigma$ median SB noises for the typical parameters for the halo search (see Section \ref{subsec:3dcog}) at $z=2.86$--$4.44$ are shown in Figure \ref{fig_sky_lambda_MUV_zs}b. Second, at $z\leq4.44$, the HST band ACS/WFC $F775W$ can capture the UV continuum of galaxies without being contaminated by $\lya$ emission  \citep[the same threshold as those used in][]{Hashimoto2017b,Kusakabe2020}. In fact, \citet{McKinney2019} reported that some local galaxies show a wide $\lya$ absorption feature, which could extend to $\lambda\simeq1260$ \AA. These features were also confirmed at high redshifts \citep[for stacked MUSE LAEs at $z\simeq2.9$--$4.6$,][]{Feltre2020}. The threshold of $z=4.44$ for rest-frame 1270 \AA\ corresponds to 6900 \AA, at which $F775W$ filter has 3\% transmission \citep[see Section 4.3 in][for more details on the conservative choice of 1270 \AA]{Matthee2021}. In addition, we limited the field to the very deep area with 100- to 140-hour integration ($r<31$ arcsec) to have a homogeneous depth. 

In order to build a spectroscopically complete sample, our parent sample is based on the HST catalog in \citet{Rafelski2015} with deep 5$\sigma$ limiting magnitudes from 27.8 to 30.1 mag. The catalog is confirmed to be complete in UV at the redshift and $M_{\rm 1500}$ ranges that we used in this paper \citep[complete for $M_{\rm 1500}\leq-17$ mag at $z=2.9$--$4.4$, see Figure 2 in][]{Kusakabe2020}. Out of a total of 9969 sources in the catalog, 797 sources are within the footprint of the more than 100-hour integration region. Among them, 142 sources are brighter than 27.5 mag in $F775W$, which corresponds to a rest-frame UV band for sources at $z=2.86$--$4.44$. This apparent magnitude cut is 0.5 mag fainter than that for the HST prior detection in the \mosaic\ field in the MUSE-HUDF Survey in \citet{Inami2017}. As mentioned above, 123 sources (87\% of 142 $F775W\leq27.5$ sources) have a reliable spec-$z$ from MUSE with ZCONF=2 or 3, and 10 sources have a possible spec-$z$ (ZCONF=1). The remaining nine sources are not in the MXDF catalog as they do not show a clear feature in their spectra, but their continua are detected in the MXDF data cube  (see below for more details).   

Among the 123 ZCONF=2 or 3 sources, 18 galaxies are located at $z=2.86$--$3.77$ and $z=3.91$--$4.44$, while 105 galaxies are outside the redshift range. Unfortunately, one galaxy, which is categorized as an LAE in the catalog of \citet{Inami2017} and shows UV lines in the MXDF data cube, has $\lya$ emission in the AO gap. Due to the limited spatial resolution of MUSE, RID=7876 and 9944 were assigned to a unique MUSE source  (MID=103 at $z_{\rm s}$=2.99). Such nonisolated sources were separately treated as described later.   

Among the 10 ZCONF=1 sources, nine sources have a possible spec-$z$ at $z_{\rm s}\leq2.86$ (0.71, 1.10, 1.25, 1.67, 1.85, 1.91, 1.99, 2.34, and 2.67). Eight sources have a spec-$z$ consistent with their photo-$z$ in \citet{Rafelski2015}, and one source with $z_{\rm s}=1.25$ has a $z_{\rm p}=1.69^{+0.13}_{-0.14}$ (see Figure \ref{fig_zp_zs} in Appendix \ref{ap:zpzs}). All of them show clear continua at the blue edge of the MUSE spectra, implying that they should not be located at $z\geq2.86$. The remaining ZCONF=1 source has $z_{\rm s}=3.94$ (RID=23135).  It has \textsf{ORIGIN}-detected $\lya$ emission with a line flux signal-to-noise ratio (S/N) higher than 5 in the catalog (6.4). However, the $\lya$ center given by \textsf{ORIGIN} is spatially offset by $0\farcs3$ from the position in \citet{Rafelski2015}, and the spec-$z$ does not match with $z_{\rm p}=0.76^{+0.11}_{-0.38}$.
Moreover, the emission is close to the edge of the AO gap. We included the source in our parent sample, but it is not selected for a subsample with a UV absolute magnitude cut used to calculate the $\lya$ halo fractions, because of its faintness.

We inspected the MXDF data for nine sources that are not in the MXDF catalog. Seven sources of those nine sources at $z_{\rm p}=0.06^{+0.04}_{-0.05}$, $0.82^{+0.07}_{-0.55}$, $1.11^{+0.14}_{-0.58}$, $1.63^{+0.12}_{-0.13}$, $2.03^{+0.15}_{-0.16}$, $2.43^{+0.13}_{-0.54}$, and $2.55^{+0.16}_{-0.17}$ show clear continua at the blue edge of the MUSE spectra, which is not the case at $z\geq2.86$. One of the remaining two sources, RID=54891, has $z_{\rm p}=2.31^{+0.18}_{-0.19}$ and does not have an \textsf{ORIGIN}-detected emission line. However, it shows weak potential $\lya$ emission at $z=2.94$ with a line flux S/N of $3.9$. Because of the low S/N, the noisy spectrum, and the missmatch between $z_{\rm s}$ and $z_{\rm p}$, it was not categorized as a ZCONF=1 source and was not included in the MXDF catalog. To minimize a sample selection bias in this work, we assigned it a tentative spec-$z$ of $z_{\rm s}=2.94$. This object is not included in the UV-bright sample described later in this Section. We confirmed that our main results can hold up with the subset. The other remaining source at $z>2.86$ that is not in the MXDF catalog (RID=6693) has $z_{\rm p}=4.21^{+0.25}_{-0.27}$ and shows a break around 6250 \AA\ in the MUSE spectrum, which can be interpreted as a Ly$\alpha$ break. The source has an emission line in the 1D spectrum, which could be interpreted as a $\lya$ emission line at $z\simeq4.17$, but is contamination from a neighboring LAE. Then, we performed a cross correlation using the \textsf{MARZ} spec-$z$ software \citep{Hinton2016,Inami2017} on different spectra extracted from 1 to 5-pixel radius apertures and did not find realistic solutions. In order to obtain our best estimate of a tentative spec-$z$, we stacked the 1D spectra centered at wavelengths of different combinations of UV absorption lines among Si\,{\sc ii}$\lambda$1260.42, O\,{\sc i}$\lambda$1302.17,Si\,{\sc ii}$\lambda$1304.37, C\,{\sc ii}$\lambda$1334.53, Si\,{\sc iv}$\lambda\lambda$1393.75,1402.77, Si\,{\sc ii}$\lambda$1526.71, C\,{\sc iv}$\lambda\lambda$1548.20,1550.78, Fe\,{\sc ii}$\lambda$1608.45, and Al\,{\sc iii}$\lambda$1670.79. It is challenging to determine a spec-$z$ for faint sources at high redshifts with this method, but we note that we can get correct spec-$z$ for brighter sources at $z>4$ with ZCONF=3, RID$=5479$ and 7067. Although we did not find a significant absorption line at any redshift solution, $z_{\rm s}=4.167$ shows 3 consecutive pixels with more than a $1\sigma$ dip compared to the continuum ($1.2\sigma$, $1.8\sigma$, and $1.6\sigma$) as a result of stacking of Si\,{\sc ii}$\lambda$1260.42, C\,{\sc ii}$\lambda$1334.53, and Si\,{\sc iv}$\lambda$1393.75 lines. This is consistent with the interpretation of the Ly$\alpha$ break. We note that Si\,{\sc iv}$\lambda$1402.77 at $z_{\rm s}=4.167$ overlaps with a skyline. Therefore, we assigned a tentative spec-$z$, $z_{\rm s}=4.167$. We did not discard these two objects (RID=54891 and 6693) to minimize a selection bias. RID=6693 can meet criteria for a UV-bright subsample described later in this Section, while RID=54891 does not, which means that it does not affect our main conclusion. We took into account the effect of uncertainties in the spec-$z$ estimation (see Section \ref{subsubsec:flah_calc}).  

In total, we have 21 galaxies at $z=2.86$--$3.77$ and $z=3.91$--$4.44$ with $F775W\leq27.5$ mag as the parent sample listed in Table \ref{table:sample}. An overview of the 21 sources is given in Figures \ref{fig_sample_cat1} to \ref{fig_sample_cat4} (see Figure \ref{fig_sample_cat1} for enlarged panels for two sources as examples). We calculated the rest-frame UV magnitude ($M_{1500}$) and the UV slope ($\beta$) using $z_{\rm s}$, $F775W$, $F850LP$, and $F105W$ in the same manner as in \citet{Kusakabe2020}. The $M_{1500}$ ranges from $-19.9$ to $-18.1$ with an average value of $-18.9$, which is fainter than $M^*$ at $z\simeq3.7$, -20.88 mag \citep{Bouwens2015b}. The average $M_{1500}$ corresponds to a typical dark matter halo mass of $M_{\rm h}\simeq1\times10^{11}$--$2\times10^{11}\ {\rm M}_\odot$, which is estimated from a $M_{1500}$--$M_{\rm h}$ relation from the GALICS semi-analytic model in \citet{Garel2015}. The UV slopes range from $-2.3$ to $0.2$, with an average value of $-1.6$. The distribution of $M_{1500}$ and $z_{\rm s}$ of the sample is shown in Figure \ref{fig_sky_lambda_MUV_zs}c. Out of the 21 galaxies, six galaxies, RID=7067, 7901, 9814, 9863, 10018, and 23124 (MID=7091, 180, 149, 106, 6700, and 7073) were included in the $\lya$-selected sample for the previous MUSE-LAH study in the HUDF, and all of them are confirmed to be LAHs \citep{Leclercq2017}. Among the 15 new sources, five galaxies, RID=5479, 7876, 22230, 23408, and 23839 (MID=7089, 103, 163, 174, and 118), are included in the MUSE-HUDF catalog in \citet{Inami2017} as well as in the updated version in Bacon et al. (in prep.), but they were not selected in the sample in \citet{Leclercq2017}, due to their low ZCONF values ($\leq1$), low S/N values of $\lya$ ($<6$), or close neighboring objects. In order to validate that our sample is not biased s the LAE selection, we checked LAE fractions ($X_{\rm LAE}$) for our sample following \citet{Kusakabe2020}. Rest-frame equivalent widths of $\lya$ emission ($EW(\lya)$) were calculated with $\lya$ fluxes at the galaxy's stellar-component scale (see Appendix \ref{ap:ew_xlae} for a description of EW measurements and $X_{\rm LAE}$ calculations). With $EW(\lya)\geq20$ \AA, which is a common criterion of LAEs, the $X_{\rm LAE}$ for our entire sample ($-20\leq M_{1500}\leq-18.0$, $z=2.9$--$4.4$) is $0.33^{+0.11}_{-0.09}$. For a fair comparison, we also calculated the $X_{\rm LAE}$ with $EW(\lya)\geq65$ \AA\ for our entire sample as $0.14^{+0.09}_{-0.06}$, which is similar to $X_{\rm LAE}=0.04^{+0.02}_{-0.01}$, $0.07^{+0.04}_{-0.02}$, and $0.11^{+0.07}_{-0.04}$ for $EW(\lya)\geq65$ \AA\ and $-21.75\leq M_{1500}\leq-17.75$ at z=3.3, 4.1, and 4.7, respectively, in \citet{Kusakabe2020}. It suggests that our sample is unbiased. The small difference in the $X_{\rm LAE}$ between the two samples could be explained by the cosmic variance due to our small survey volume in the MXDF ($4\times10^3$ cMpc$^3$).

As in the case of RID=7876 and 9944 (MID=103), the MUSE spatial and spectral resolutions are not always high enough to disentangle the spec-$z$ assignment. It can also make it difficult to assign extended $\lya$ emission to close HST sources. Following \citet{Inami2017}, we checked if galaxies have close projected neighbors within $0\farcs6$ using the catalog of \citet{Rafelski2015}. We did not consider MUSE LAEs as neighbors here, because extended $\lya$ emission from satellites is one of the candidates for the powering sources of $\lya$ haloes. In our sample, 6 sources (RID=4764 , 4838, 5479, 7876, 9944, and 10018) have one HST-detected galaxy within $0\farcs6$. We used the MXDF catalog to check the spec-$z$ of the HST neighbors. The HST neighbors of RID=4764 and 10018 are located at different redshifts and do not contaminate the $\lya$ NB used in the halo tests (RID=10516 at $z_{\rm s}=1.42$ and RID=10046 at $z_{\rm s}=2.59$, respectively). Unfortunately, the HST neighbor of RID=5479, RID=5498, does not have a spec-$z$, but the photo-$z$, $z_{\rm p}=2.99^{+0.27}_{-0.28}$, is not close to the spec-$z$ of RID=5479, $z_{\rm s}=4.16$. Therefore, we regarded these three sources as isolated sources. Meanwhile, the HST neighbor of RID=4838 (RID=6666 at $z_{\rm s}= 3.06$) has a velocity offset of $\Delta V=-25.4$ km s$^{-1}$ from RID=4838, which indeed contaminates the $\lya$ NB used in the halo test (see Figures \ref{fig_sample_cat2} and \ref{fig_group1_NB}). They belong to a group of galaxies within the cosmic web detected with $\lya$ emission in \citet{Bacon2021}, see Appendix \ref{ap:nonisolated} for more details. Since both RID=7876 and 9944 are assigned to counterparts of MID=103 at $z_{\rm s}=2.09$, they are interpreted to share $\lya$ emission in their $\lya$ NBs. In addition, MID=103 has another neighbor within $0\farcs6$, RID=7847 at $z_{\rm s}=3.00$ in our sample ($\Delta V=458$ km s$^{-1}$ from RID=7876 and 9944), which leads to  mutual contamination of their $\lya$ NBs (see Figures \ref{fig_sample_cat2} and \ref{fig_sample_cat3}). Although RID=7847 is located further than $0\farcs6$ from the positions of RID=7876 and 9944, these three sources also belong to a larger structure of cosmic web found in \citet{Bacon2021}, as shown in Figure \ref{fig_group2_NB}. Therefore, RID=4838, 7847, 7876, and 9944 might live in a different kind of environment from the rest of our sample, which may affect the existence of a $\lya$ halo and $\lya$ halo properties. Additional details of four nonisolated sources are given in Appendix \ref{ap:nonisolated}. We calculated $\lya$ halo fractions for the isolated sources and the nonisolated sources as well as all sources in Section \ref{subsec:flah}. 

In Figure \ref{fig_sky_lambda_MUV_zs}c, we introduce an unbiased subsample, the "UV-bright sample", with a redshift range of $z=2.86$--$4.44$ with $-20.0\leq M_{1500}\leq-18.7$ mag. It contains 12 galaxies, of which 10 are isolated sources and two are nonisolated galaxies. 

\clearpage

In summary, our sample was built in two steps. First, we selected galaxies brighter than $F775W = 27.5$ in the catalog of \citet{Rafelski2015}, which corresponds to rest-UV for galaxies at $z\sim3$--$4$. Second, we applied a selection on the spectroscopic redshift of these objects to keep only galaxies with $z=2.86$--$4.44$. We note that all galaxies except nine have a spectroscopic redshift from the MUSE catalog. These nine galaxies have a photometric redshift and their continuum is detected in the MUSE cube. Seven of them have $z_{\rm p}=0$--$2.5$ with clear continuum detection at the blue edge of the MUSE cube: there is no sign of a Lyman break, which is very unlikely for sources at $z>2.86$. We did not include these sources in our sample. The remaining two sources were included in our sample, with ZCONF=0, in order to minimize a possible selection bias in favor of $\lya$ emitters. Therefore, we stress that our sample of 21 galaxies is rest-UV selected.

\subsection{Continuum subtraction}\label{subsec:consub}
For the following analysis, we provided $15\farcs0\times15\farcs0$ cutouts of the MUSE cube (minicube), the HST/ACS/WFC $F775W$ image \citep{Beckwith2006,Illingworth2013} and the HST segmentation map \citep{Rafelski2015} for each source. We used continuum-subtracted minicubes for most of the analysis in this paper, such as the $\lya$ narrow bands explained in Section \ref{sec:uvlah}.

The continuum subtraction is useful not only to investigate line emission, but also to remove neighboring sources around a targeted source. The continuum minicubes were provided from a spectral median filtering on the original minicubes in a 100-pixel spectral window ($\pm50$ pixels) in a similar manner to those in previous MUSE papers \citep[with a 200-pixel window, e.g.,][see also \citealt{Herenz2017} for a 150-pixel window]{Leclercq2017} and in addition excluded $\pm400$ km s$^{-1}$ around the $\lya$ wavelength. Although the method with a 200-pixel window was validated for the MUSE-HUDF and MUSE Wide data sets, in particular for LAEs, it can overestimate the continuum around a $\lya$ break ($\lya$ absorption) for UV-bright galaxies in the MXDF, since the MXDF data are deep enough to detect the UV continuum and the break. The oversubtraction due to the general 200-pixel window leads to artificial absorption at the position of a targeted UV source on a $\lya$ narrow-band image for some sources. We examined different settings for the continuum minicubes: 1) the general setting with a 200-pixel window, 2) a 200-pixel window with a mask around the $\lya$ wavelength ($\pm400$ km s$^{-1}$), 3) a 100-pixel window with a mask around the $\lya$ wavelength, and 4) a 60-pixel window with a mask around the $\lya$ wavelength. The $\lya$ masks cover 10 to 14 spectral slices at $z=3$ to $4.4$, respectively. 

We found a trend that the continua around $\lya$ were increasingly overestimated from settings 4) to 1) (see Figure \ref{fig_consub} in Appendix \ref{ap:consub}). This trend becomes stronger for galaxies showing a $\lya$ absorption feature. We also found that the 60-pixel window was too narrow to derive an accurate continuum. We confirmed that the continuum at $\lambda>1270$ \AA\ in the rest frame did not change among the four settings. Therefore, we adopted the third setting with a 100-pixel window, with a mask around $\lya$. In Appendix \ref{ap:consub}, we discuss two examples of the difference in the radial SB profiles for four different settings as described below (see Figure \ref{fig_consub}).

The obtained continuum minicube was subtracted from the original minicube to provide the continuum-subtracted minicube. Even with this optimized choice, we have a potential uncertainty in the continuum estimation around the $\lya$ wavelength at the position of a targeted UV-selected source, which however does not affect the extended $\lya$ emission beyond the galaxy's stellar component. For this reason, we did not use the spatial pixels on the main part of the galaxy's stellar component when we tested for the existence of $\lya$ haloes in the third step (see Figure \ref{fig_flow}).

\section{$\lya$ haloes around UV-selected galaxies}\label{sec:uvlah}

\begin{figure}
   \centering
   \includegraphics[width=0.95\hsize]{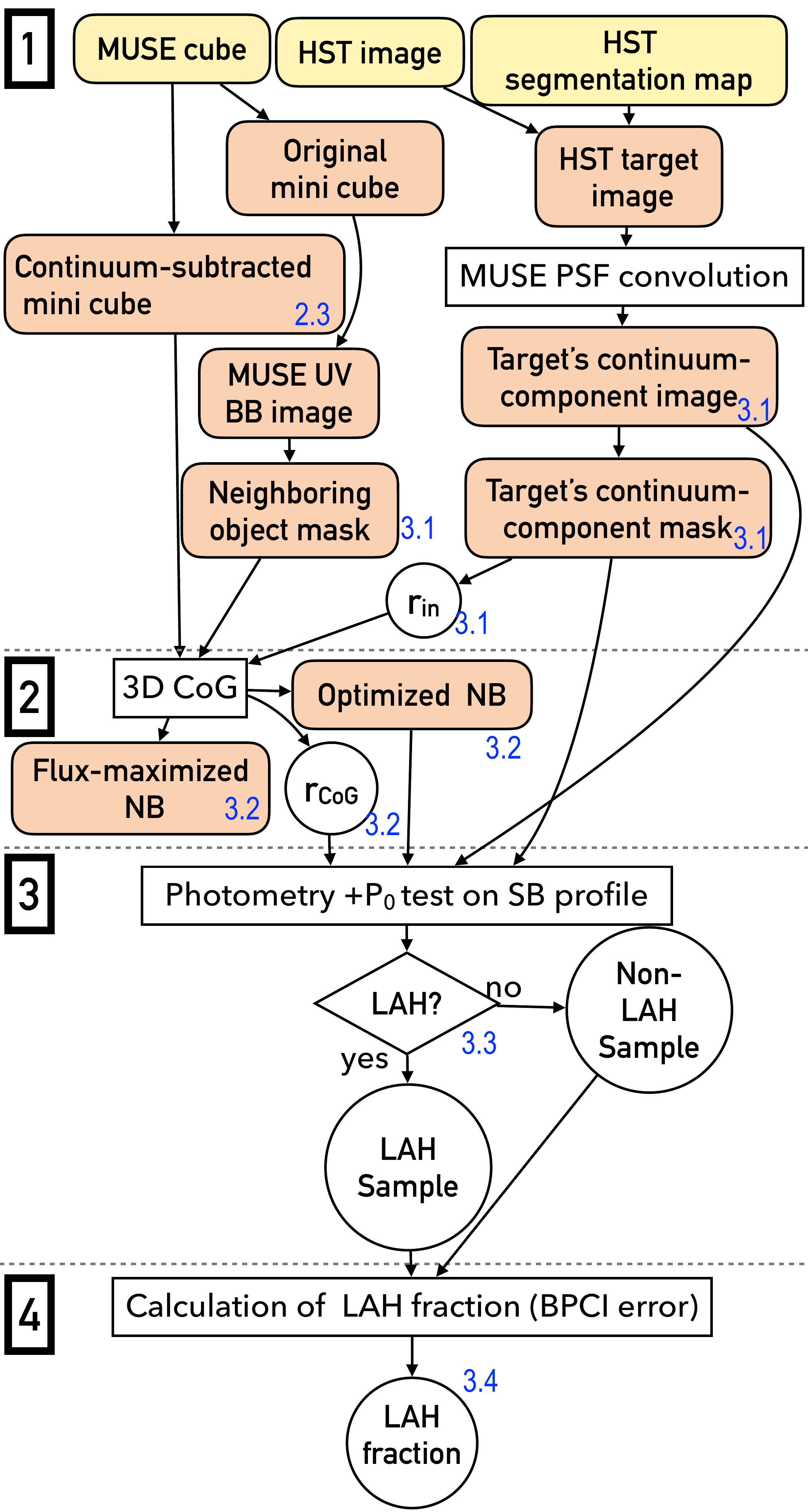}
      \caption{Flowchart of the main part of our analysis with the four steps. The input data are indicated by the yellow rounded boxes at the top of the panel, while output data are indicated by the orange rounded boxes. The analysis and classification are shown by rectangles and diamonds, respectively. The obtained samples and parameters are presented by circles. The details of each step are provided in Sections \ref{subsec:consub}, \ref{subsec:mask}, \ref{subsec:3dcog}, \ref{subsec:testhalo}, and  \ref{subsec:flah}. The numbers colored blue indicate the Sections as a reference. The flux-maximized NBs were used to obtain the average SB profiles in Section \ref{subsubsec:SBcorrection}. 
              }
         \label{fig_flow}
\end{figure}

The main part of our analysis consists of four steps. First, we created a mask for the continuum-like component (target's continuum-component mask) and a neighboring object mask for neighbors for each source (Section \ref{subsec:mask}). The target's continuum-component mask was used to obtain an inner radius. Second, we provided a flux-maximized $\lya$ NB with a curve-of-growth radius by tuning the width of the NB and  the radius to maximize the halo flux and then provided optimized NBs only with high-S/N spectral slices (Section \ref{subsec:3dcog}). Third, we tested for the existence of a $\lya$ halo around each source (Section \ref{subsec:testhalo}). Fourth, we derived the $\lya$ halo fractions (Section  \ref{subsec:flah}). The flow of the analysis is illustrated in Figure \ref{fig_flow}.

\subsection{Masks and inner radii}\label{subsec:mask} 

We used a target's continuum-component mask to extract the 1D spectrum within the galaxy's stellar-component scale and to define the inner radius used to adjust the $\lya$ NB (Section \ref{subsec:3dcog}). The continuum-component masks were created with the HST segmentation maps in a similar manner to those used to create object masks for HST prior extractions in \citet{Inami2017}. The HST segmentation map indicates areas in which galaxies are detected and defines the boundaries of the objects (more details are given in Appendix \ref{ap:hstmask}, see also Figures \ref{fig_hstmask1} and \ref{fig_hstmask2}). If a pixel does not belong to the target on the cutout segmentation map, we replaced the corresponding pixel of the $F775W$ cutout with zero and provided an HST target image. We convolved the HST target image with the MUSE PSF at the $\lya$ wavelength. Then, we resampled the PSF-convolved HST target image with the spatial resolution of MUSE, aligning it with the MUSE sky coordinate. We normalized the PSF-convolved image so that the peak became 1. To provide the target's continuum-component mask for each source, we applied a threshold value of 0.2 to the normalized convolved image (see Figures \ref{fig_hstmask1} and \ref{fig_hstmask2}). This threshold was used to extract the spectra from non-AO MUSE cubes for the HST-prior sources in the catalog in \citet{Inami2017}. For the MXDF data, the target's continuum-component masks typically include 66\% of the total fluxes in the target's continuum-component images. We defined an inner radius, $r_{\rm in}$, beyond which the normalized PSF-convolved profile is below 0.2. We used $r_{\rm in}$ in the SB profile measurements for the $\lya$ halo tests later, to exclude the SB inside the galaxy's stellar-component scale. The $r_{\rm in}$ ranges from 0.8 arcsec to 1.0 arcsec with an average value of 0.81 arcsec (in physical scales at different redshifts, 5.5 to 7.0 kpc with an average of 6.2 kpc).

We used a neighboring object mask to exclude pixels that might be affected by bright neighbors on a MUSE NB when we adjusted the $\lya$ NB and measured the $\lya$ radial SB profile. To mask the bright-continuum emission from the neighbors, we created a broad band (BB) image from the original MUSE minicube without continuum subtraction at rest-frame 1300 to 1800 \AA\ for the target (see Figures \ref{fig_sample_cat1} to \ref{fig_sample_cat4}). Then we clipped pixels whose S/N are higher than 20. With this threshold, we can mask the main part of continuum-bright galaxies on the optimized NBs and the flux-maximized NBs. A lower S/N threshold for the BBs, 10, changed the S/N values for extended $\lya$ emission on the optimized $\lya$ NBs used in the halo tests (see Section \ref{subsubsec:testhalo_method_sn}) because we lost the area with $\lya$ emission. However, it did not change the results for the tests for haloes for any sources, except for one (RID=23408). Since we used continuum-subtracted minicubes for most of the analysis and would like to keep as many spatial pixels as possible, the higher threshold of 20 was adopted. The two masks for each source are shown in Figures \ref{fig_hstmask1} and \ref{fig_hstmask2}. We note that the emission rings showing up at the left bottom of the NBs of RID=7876 and 9944 in Figures \ref{fig_sample_cat2} and \ref{fig_sample_cat3} are caused by a source whose emission extends beyond the neighboring-object masks. The main conclusion about the $\lya$ halo fraction is not affected by the ring features, since the objects are not included in the isolated sample.

\subsection{3D curve-of-growth for narrow bands}\label{subsec:3dcog}

To define a 3D volume in a minicube that includes all the potential flux of a $\lya$ halo, we adjusted the NB width and the annulus used for the halo photometry simultaneously, using the spec-$z$ and the spatial coordinate of the HST targets. The inner radius of the annulus is $r_{\rm in}$. The outer radius of the annulus is a curve-of-growth radius (CoG; $r_{\rm CoG}$), which is commonly used to derive the total $\lya$ flux from a NB image \citep[e.g.,][]{Drake2017b,Leclercq2017}. It is the minimum radius among consecutive 1pix-width annuli with increasing radius around a target, whose flux reaches or dips below zero \citep[see][for more details]{Drake2017b}.   

First, we created 400 NBs from a minicube with a combination of widths to redder and bluer wavelengths from 0 to 20 adjacent pixels around the $\lya$ wavelength. We applied the 2D neighboring object mask to the created NBs. Then, we derived $r_{\rm CoG}$ by annulus photometry with \textsf{PHOTUTILS} \citep{Bradley2021} on each NB and measured the flux between $r_{\rm in}$ and $r_{\rm CoG}$ centered at the spatial coordinate of the HST target. Finally, we chose the combination of a NB width and $r_{\rm CoG}$ that gives the highest $r_{\rm in}$--$r_{\rm CoG}$ flux. The flux-maximized NB widths range from 21.25 \AA\ to 48.75 \AA\ (4.1 \AA\ to 12.2 \AA, or 1014.9 km s$^{-1}$ to 3012.1 km s$^{-1}$) with the average of 31.8 \AA\ (7.5 \AA, or 1848.0 km s$^{-1}$) in observed wavelength (rest-frame wavelength or velocity difference). The $r_{\rm CoG}$ ranges from 1.4 arcsec to 7.2 arcsec (10.5 to 56.8 kpc) with the average of 4.2 arcsec (32.0 kpc). The $r_{\rm CoG}$ and NB width are shown in Figures \ref{fig_sample_cat1} to \ref{fig_sample_cat2}. The $r_{\rm CoG}$ and the flux-maximized NB were used to measure a surface brightness profile in the third step and in Section \ref{subsubsec:SBcorrection}, respectively.

We optimized the $\lya$ NBs by selecting spectral slices with S/N$>1.5$ for the flux between $r_{\rm in}$ and $r_{\rm CoG}$. We inspected the spectral slices visually and discarded the pixels whose signals might be enhanced by noise peaks. Out of 21, 18 sources have at least five spectral slices (6.25 \AA) with S/N$>1.5$. The total wavelength width ranges from 8.75 \AA\ up to 22.5 \AA, with an average value of 14.5 \AA\ in the observed frame (corresponding to 539.7 km s$^{-1}$ to 1298.2 km s$^{-1}$, and an average of 838.5 km s$^{-1}$). The remaining three sources (RID=6693, 23135, and 54891) do not have a sufficient number of high S/N pixels. In particular, RID=23135 and 54891 do not show a realistic line profile. Their pixels with S/N$>1.5$ seem to be affected by noise. Therefore, we created the optimized NBs for these sources using the 1D spectra inside the target's continuum-component mask. RID=23135 has six S/N$>1.5$ spectral slices, which were used to create the optimized NB. However, RID=6693 and 54891 have only four and two S/N$>1.5$ spectral slices, and the optimized NB was created from the five highest-S/N spectral slices. The chosen spectral slices are shown by yellow or orange shaded areas on the spectra in Figures \ref{fig_sample_cat1} to \ref{fig_sample_cat4}. With the average width of optimized NB and the average unmasked area between $r_{\rm in}$ and $r_{\rm CoG}$, the median 5$\sigma$ SB limit is estimated to be $4.1\times10^{-19}$ \sbl\ over the redshift range (see Figure \ref{fig_sky_lambda_MUV_zs}b), and the 5$\sigma$ flux limit is $3.1\times10^{-18}$ \flcgs\ over the redshift range. The optimized NB were used to measure a surface brightness profile in the third step.

\subsection{Test for the presence of $\lya$ haloes}\label{subsec:testhalo}
\begin{table*}
\caption{Summary of the tests for the existence of $\lya$ haloes.}\label{table:halotest}     
\begin{tabular}{rrrrrrrrr}
\hline
   RID &  $r_{\rm in}$ &  $r_{\rm CoG}$ &     S/N($r_{\rm in}$-$r_{\rm CoG}$) &       $p_{0}$ & DoF & R. $\chi^2$ & R. $\chi^2$ ($p_{0}$=0.05) & LAH \\
& (arcsec/kpc) & (arcsec/kpc) & & & & & & \\
\hline
  4587 &           0.8/6.31 &            3.0/23.65 &   5.75 &  6.04e-03 &   5 &        3.26 &                2.21 &  Yes \\
  4764 &           0.8/6.19 &            5.0/38.67 &  16.44 &  2.84e-49 &  17 &       16.39 &                1.62 &  Yes \\
  4838 &           0.8/6.27 &            6.0/47.01 &  11.94 &  4.65e-19 &  19 &        6.98 &                1.59 &  Yes \\
  5479 &           1.0/7.0  &            3.2/22.39 &  12.94 &  3.78e-01 &   6 &        1.07 &                 2.1 &   No \\
  6693 &           0.8/5.59 &            7.2/50.31 &   4.84 &  2.18e-22 &   8 &       15.11 &                1.94 &  Yes \\
  7067 &           0.8/5.45 &            6.6/44.99 &  13.63 &  6.25e-30 &  14 &       12.51 &                1.69 &  Yes \\
  7847 &           0.8/6.31 &            3.4/26.81 &   9.23 &  4.39e-06 &  10 &        4.33 &                1.83 &  Yes \\
  7876 &           0.8/6.31 &            5.0/39.45 &   9.13 &  1.39e-12 &  13 &        6.52 &                1.72 &  Yes \\
  7901 &           0.8/6.02 &            2.8/21.06 &  17.48 &  1.09e-05 &   7 &        5.01 &                2.01 &  Yes \\
  9814 &           0.8/5.86 &            3.6/26.35 &   20.0 &  1.28e-16 &   9 &       10.61 &                1.88 &  Yes \\
  9863 &           0.8/6.13 &            5.2/39.86 &  14.46 &  4.08e-05 &   7 &        4.57 &                2.01 &  Yes \\
  9944 &           0.8/6.31 &            7.2/56.81 &   9.22 &  1.12e-13 &  17 &        5.85 &                1.62 &  Yes \\
 10018 &           0.8/6.31 &            2.4/18.93 &  27.98 &  7.44e-03 &   5 &        3.16 &                2.21 &  Yes \\
 22230 &           0.8/6.02 &            1.4/10.53 &   6.72 &  2.73e-01 &   2 &         1.3 &                 3.0 &   No \\
 22386 &           0.8/6.35 &            3.2/25.41 &  16.05 &  1.32e-39 &   7 &       28.53 &                2.01 &  Yes \\
 22490 &           0.8/6.31 &            6.6/52.04 &  13.01 &  1.24e-38 &  16 &       13.98 &                1.64 &  Yes \\
 23124 &           0.8/5.93 &            3.2/23.74 &   14.9 &  1.76e-16 &   8 &       11.51 &                1.94 &  Yes \\
 23135 &           0.8/5.72 &            1.6/11.44 &    <0.0 &       Nan &  Nan &         Nan &                 Nan &   No \\
 23408 &           0.8/6.31 &            3.4/26.83 &   6.72 &  3.54e-02 &   4 &        2.58 &                2.37 &  Yes \\
 23839 &           0.8/6.3 &            6.2/48.79 &  14.21 &  2.99e-68 &  13 &       27.44 &                1.72 &  Yes \\
 54891 &           0.8/6.35 &            2.2/17.46 &   1.29 &       Nan &  Nan &         Nan &                 Nan &   No \\
\hline
\end{tabular}
\tablefoot{   RID: ID in \citet{Rafelski2015},  $r_{\rm in}$: inner radius for the test defined in Section \ref{subsec:mask},  $r_{\rm CoG}$: curve of growth radius used as the outer radius for the test (Section \ref{subsec:3dcog}),  S/N($r_{\rm in}$-$r_{\rm CoG}$): S/N of the flux measured at $r=r_{\rm in}$-$r_{\rm CoG}$ (halo candidates have a S/N($r_{\rm in}$-$r_{\rm CoG})\geq3$),  DoF: degrees of freedom for the test (Section \ref{subsubsec:testhalo_method}), $p_{0}$: probability of the SB profile at $r=r_{\rm in}$-$r_{\rm CoG}$ to match that of the target's continuum-component image ($\lya$ haloes are confirmed if $p_{0}$<0.05), R. $\chi^2$: reduced $\chi^2$ of the SB fitting test, which is converted to a $p_{0}$ value, R. $\chi^2$ ($p_{0}$=0.05): reduced $\chi^2$ corresponding to the threshold $\hat{p_{0}}$=0.05,  LAH: Whether the object has a $\lya$ halo or not. The SB profiles of RID=23124 and 54891 were not tested due to the low S/N($r_{\rm in}$-$r_{\rm CoG}$). We note that RID=23135 has a negative S/N($r_{\rm in}$--$r_{\rm CoG}$).}
\end{table*}

We tested for the existence of $\lya$ haloes with two steps. First, we checked the S/N of fluxes in the annulus from $r_{\rm in}$ to $r_{\rm CoG}$ to select LAH candidates (Section \ref{subsubsec:testhalo_method_sn}). Second, we tested for the existence of $\lya$ haloes with the radial SB profiles at $r=r_{\rm in}$--$r_{\rm CoG}$ (Section \ref{subsubsec:testhalo_method}), the results of which are described in Section \ref{subsubsec:testhalo_results}. We investigated non-LAH objects in Section \ref{subsubsec:testhalo_nolahs} and Appendix \ref{ap:comp} with completeness simulations of the halo tests.

\subsubsection{Selection of LAH candidates based on the S/N at $r=r_{\rm in}$--$r_{\rm CoG}$}\label{subsubsec:testhalo_method_sn}

We measured the S/N of fluxes in the annulus from $r_{\rm in}$ to $r_{\rm CoG}$, S/N($r_{\rm in}$--$r_{\rm CoG}$), on the optimized NBs using \textsf{PHOTUTILS} with the two masks. The sources that have a S/N higher than or equal to three were regarded as $\lya$ halo candidates, which were also visually inspected. The top panel of Figure \ref{fig_halotest} shows the histogram of S/N($r_{\rm in}$--$r_{\rm CoG}$). Among 21 galaxies, 19 are LAH candidates. The two objects with low S/N values are RID=23135 and 54891, whose NBs were optimized based on the 1D spectra at $r\leq r_{\rm in}$ in Section \ref{subsec:3dcog}. They have low S/Ns on the 1D spectra at $r=r_{\rm in}$--$r_{\rm CoG}$, and the low values of S/N($r_{\rm in}$--$r_{\rm CoG}$) on 2D NBs were expected.

\subsubsection{Test with $\lya$ radial SB profiles}\label{subsubsec:testhalo_method}

In order to confirm the presence of extended $\lya$ emission, we checked if the emission is {\it not} from the outer wing of bright central $\lya$ emission. To do so, we assessed the null hypothesis that the source does not have $\lya$ emission more extended than the PSF-convolved continuum-like component. In other words, we measured the significance of the deviation of the observed radial SB profiles from those of the target's continuum-component images provided in Section \ref{subsec:mask} ($p_0$ test). Our method carefully prevents potential uncertainties of SB profiles at $r< r_{\rm in}$ due to the continuum subtraction. We measured the observed SB profile on the optimized NB using \textsf{PHOTUTILS} with two masks. The default width of the radial bins was one MUSE spatial pixel, but the bins were combined with the outer bins to have $S/N\geq2$. If spatial pixels at large radii close to $r_{\rm CoG}$ had too low S/Ns, which reduced the combined S/N and made it lower than two, we did not use the pixels in the fitting. We note that it did not change the results of the halo test for the sample but gives lower $p_0$ values (higher reduced $\chi^2$), which helps detect diffuse haloes combined with other conditions. 
We measured the effective area considering the masks and derived the SB profiles by dividing the fluxes with the effective area. Then, we fit the observed SB profile with that of the target's continuum-component in the range of $r_{\rm in}$ to $r_{\rm CoG}$ and measured the reduced $\chi^2$, which was converted into the probability of the null hypothesis, $p_0$, with \textsf{scipy.stats.chi2.sf}. Here, the number of degrees of freedom (DoF) is 1 smaller than the number of the combined radial bins. The free parameter is the amplitude of the SB (i.e., flux). If a source has a $\lya$ halo, the SB profile from $r_{\rm in}$ to $r_{\rm CoG}$ is significantly different from that of the target's continuum-component, which makes $p_{0}$ small. Our threshold for the probability of the null hypothesis, $\hat{p_0}$, is 0.05, which is the same as that used in \citet{Leclercq2017}. 

We visually investigated the sources that have a larger reduced $\chi^2$ than the threshold (see Table \ref{table:halotest}). We note that the results of the null hypothesis tests were the same if we used different thresholds of S/N for the radial binning, 1.0, 1.5, and 2.0, and when we adopted different $\hat{p_0}$ from 0.04 to 0.1. We also checked the results for the four different settings in the continuum subtraction. The results were the same for all the sources except for one object (RID=23408). These checks imply that our tests for the existence of $\lya$ haloes are robust. The results of the tests are described in Section \ref{subsubsec:testhalo_results} (see also Table \ref{table:halotest} and Figures \ref{fig_halotest} and \ref{fig_SBtest}).

With this test, we could miss very faint or diffuse $\lya$ haloes below the detection limits and $\lya$ haloes whose shapes are very different from our circular assumption (see Section \ref{subsubsec:testhalo_nolahs} for more details). The S/N$\geq3$ criterion corresponds to SB$\geq3.2\times10^{-19}$ \sbl\ and flux $\geq2.2\times10^{-18}$ \flcgs, in the case of the average width of the optimized NB and the average unmasked area between $r_{\rm in}$ and $r_{\rm CoG}$ (see Section \ref{subsec:3dcog} and Figure \ref{fig_sky_lambda_MUV_zs}b).

As a final note, our method is different from \citet{Leclercq2017}. Their method was designed for galaxies with strong central $\lya$ emission which can {\it de facto} always be fit by the same exponential profile as the continuum. Our galaxies are UV-selected and do not always show strong central $\lya$ emission. In practice, their inner \lya{} emission is not well modeled by the continuum so that the method of \citet{Leclercq2017} is not applicable to our sample. The higher signal-to-noise ratio of our data also makes their modeling assumptions more difficult in practice in our sample. We thus use a more general method, which we verified recovers the LAHs of \citet{Leclercq2017} for the six objects we have in common (and which thus all have strong central \lya{} emission).

\subsubsection{Results of the halo test}\label{subsubsec:testhalo_results}

\begin{figure} 
   \centering
   \includegraphics[width=\hsize]{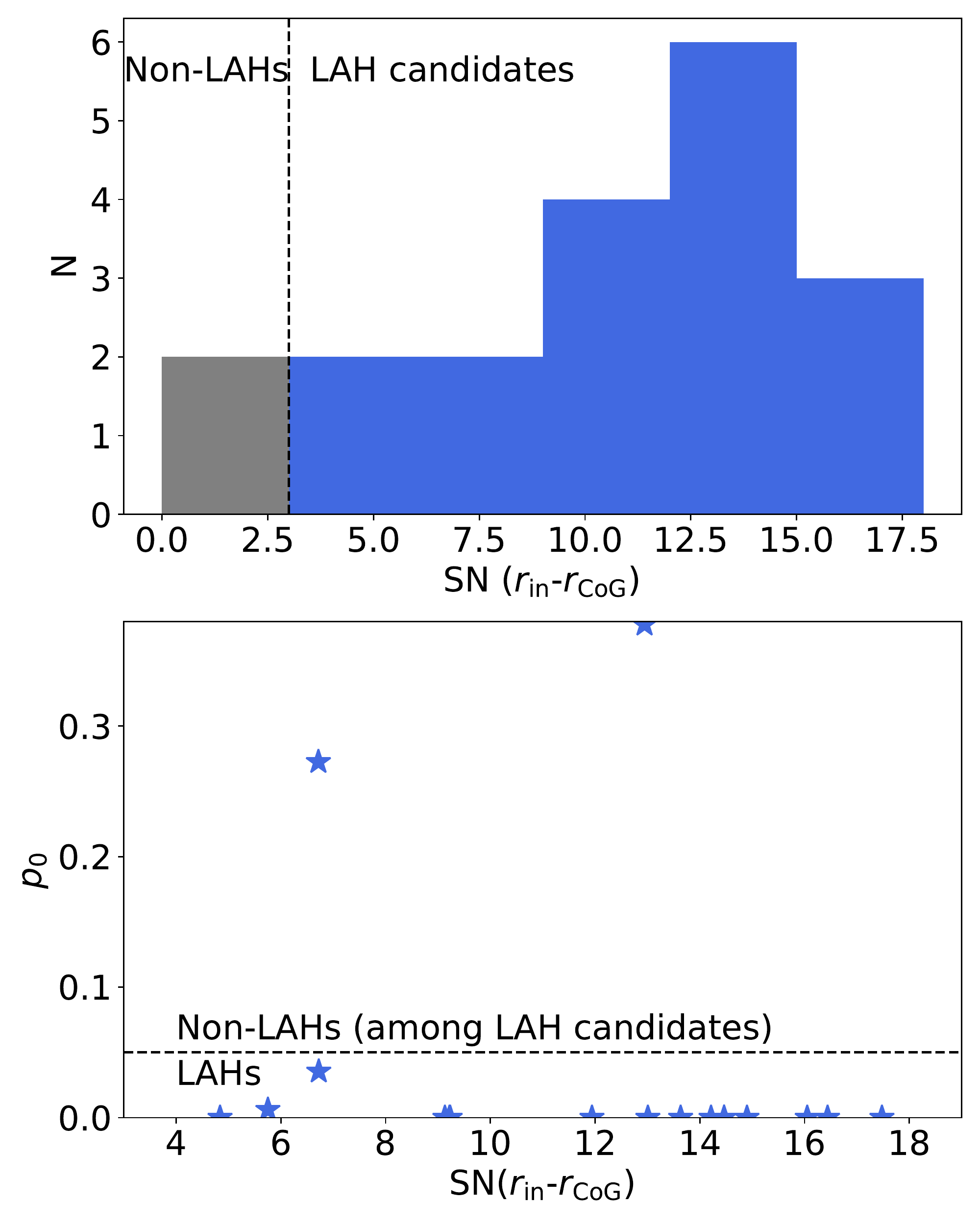}
      \caption{Results of the tests for the existence of a $\lya$ halo. {\it Top panel:} histogram of the flux S/Ns between $r=r_{\rm in}$ and $r_{\rm CoG}$ measured on the optimized $\lya$ NBs. Blue and gray histograms indicate galaxies above and below the threshold of S/N=3 (black dashed line), Non-LAHs and LAH candidates, respectively. RID=23135 has a negative S/N($r_{\rm in}$--$r_{\rm CoG}$) and is counted as S/N($r_{\rm in}$--$r_{\rm CoG}$)$=0$ in the plot.  {\it Bottom panel:} probability of the null hypothesis that the SB profile at $r=r_{\rm in}$-$r_{\rm CoG}$ matches that of the target's continuum-component profile vs. flux S/N between $r=r_{\rm in}$ and $r_{\rm CoG}$ for the LAH candidates. A black dashed line represents the border of LAHs and Non-LAHs.
      }
         \label{fig_halotest}
\end{figure}

\begin{figure*} 
   \centering
       \includegraphics[width=1.0\hsize]{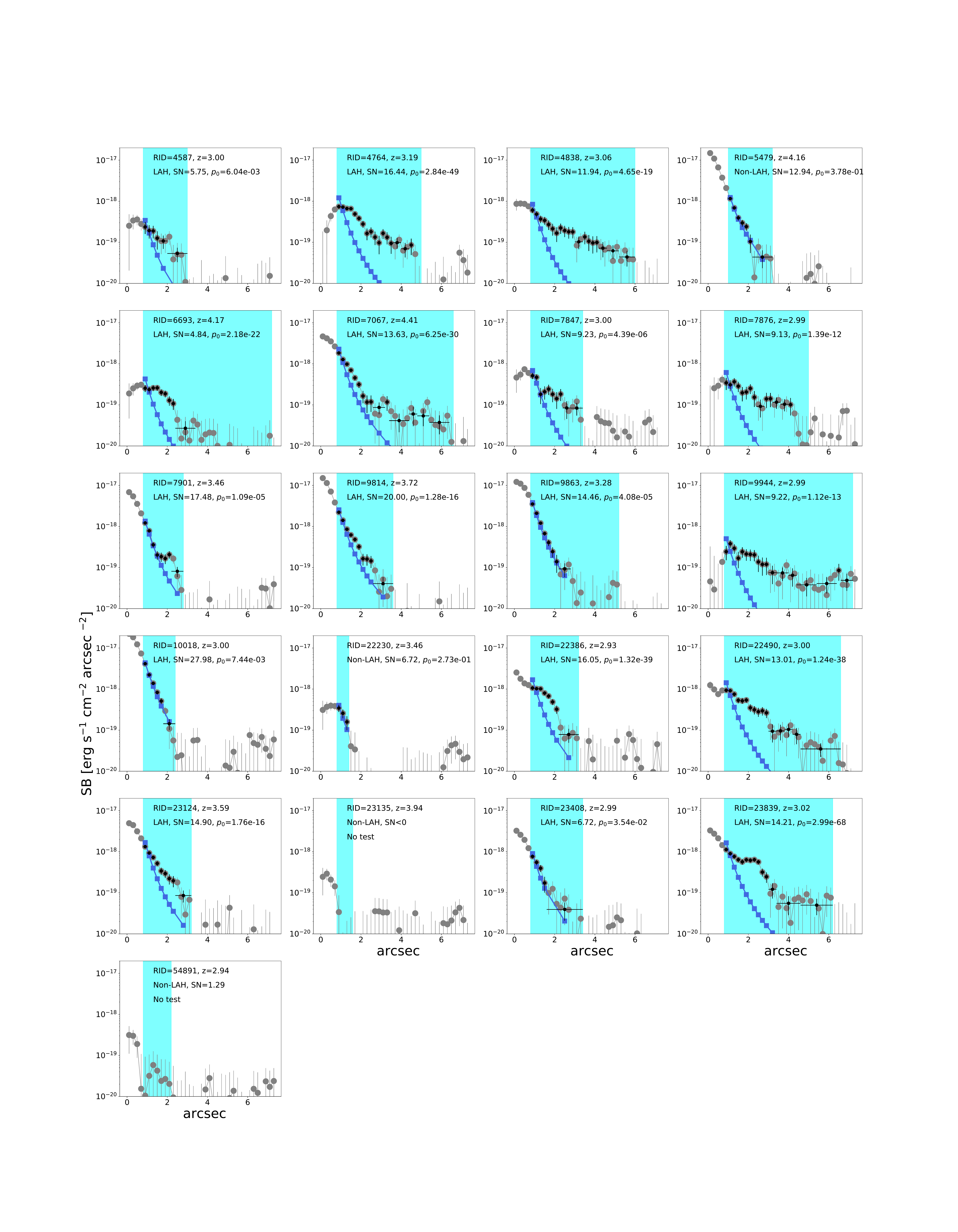}
      \caption{SB profiles from the optimized NBs with results of the $\lya$ halo tests. The gray, black, and blue points indicate the observed radial profiles, those used for the fit (with binning if needed), and the best-fit SB profiles of the PSF-convolved continuum-like component, respectively. The x error bars of the black points mark the range for each (combined) radial bin. The cyan shaded areas represent the radial ranges, $r=r_{\rm in}$--$r_{\rm CoG}$ (see Section \ref{subsec:testhalo}). The classification of LAH or non-LAH, the S/N of the flux at $r=r_{\rm in}$--$r_{\rm CoG}$, and the probability of the null hypothesis that the SB profile at $r=r_{\rm in}$-$r_{\rm CoG}$ matches that of the target's continuum-component profile are shown at the top of each panel. 
      }
         \label{fig_SBtest}
\end{figure*}

Among the 19 LAH candidates, 17 $\lya$ haloes around the galaxies are confirmed with $p_0\leq0.05$ (see the bottom panel of Figure \ref{fig_halotest}). All LAH sources have small $p_0$ values, ranging from $3.0\times10^{-68}$ to 0.035 (corresponding to reduced $\chi^2$ values from 28.5 to 2.6), which indicate significant deviations of the radial SB profiles from those of the  PSF-convolved continuum-component profiles at $r\geq r_{\rm in}$. However, the remaining two sources, RID=5479 and 22230, are not confirmed to have a $\lya$ halo due to their high $p_0$ values, 0.37 and 0.27, respectively, which are higher than the thresholds. Interestingly, both sources have bright $\lya$ fluxes and high S/N($r_{\rm in}$--$r_{\rm CoG}$) values, but the radial SB profiles at $r\geq r_{\rm in}$ are statistically consistent with those of the target's continuum-component images (see Section \ref{subsubsec:testhalo_nolahs} and Appendix \ref{ap:22230}). The radial SB profiles and the results of the tests for individual sources are summarized in Figure \ref{fig_SBtest} and Table \ref{table:halotest}.

In total, we have 17 LAHs and four non-LAHs. The results of this LAH test are consistent with those in \citet{Leclercq2017} for the six common sources with RID=7067, 7901, 9814, 9863, 10018, and 23124, which are classified as LAHs in both studies, though the methods used are different. Interestingly, we found LAHs around non-LAEs with net negative $EW(\lya)$: for instance, RID=4587 and 4764, which have $EW(\lya)_{\rm net}\leq-10.0$ \AA\ and -14.2 \AA, respectively, as demonstrated in Appendix \ref{ap:ew_xlae} and Figure \ref{fig_ew} (see also the central dip on the SB profiles in Figure \ref{fig_SBtest}). RID=7876 and 9944 might be also non-LAEs with LAHs. However, unfortunately, they are nonisolated sources, and we cannot decline a possibility that their LAHs are mainly contributed from other sources in the large-scale structure like RID=7847 (see Figure \ref{fig_group2_NB}). 
This is the first time to confirm the presence of LAHs around non-LAEs. Further discussion of the $\lya$ line properties and galaxy populations is beyond the scope of this paper but it could be interesting to investigate in the future.

\subsubsection{Non-LAH objects}\label{subsubsec:testhalo_nolahs}

\begin{figure} 
   \centering
   \includegraphics[width=\hsize]{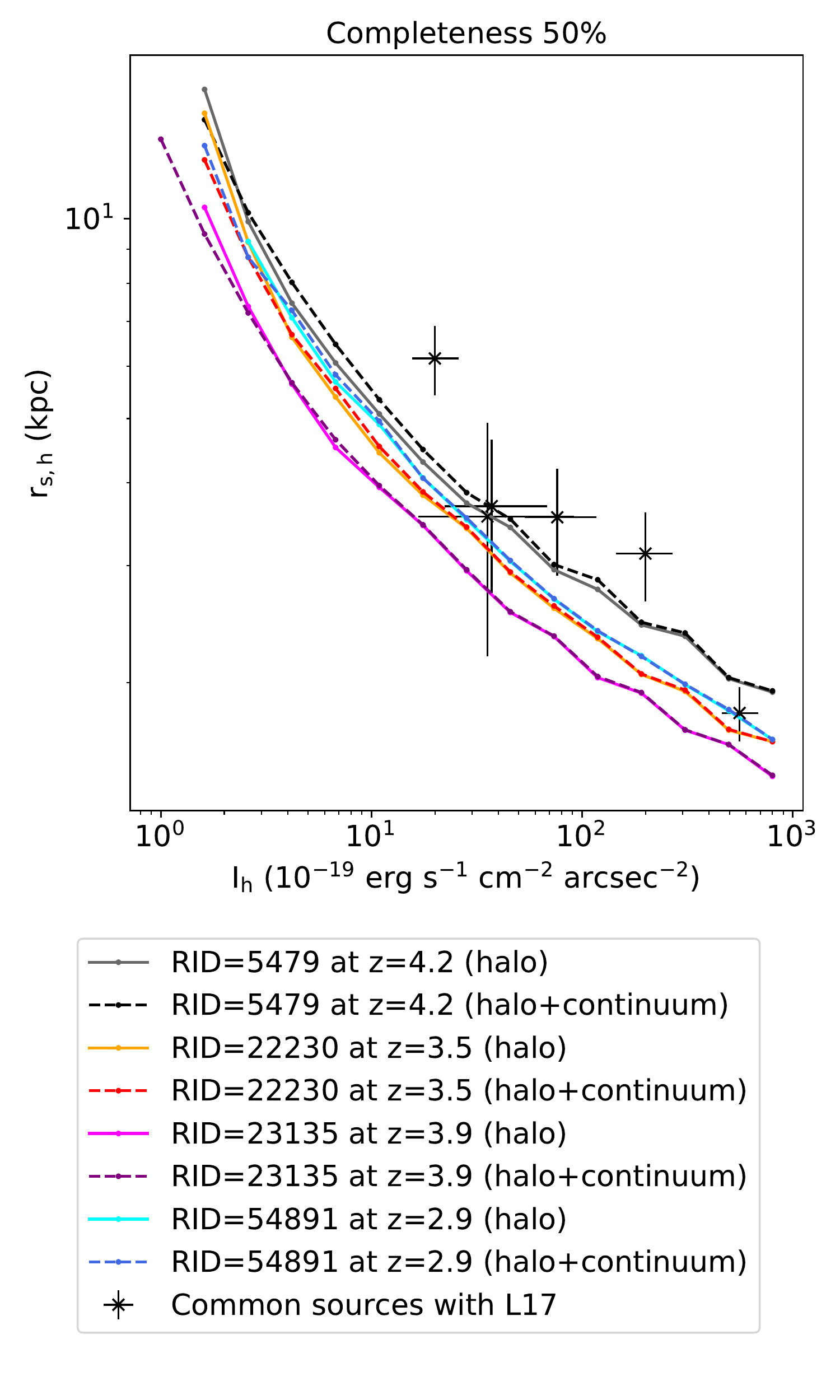}
      \caption{
       Halo parameter set ($I_{\rm h}$ and $r_{\rm s, h}$) showing 50\% completeness for four non-LAH objects. We explored potential halo parameters for four non-LAHs by simulating completeness values of the halo tests. We assumed exponential profiles for the halo and continuum-like component models (h+c) shown by dashed lines and for halo only models (h) shown by solid lines (see Appendix \ref{ap:comp}). The gray solid (black dashed), orange solid (red dashed), magenta solid (violet dashed), and cyan solid (blue dashed) lines represent RID=5479, RID=22230, RID=23135, and RID=54891 with halo only models (halo and continuum models), respectively. The black crosses indicate the halo parameters measured in \citet{Leclercq2017}  for the common sources between \citet{Leclercq2017} and ours. 
      }
         \label{fig_comp4to1}
\end{figure}

Faint diffuse $\lya$ haloes could be undetected because they would be hidden in the noise. We investigated the four non-LAH objects and simulated their completeness individually. Here we briefly explain the simulations, introduce non-LAHs, and comment on a non-LAH included in the UV-bright sample (Section \ref{subsec:sample}). The details of the simulations and the remaining three non-LAHs are described in Appendix \ref{ap:comp}.

The completeness of our halo test is sensitive to the SB profiles of haloes. Assuming an exponential SB profile, as used in \citet{Wisotzki2016} and \citet{Leclercq2017}, we applied the same halo test to mock NBs of noise and PSF-convolved halo models (see Appendix \ref{ap:comp} for more details). We also simulated haloes with a $\lya$ continuum-like component, whose flux was matched to the central Ly$\alpha$ flux of the observed non-LAH objects. Figure \ref{fig_comp4to1} summarizes the parameter sets with a 50\% completeness of the four non-LAHs for both cases. Generally, the completeness decreases when the central surface brightness $I_{\rm h}$ decreases or the scale length for haloes $r_{\rm s, h}$ decreases. The profiles are similar among the four sources but depend on the noise levels. If a galaxy has a bright $\lya$ continuum-like component, the completeness can be enhanced or suppressed depending on the parameters (see Figure \ref{fig_comp_rel}). The completeness for individual sources is shown in Figures \ref{fig_comp_4a} and \ref{fig_comp_4b} in Appendix \ref{ap:comp}, and our halo tests are found to be complete when the halo fluxes are equal to or brighter than $\simeq 4\times10^{-18}$ erg s$^{-1}$ cm$^{-2}$ to $6\times10^{-18}$ erg s$^{-1}$ cm$^{-2}$ in most of the parameter ranges. We checked the parameters measured in \citet{Leclercq2017} for the six common sources with $\lya$ haloes. These sources are mostly located above the 50\% completeness lines for non-LAH sources.  

Out of the four non-LAHs, one object has a high S/N($r_{\rm in}$--$r_{\rm CoG}$) (>10), but no extended emission is significantly confirmed (RID=5479). Another object with S/N($r_{\rm in}$--$r_{\rm CoG}$)=6.7 has anisotropic spatially offset $\lya$ emission (RID=22230). Its extended emission cannot be confirmed with the current method and data set. The remaining two objects do not have detectable emission at $r\geq r_{\rm in}$. The last three objects are not included in the UV bright sample, and therefore the results of the halo tests for the last three sources do not affect the main conclusions of this paper. 

RID=5479 has a high confidence level of ZCONF=3 for $z_{\rm s}=4.16$ in the MXDF catalog and is UV bright, $M_{1500}=-19.34$. It clearly shows $\lya$ emission lines on the 1D spectra for both $r\leq r_{\rm in}$ and $r= r_{\rm in}$--$r_{\rm CoG}$ as shown in Figure \ref{fig_sample_cat2}, with $r_{\rm in}=1''$ and $r_{\rm CoG}=3\farcs2$. The S/N($r_{\rm in}$--$r_{\rm CoG}$) is 12.94. However, the radial SB profile at $r\geq r_{\rm in}$ follows that of the MUSE PSF-convolved HST profile (continuum-like component), and the $p_0$ value is as high as 0.38 (see Figure \ref{fig_halotest}). The radial SB profiles are also consistent at $r\leq r_{\rm in}$. According to the completeness simulations, the completeness is approximately 50\%, for instance, when the parameters are $r_{\rm s, h}$=5.5 kpc and $I_{\rm h}=8.8\times10^{-19}$ erg s$^{-1}$ cm$^{-2}$ arcsec$^{-2}$ or $r_{\rm s, h}$=10.3 kpc and $I_{\rm h}=2.4\times10^{-19}$ erg s$^{-1}$ cm$^{-2}$ arcsec$^{-2}$ (see Figure \ref{fig_comp_4a}). As discussed in Appendix \ref{ap:comp}, the completeness can be lowered when a source has a bright continuum-like $\lya$ component at the center, which can hide a halo feature. However, this only happens in a narrow parameter range, and the bright continuum-like $\lya$ component can also enhance the completeness for some cases (Figure \ref{fig_comp_rel}). We conclude that it does not constitute a serious bias in our tests. From the distribution of the median $r_{\rm CoG}$ in the simulations, we also find that the measured $r_{\rm CoG}$ prefers the halo parameters with completeness values lower than 50\%. The limitation of our data and method for the halo detection for RID=5479 is illustrated in Figure \ref{fig_comp_4a}.

If deeper IFU data with a higher spatial resolution than the MXDF data were available, it could be possible to detect faint and compact $\lya$ haloes among them. However, we would like to recall that our MUSE data were taken with AO and have an unprecedented depth of 100 to 140-hour integration, the longest exposure times on the VLT so far. Considering the small sample size and the small number of non-LAHs, we did not correct for the incompleteness below.

\begin{figure*} 
   \centering
   \includegraphics[width=\hsize]{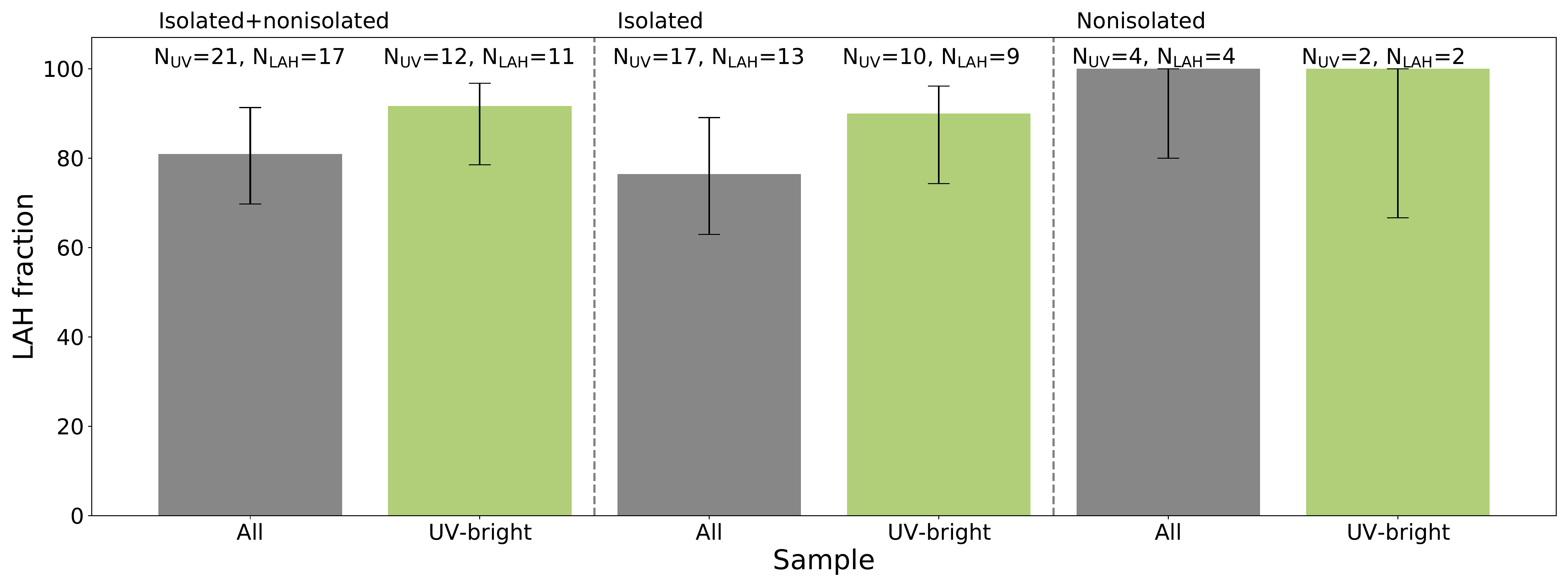}
      \caption{
      $\lya$ halo fractions for the complete sample (gray bars) and the UV-bright sample ($-20.0\leq M_{1500}\leq-18.7$; green bars). The $X_{\rm LAH}$ for isolated sources and that for nonisolated sources are shown in the second and third columns, respectively. The black error bars indicate 68.3\% confidence intervals. 
      }
         \label{fig_xlah_all}
\end{figure*}

\subsection{$\lya$ halo fraction}\label{subsec:flah}
\subsubsection{Calculation of the $\lya$ halo fraction and its uncertainties}\label{subsubsec:flah_calc}

The fraction of $\lya$ haloes around UV-selected galaxies, $X_{\rm LAH}$, is defined as follows: 
\begin{equation}
\ 
X_{\rm LAH}=\frac{N_{\rm LAH}}{N_{\rm LAH}+N_{\rm Non-LAH}}=\frac{N_{\rm LAH}}{N_{\rm UV}},
\ 
\end{equation}
where $N_{\rm LAH}$, $N_{\rm Non-LAH}$, and $N_{\rm UV}$ are the number of objects in a LAH sample, a non-LAH sample, and a parent sample, respectively. We calculated $X_{\rm LAH}$ for isolated sources and nonisolated sources separately, and for both subsets, for the UV-bright sample and the whole sample. 

We calculated the statistical uncertainty in $X_{\rm LAH}$. Statistical uncertainties for fractions (i.e., Bernoulli trials) are given by a binomial proportion confidence interval (BPCI). Following our previous paper on LAE fractions \citep{Kusakabe2020}, we derived the statistical uncertainties in $X_{\rm LAH}$ using \textsf{astropy.stats.binom\_conf\_interval}, with the Wilson score interval as an approximation formula \citep{WIlson1927}. Our $1\sigma$ uncertainties correspond to 68.3\% confidence intervals. For samples including RID=6693 or 54891, which has a low ZCONF=0, we adopted fractions including them in the $\lya$ fractions but took conservative uncertainties consisting of the minimum and maximum values of 68.3\% confidence intervals of fractions with and without RID=6693 or 54891. 

There are three other potential contributions to the uncertainty in $X_{\rm LAH}$. The first are the uncertainties in $N_{\rm LAH}$ due to the incompleteness of $\lya$ halo confirmations. As discussed in Section \ref{subsubsec:testhalo_nolahs}, the completeness of haloes sensitively depends on the halo SB profiles. Although spatial and spectral profiles of individual $\lya$ haloes around LAEs have been investigated \citep[e.g.,][]{Leclercq2017,Leclercq2020,Claeyssens2019}, we do not know the true distribution of the parameters of the halo profiles below the detection limits. Therefore, we did not correct for a completeness factor for $N_{\rm LAH}$ in this study. The number of non-LAHs is small for our samples and the completeness correction would thus not change our main conclusions. The second contribution comes from the uncertainty in the PSF estimation for the MXDF (Bacon et al. in prep.). Since the survey area of the MXDF is not large, and the MXDF data cubes were obtained by stacking many different observations, the PSF is smoothed and homogenized. Therefore, the uniform formula for the PSF applied to the entire field should not have a significant effect on our $\lya$ halo tests. Third, we have the field-to-field variance. Since the survey volume of MXDF is limited to $4.0\times10^3$ cMpc$^3$ ($z=2.86$--$4.44$, excluding the AO gap), our measurement of $X_{\rm LAH}$ may be different from the cosmic average. There is a possibility that $X_{\rm LAH}$ depends on the environment. Therefore, we calculated $X_{\rm LAH}$ for both the sample of all the sources and of isolated sources as precisely as possible.

\subsubsection{High $\lya$ halo fractions}\label{subsubsec:XLAH}

\begin{table}
\caption{$\lya$ halo fractions for all the sources and the UV-bright sample}\label{table:xlah_allsub}     
\begin{tabular}{llll}
\hline
                Sample &             $X_{\rm LAH}$ & 6693? & 54891? \\
\hline
                   All & $80.95^{+10.35}_{-11.23}$ &       Yes &        Yes \\
          All/isolated & $76.47^{+12.59}_{-13.55}$ &       Yes &        Yes \\
      All/nonisolated &   $100.0^{+0.0}_{-20.02}$ &        No &         No \\
             UV-bright &  $91.67^{+5.11}_{-13.15}$ &       Yes &         No \\
    UV-bright/isolated &   $90.0^{+6.11}_{-15.68}$ &       Yes &         No \\
UV-bright/nonisolated &   $100.0^{+0.0}_{-33.36}$ &        No &         No \\
\hline
\end{tabular}
\tablefoot{The $X_{\rm LAH}$ values for the ``All'' sample and the UV-bright sample. It also shows whether a sample includes RID=6693 or 54891, whose low ZCONF have to be considered in the uncertainty calculation. }
\end{table}

Figure \ref{fig_xlah_all} shows a high $\lya$ halo fraction for the sample including all sources, $80.1^{+10.1}_{-11.2}$\% (the dark gray bar in the first column). We note that all the non-LAHs are categorized as isolated galaxies (Section \ref{subsec:sample}). However, the fraction remains high even when the sample is limited to the 17 isolated sources, $76.5^{+12.6}_{-13.6}$\%. The $X_{\rm LAH}$ increases to $100.0^{+0.0}_{-20.0}$\% for nonisolated galaxies, though it is consistent within the $1\sigma$ error bars. Since all the sources were selected from the apparent magnitude cut of $F775W$, we also check $X_{\rm LAH}$ for the UV-bright sample with $-20.0\leq M_{1500}\leq-18.7$. The $\lya$ halo fraction is as high as $91.7^{+5.1}_{-13.2}$\% (the green bar in the first column). Even when we split them into isolated galaxies and nonisolated galaxies, the $X_{\rm LAH}$ values remain high, $90.0^{+6.1}_{-15.7}$\% and $100.0^{+0.0}_{-33.4}$\%, respectively. The numerical values of LAH fractions are listed in Table \ref{table:xlah_allsub}.

The detection of extended $\lya$ emission with stacked $\lya$ narrow bands for the UV-bright galaxies cannot provide information on the individual presence of a $\lya$ halo \citep[e.g., ][]{Steidel2011}. The general presence of $\lya$ haloes is confirmed for individual UV-selected galaxies with observations for the first time in this work. It is similar to the high fraction of $\lya$ haloes around high-z MUSE LAEs \citep[e.g., about 80\% in][Saust et al. in prep.]{Leclercq2017}, though the methods are different. We discuss the cause of the high $\lya$ halo fractions in the next section.

\section{Discussion}\label{sec:discussion}
We discuss implications from the high $\lya$ halo fractions, incidence rates of $\lya$ emission, and implications from four non-LAHs in Sections \ref{subsec:whyhighXLAH}, \ref{subsec:dndz}, and \ref{subsec:imp_nonLAH}, respectively.

\subsection{Why do most of galaxies have a LAH?}\label{subsec:whyhighXLAH}

\begin{figure*}
   \centering
   \includegraphics[width=\hsize]{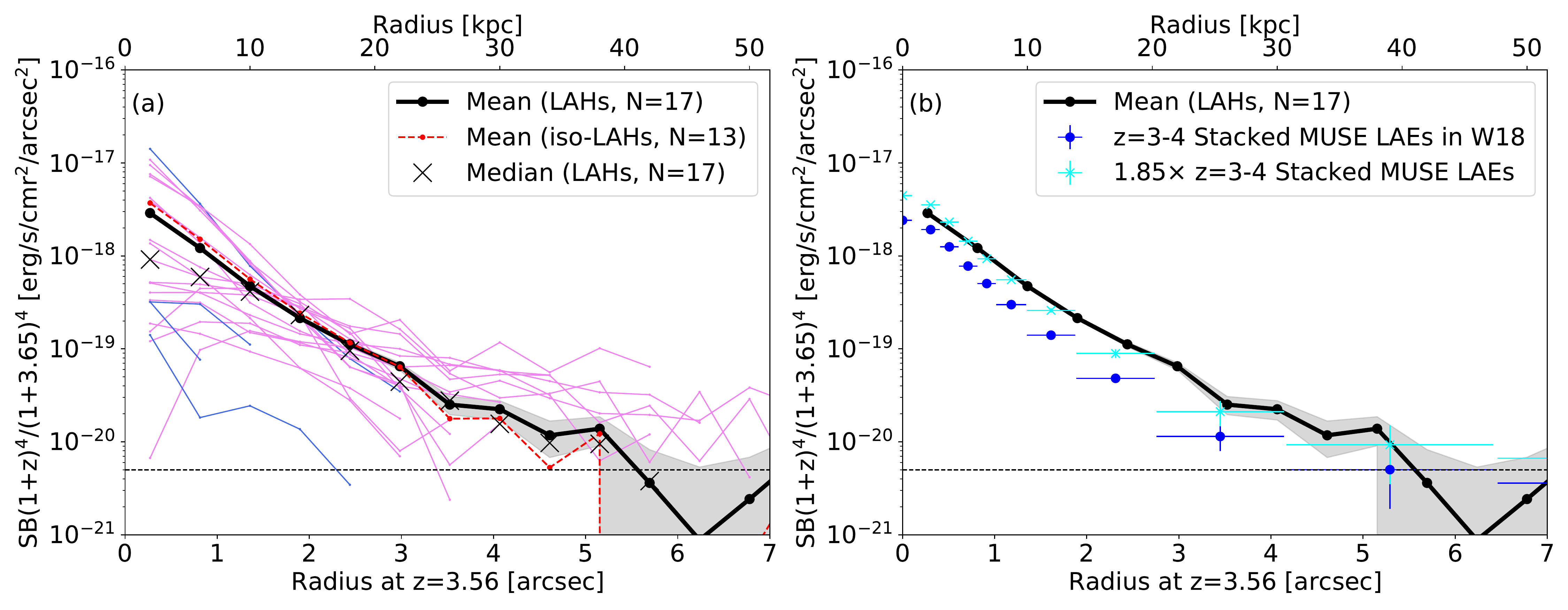}
      \caption{ Typical $\lya$ SB profiles with cosmic dimming corrections scaled to $z=3.65$. {\it (a)}: The violet and blue solid lines show individual SB profiles of LAHs and non-LAHs, respectively. The black solid line and the gray shaded area indicate the mean SB profile of 17 LAHs and the $1\sigma$ uncertainty, respectively, while the red dashed line indicates the mean SB profile of 13 isolated LAHs. The black crosses represent the median profile of the 17 LAHs. The black dashed holizontal line shows the typical $1\sigma$ SB limit of the mean LAH stack. {\it (b)}: The blue circles and cyan crosses show the median SB profile of stacked MUSE LAEs at $z=3$--$4$ \citep[][W18]{Wisotzki2018} and the same data multiplied by 1.85 to compare the profile shapes.
       }
         \label{fig_SB}
\end{figure*}

\begin{figure}
   \centering
   \includegraphics[width=\hsize]{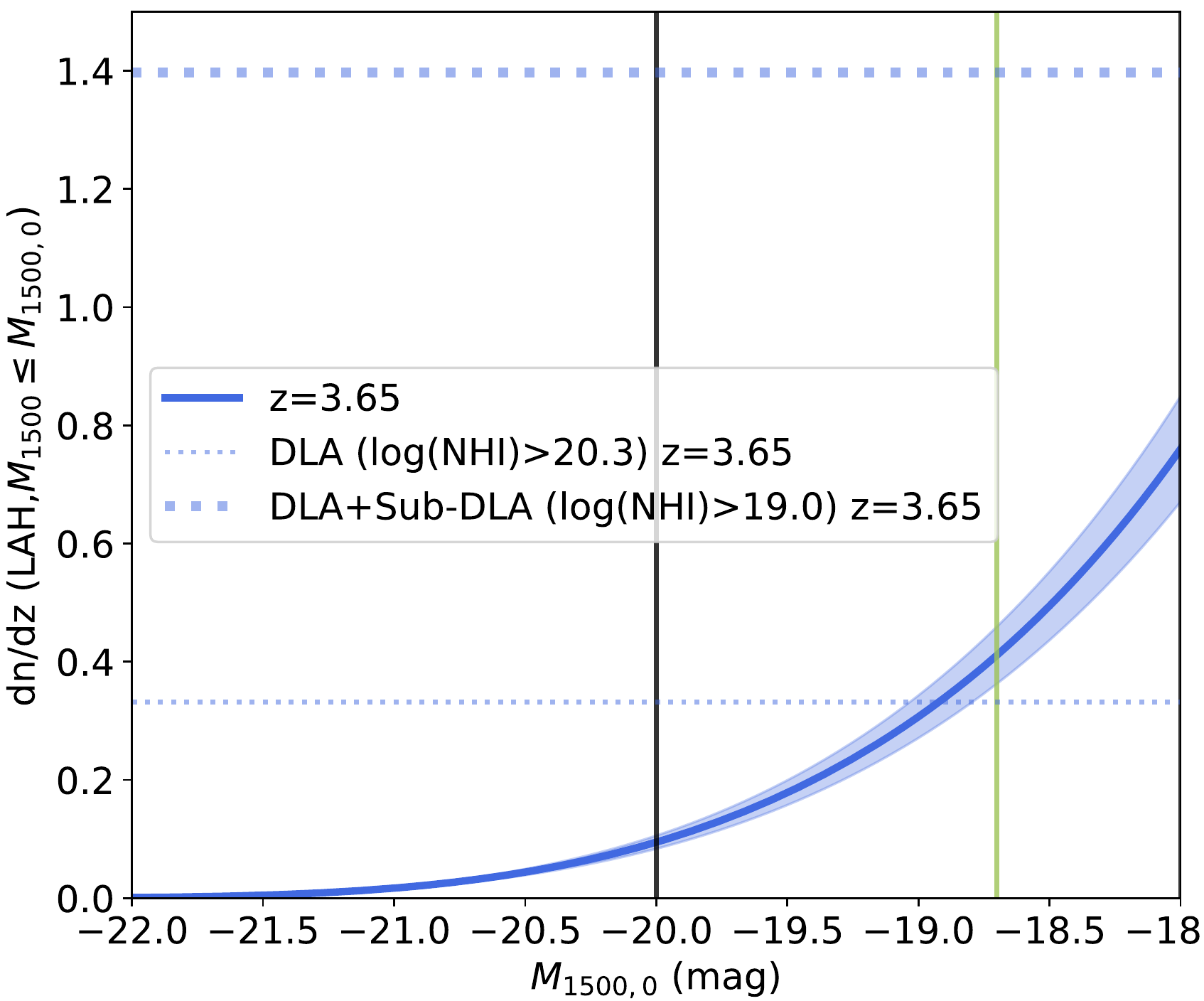}
      \caption{Cumulative incidence rate of LAHs, $dn/dz({\rm LAH},\,M_{1500} \leq M_{1500,0})$, as a function of $M_{1500,0}$. The blue solid line and shading show the $dn/dz({\rm LAH},\,M_{1500} \leq M_{1500,0})$ at $z\simeq3.65$ and the uncertainty, respectively. The uncertainty is propagated from the 1$\sigma$ error of the $X_{\rm LAH}$ for all the sources. The solid black and green lines indicate $M_{1500}$ of the brightest source in our sample and the upper limit of $M_{1500}$ for the  UV-bright sample, respectively (the $M_{1500}$ of the faintest source is -18 mag). The blue dashed thin and thick lines represent the  $dn/dz$ for DLAs and DLAs+sub-DLAs at $z=3.65$ calculated from the best-fit formulae of the z evolution of absorbers in \citet{Zafar2013}.
  }
         \label{fig_dndz_UVz3}
\end{figure}

\begin{figure*}
   \centering
   \includegraphics[width=\hsize]{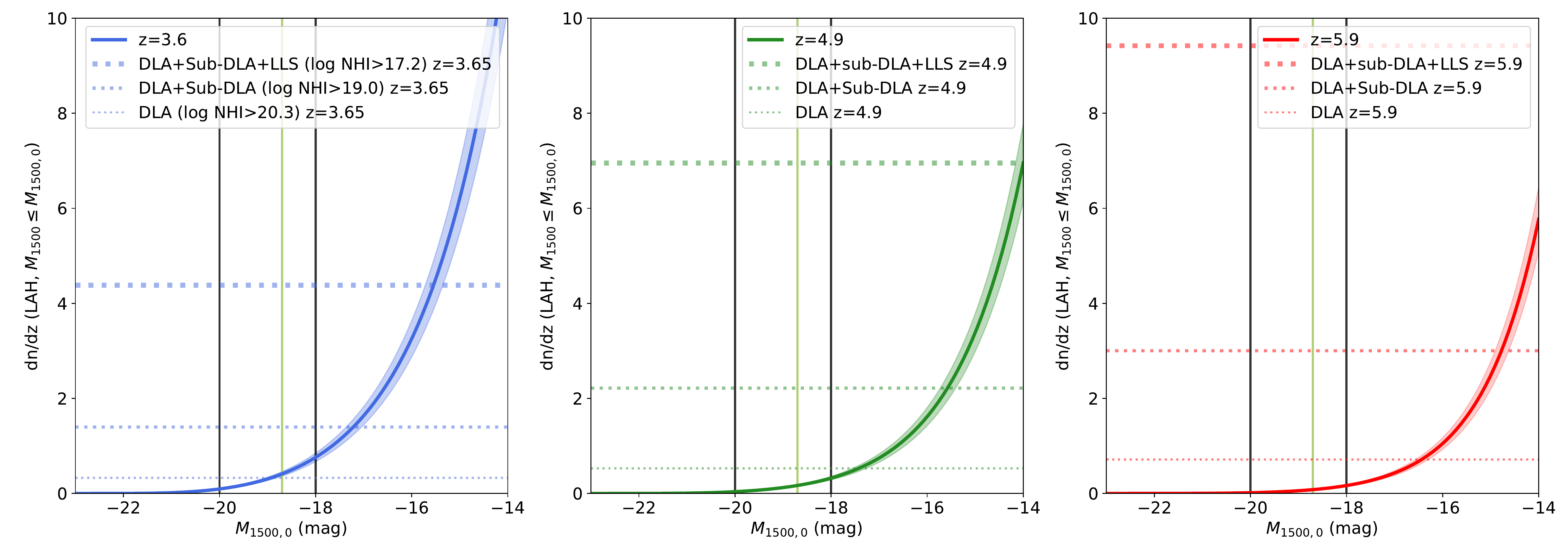}
      \caption{ Cumulative incidence rate of LAHs, $dn/dz({\rm LAH},\,M_{1500} \leq M_{1500,0})$, as a function of $M_{1500,0}$ at different redshifts. The different colors (blue, green, and red) indicate different redshifts ($z\simeq3.65$, 4.9, and 5.9 in panels a to c, respectively). The solid black vertical lines show the $M_{1500}$ of the brightest and faintest sources in our sample. The green vertical line indicates the upper limit of $M_{1500}$ for the UV-bright sample.       {\it (a)}: At $z\simeq3.65$. The blue solid line and shading show the $dn/dz({\rm LAH},\,M_{1500} \leq M_{1500,0})$ and the uncertainty, respectively. The blue-dashed thin line, medium-thick line, and thick line represent the $dn/dz$ for DLAs, DLAs+sub-DLAs, and DLAs+sub-DLAs+LLSs at the same redshift, respectively. 
      {\it (b)}: At $z\simeq4.9$. {\it (c)}: At $z\simeq5.9$. 
      }
         \label{fig_dndz}
\end{figure*}

It is now well established that there is extended \lya\ emission around individual LAEs \citep{Wisotzki2016,Leclercq2017}. The key result of the present study is that not only LAEs have \lya\ haloes: we found a very high $\lya$ halo fraction of $\simeq80$--90\% for our UV-selected galaxies at $z=2.9$--$4.4$, in other words, star-forming galaxies at high redshifts. It implies that UV-selected galaxies generally have a significant amount of cool/warm gas in the CGM. 

As introduced in Section \ref{sec:intro}, various mechanisms have been proposed to power $\lya$ haloes: 1) scattering of $\lya$ from star-forming regions \citep[e.g.,][]{Laursen2007,Zheng2011}, 2) gravitational cooling radiation \citep[e.g.,][]{Haiman2000,Fardal2001}, 3) star formation in satellite galaxies \citep[e.g.,][]{Zheng2011,Mas-Ribas2017a,Mas-Ribas2017b,Mitchell2021}, and 4) fluorescence \citep[e.g., ][]{Furlanetto2005, Cantalupo2005,Kollmeier2010,Mas-Ribas2016,Mas-Ribas2017b}. In process 1), $\lya$ photons are produced in star-forming regions inside galaxies and then scattered by \HI\ gas in the ISM and the CGM \citep[e.g.,][see also \citealt{kakiichi2018a} and \citealt{Garel2021}]{Barnes2010,Dijkstra2012,Verhamme2012}. From the spectral shapes of $\lya$ emission with spatial information, previous studies suggested that the scattering process happened in outflowing media \citep[e.g.,][see also \citealt{Verhamme2006,Verhamme2018}]{Claeyssens2019,Leclercq2020,Chen2021}. Resonant scattering in outflowing media can happen in and around galaxies, if they have a significant amount of surrounding outflowing gas \citep[see also][]{Kusakabe2019}. In process 2), $\lya$ photons are emitted by collisionally excited inflowing gas, which releases gravitational energy \citep[CGM in-situ emission, e.g.,][]{Rosdahl2012}. In the process 3), $\lya$ photons are produced through star formation in satellite galaxies, which can appear as a Ly$\alpha$ halo if individual LAEs are clustered on small scales \citep{Mas-Ribas2017a}. Finally, the $\lya$ photons in process 4) are produced by recombination of the CGM gas photo-ionized by the UV background, near-by bright objects, or the galaxies \citep[e.g.,][]{Mas-Ribas2016}. These $\lya$ emission mechanisms are difficult to distinguish from one another, although they do not necessarily trace the same gas phases and kinematics. 

\citet{Mitchell2021} present a zoom-in cosmological radiation hydrodynamics simulation of a single LAE at $z=3-6$ to study the origin and dynamics of the CGM as well as their $\lya$ signature. Their models can almost reproduce the average $\lya$ SB of stacked MUSE LAEs \citep{Wisotzki2018}, implying that the $\lya$ haloes are driven by the processes 1 to 3 above depending on the distance to the galaxy: CGM scattering of galactic $\lya$ emission, in-situ emission of CGM gas (mostly infalling), and $\lya$ emission from satellite galaxies (see their Figures 6 and 7 for the contribution of these processes to the $\lya$ halo). \citet{Byrohl2021} predict $\lya$ SB profiles at $z=2$--$5$ with illustris TNG50 simulations, which agree with those of the individual MUSE LAEs in \citet{Leclercq2017}. They find scattered photons from star-forming regions to be the major source of $\lya$ haloes in their simulations.

In fact, a cool/warm gas reservoir around high-$z$ galaxies is expected from both observations and simulations. Numerous observational campaigns revealed the existence of multiphase extended gas reservoirs around galaxies at low redshifts \citep[e.g., summarized in Figure 7 of][]{Tumlinson2017}. \citet{Werk2014} use the transverse absorption-line technique and measure the hydrogen column densities of 33 $L\simeq L_\star$ galaxies at $z\simeq0.2$ in the COS-Halos survey. They constrain the lower limit of the cool, highly ionized CGM mass as $6.5\times10^{10}\,  M_{\odot}$ and the extension as 300 kpc \citep[see also e.g.,][for low-z observations]{Zhang2016,Zabl2019,Schroetter2021, Beckett2021arXiv}. Neutral hydrogen gas is also detected up to the virial radius and beyond in the CGM of star-forming field dwarf galaxies at z$\lesssim0.2$ \citep[$L\lesssim 0.1 L^\star$, $M_\star\simeq10^{8}$--$10^{9}$ M$_\odot$,][]{Johnson2017}, which are more representative of our sample. At $z=2$--$3$, \citet{Rudie2012} find covering fractions of $90\pm9$\% and $30\pm14$\% for $N(\HI)>10^{15.5}\ {\rm cm^{-2}}$ and $N(\HI)>10^{17.2}\ {\rm cm^{-2}}$, respectively, within the viral radius of massive LBGs \citep[see also][]{Prochaska2013b}. The number densities (or incidence rates) of absorbers imprinted on quasar spectra are found to increase from $z=0$ to $z=5$ \citep[e.g.,][for DLAs, LLSs, and Sub-DLAs, respectively]{Prochaska2005,Songaila2010,Zafar2013a}. It implies the presence of denser and more neutral hydrogen CGM gas at higher redshifts. Simulations of galaxy formation and evolution generally predict that the CGM contains a significant amount of neutral hydrogen gas \citep[e.g.,][]{Fumagalli2011,VandeVoort2012}, though they make different predictions for the mass, extent, and physical state of the gas \citep[e.g., ][]{Tumlinson2017}. Until recently, it has been very challenging for simulations to reproduce the observed high covering fractions of \HI\ at high redshifts. \citet{Rahmati2015} use the EAGLE cosmological, hydrodynamical simulations \citep{Schaye2015} and predict the \HI\ gas distribution around high-$z$ massive galaxies. They show a high covering fraction of $N(\HI)\geq 10^{17.2}\ {\rm cm^{-2}}$ gas of $\simeq60$--$70$\% at $z=4$ for halo masses of $M_{\rm h}>10^{11}\ M_\odot$, which is found to be driven by stellar and AGN feedback. Interestingly, massive haloes with $M_{\rm h}>10^{12}\ M_\odot$ are predicted to have nearly scale-invariant profiles of the \HI\ gas covering fraction at a given redshift, implying the generality of a rich content of \HI\ gas in the CGM in high-$z$ massive haloes. The covering fractions increase rapidly with redshifts at a given $M_{\rm h}$ in the simulations, as a result of increasing rates of accretion and the higher mean density of the Universe (see their Figure 5). This redshift evolution is also predicted by the FIRE-2 cosmological zoom simulations in \citet{Stern2021arXiv2}, which also show that a large fraction of the inner CGM volume is occupied by cool neutral gas in haloes with $M_{\rm h}\leq 10^{12}\,  M_{\odot}$ and that it extends to the outer region of haloes with $M_{\rm h}\lesssim10^{11}$ $M_{\odot}$ (i.e., comparable to the typical $M_{\rm h}$ of our sample, $\simeq1\times10^{11}$--$2\times10^{11}\ {\rm M}_\odot$, see Section \ref{subsec:sample}). In their simulations, neutral hydrogen gas is maintained by a shorter cooling time than the free-fall time and a gas density high enough to be self-shielded from photoionizing radiation. As different simulations predict different dependences of the \HI\ gas covering fractions on $M_{\rm h}$, which could be caused by the use of different feedback models and resolutions \citep[][see also \citealt{Faucher-Giguere2015}; \citealt{Peeples2019}]{Rahmati2015,Stern2021arXiv2}, it is challenging to infer the physics behind our high $\lya$ halo fractions from them. Nevertheless, the state-of-the-art simulations suggest a rich cool/warm gas content in the high-$z$ CGM.

To better understand the role of the CGM gas in galaxy evolution and the physical mechanisms at play in CGM, it is important to unveil the link between absorbers and host galaxies. So far, significant efforts have been made to investigate impact parameters, \HI\ covering fractions, metallicities of absorbers, and their dependence on properties of the hosts such as luminosities, masses and star formation rates \citep[e.g.,][see also \citealt{Turner2017a}]{Rudie2012,Turner2014,Rubin2015,Fumagalli2015,Krogager2017,DeCia2018,Mackenzie2019}. \citet{Muzahid2021arXiv2} succeed in characterizing the gaseous CGM of 96 LAEs at $z=2.9$--$3.8$ with background quasars from impact parameters of 16 to 315 kpc in the MUSE Quasar-field Blind Emitters Survey (MUSEQuBES). As introduced in Section \ref{sec:intro}, mapping gas with emission is another ideal way to investigate it, but the connection between the gas detected in absorption and that in emission has not been uncovered yet, in particular at high redshifts. At $z\simeq3.25$, \citet{Fumagalli2017b} detect extended $\lya$ emission for a counterpart DLA with a $\simeq40$ kpc extend at a projected distance of $\simeq30$ kpc from the quasar sightline \citep[see also e.g.,][for $\lya$ haloes of DLAs at $z\sim3$]{Christensen2004,Kashikawa2014}. Very recently, \citet{Zabl2021arXiv} reported an \MgII\ emission halo around a star-forming galaxy at $z=0.7$ near a quasar sightline for the first time thanks to deep MUSE data reaching a SB limit of $\simeq1\times10^{-18}$ erg s$^{-1}$ cm$^{-2}$ arcsec$^{-2}$ ($2\sigma$). As a complementary approach to the individual studies \citep[e.g.,][]{Fumagalli2017b,Zabl2021arXiv} and a practical strategy for high redshifts, \citet{Wisotzki2018} estimate the incidence rate of extended $\lya$ emission of MUSE LAEs and compare it with the incidence rates of $\lya$ absorbers \citep[][]{Rauch2008}. Using our unbiased sample, we investigate whether LAHs of star-forming galaxies can account for absorber statistics at high redshifts in the next Section, following \citet{Wisotzki2018}.

\subsection{Incidence rate of the CGM gas visible in $\lya$ emission}\label{subsec:dndz}

As mentioned in Section \ref{subsec:whyhighXLAH}, the typical relations between the gas detected in absorption and emission have not been revealed. The general presence of LAHs allows us to estimate the incidence rate of the CGM gas detected in $\lya$ per line of sight per unit redshift, $dn/dz$, under the assumption that this extended $\lya$ emission traces the hydrogen content in the CGM. First, we derived the mean SB profile for our LAH subsample with cosmic dimming corrections and obtained the average area visible in $\lya$ (Section \ref{subsubsec:SBcorrection}). Second, we calculated the $dn/dz$ for LAHs assuming number densities from UV luminosity functions (Section \ref{subsubsec:dndz_calc}). Third, we compared them with $dn/dz$ for LLSs, sub-DLAs, and DLAs (Section \ref{subsubsec:dndz_comp}).

\subsubsection{Mean $\lya$ radial SB profile with cosmic dimming correction and $\lya$  size}\label{subsubsec:SBcorrection}

In order to compute the mean $\lya$ radial SB profile of our sample at different redshifts, we corrected for the cosmological dimming effect as follows. We used \textsf{PHOTUTILS} with the neighboring object masks, and measured $\lya$ fluxes from the flux-maximized NBs with a fixed aperture and annuli of fixed physical scale, from 1 kpc to 56 kpc, in 15 linear steps. Then, we converted the SB profile at each redshift and each radius, $SB_{\rm (z)}$(r [kpc]), in units of erg s$^{-1}$ cm$^{-2}$ arcsec$^{-2}$ with arcsec corresponding to kpc at the given $z$, to the SB profile at $z=3.56$ (the midpoint of $z=2.96$--$4.44$), $SB_{\rm (z=3.56)}$(r [kpc]):
\begin{equation}
SB_{\rm (z=3.56)} (r {\rm [kpc]})=SB_{\rm (z)} (r {\rm [kpc]}) \frac{(1+z)^4}{(1+3.56)^4}, 
\end{equation}
with units of erg s$^{-1}$ cm$^{-2}$ arcsec$^{-2}$ with arcsec at $z=3.56$. We took the means of the SB in each radial bin among our LAHs. To estimate $1\sigma$ uncertainties on the mean SBs, we summed the variances of the SB in a given radial bin over the LAH sample, took the square root, and divided it by the sample size. We also computed the mean SB profile for only isolated LAHs.

Figure \ref{fig_SB}a shows the mean $\lya$ SB of 17 LAHs. The $\lya$ emission is found to extend to 5.4 arcsec (40 kpc), above the SB limit of the typical $1\sigma$ uncertainty of $5\times10^{-21}$ erg s$^{-1}$ cm$^{-2}$ arcsec$^{-2}$. The mean SB profile of all the LAHs (black line) is consistent with the median for 17 LAHs (black crosses) and  the mean for 14 isolated LAHs (red dashed line). We compared it with that for median-stacked MUSE LAEs at $z=3$--$4$ from \citet{Wisotzki2018} in panel (b). Although individual SB profiles show a wide diversity (Figure \ref{fig_SBtest}), it is remarkable that our mean SB profile looks very similar to the stack of \citet{Wisotzki2018} in terms of typical extent and slope. The factor two difference in amplitude may be fully explained by the different samples: galaxies in \citet{Wisotzki2018} are mostly extremely faint LAEs. Interestingly, our halo \emph{size} is nevertheless similar to that of the faint LAEs with the SB limits, irrespective of our stacking methods described above. 

 In an ideal case with higher S/N values, we would be able to estimate the halo size (for instance, 90\% luminosity radius) by fitting the SB with two-component Sersic profiles. This would allow us to also probe the halo-size dependence on $M_{1500}$, which might exist. However, even with the MXDF data set, the S/N is not high enough to constrain the area detected with $\lya$ in such a sophisticated way. Therefore, following \citet{Wisotzki2018}, we calculated the $\lya$ size corresponding to the typical $1\sigma$ SB limit, $5\times10^{-21}$ erg s$^{-1}$ cm$^{-2}$ arcsec$^{-2}$.

\subsubsection{The incidence rate of LAHs}\label{subsubsec:dndz_calc}
Assuming that LAHs have the same area ($A_{\rm LAH}$) on average, we calculated the incidence rate of LAHs, $dn/dz({\rm LAH})$, for our sample. We use the UV luminosity function (LF) from \citet{Bouwens2015b} at redshift $z\simeq3.8$, similar to the midpoint redshift of our sample. This LF has a characteristic magnitude of $M^\star$=-20.88 mag, a normalization $\phi^*=1.97$ Mpc$^{-3}$, and a faint-end slope of $\alpha=-1.64$\footnote{We note that \citet{Bouwens2015b} derive the UVLFs for a longer wavelength (rest-frame 1600 \AA) than that used in this study (1500 \AA), and we assumed that the wavelength difference can be ignored.}. A cumulative $dn/dz({\rm LAH},\,M_{1500} \leq M_{1500,0})$ for a certain $M_{1500,0}$ limit is given as follows:
\begin{equation}
\frac{dn}{dz}({\rm LAH},M_{1500}\leq M_{1500,0})=\frac{A_{\rm LAH} n(M_{1500,0}) V_{\rm MXDF} X_{\rm LAH}}{A_{\rm MXDF} \Delta{z}} ,
\end{equation}
where $n(M_{1500,0})$, $V_{\rm MXDF}$, $A_{\rm MXDF}$, and $\Delta{z}$ are the number density for $M_{1500}\leq M_{1500,0}$, the survey volume, the survey area, and the redshift bin width, respectively. We integrated the LF from -23 mag to -18 mag, which is the faintest magnitude in our sample, see Figure \ref{fig_sky_lambda_MUV_zs}c. The obtained $dn/dz({\rm LAH},\,M_{1500} \leq -18)$ is $0.76^{+0.09}_{-0.09}$, whose uncertainty is propagated from the 1$\sigma$ error of $X_{\rm LAH}$ for all the sources. The lower $M_{1500}$ limit of -23 mag is brighter than our brightest LAH with  $M_{1500}\simeq-20$ mag, but the contribution from galaxies with $M_{1500}=-23$ to $-21$ mag to our $dn/dz({\rm LAH},\,M_{1500} \leq -18)$ is small. The high incidence rate of LAHs suggests that about 80\% of the sky is covered by the $\lya$ emission from galaxies with $M_{1500}\leq-18$ mag at $z=3$--$4$. The value is similar to that for MUSE LAEs at $z=3$--$4$ with their SB limit of $\simeq1.5\times10^{-20}$ erg s$^{-1}$ cm$^{-2}$ arcsec$^{-2}$, $1.0\pm0.4$, but is lower than that for the similar SB limit of $\simeq5\times10^{-21}$ erg s$^{-1}$ cm$^{-2}$ arcsec$^{-2}$, $2.3\pm 1.3$ \citep{Wisotzki2018}. The difference could be caused by the sample selections as we discussed above for the difference in the SB profiles. We also calculated the $dn/dz({\rm LAH},\,M_{1500} \leq M_{1500,0})$ with different $M_{1500,0}$, which is shown in Figure \ref{fig_dndz_UVz3} and used in the comparison in Section \ref{subsubsec:dndz_comp}.

\subsubsection{The incidence rate of LAHs compared with those of strong $\lya$ absorbers}\label{subsubsec:dndz_comp}

We compare the incidence rate of LAHs with those of DLAs and sub-DLAs, $dn/dz({\rm DLA})$ and $dn/dz({\rm sub-DLA})$, at our median redshift ($z=3.65$) from \citet{Zafar2013}. They show the best-fit redshift evolution of the observed $dn/dz$ for DLAs and sub-DLAs at $z=0$--$5$ by scaling the best-fit relation for LLSs in \citet{Songaila2010}. As shown in Figure \ref{fig_dndz_UVz3}, our $dn/dz({\rm LAH},\,M_{1500} \leq -18)=0.76^{+0.09}_{-0.09}$ is in between $dn/dz({\rm DLA})=0.33$ (blue dashed thick line) and $dn/dz({\rm DLA})+dn/dz({\rm sub-DLA})=1.4$ (blue dashed thin line). Our incidence rate for the UV-bright sample, $dn/dz({\rm LAH},\,M_{1500} \leq -18.7)$ (green vertical line), is closer to  $dn/dz({\rm DLA})$ but is also located in between them. It suggests that $\lya$ haloes are counterparts of DLAs and sub-DLAs. 

It is interesting to extend the exercise to higher redshifts and fainter $M_{1500}$ in order to discuss the redshift evolution. We assumed that galaxies have the same $A_{\rm LAH}$ in comoving kpc$^2$ at redshifts of $z\simeq4.9$ and $z\simeq5.9$ as at $z\simeq3.65$ \citep[e.g.,][]{Wisotzki2016,Leclercq2017}. We used UV LFs at $z\simeq4.9$ ($z\simeq5.9$) from \citet{Bouwens2015b}, $M^\star$=-21.10 mag, $\phi^*=0.79$ Mpc$^{-3}$, and $\alpha=-1.76$ ($M^\star$=-21.10 mag, $\phi^*=0.39$ Mpc$^{-3}$, and $\alpha=-1.90$). The $dn/dz({\rm LAH},\,M_{1500} \leq M_{1500,0})$ were calculated down to $M_{1500,0}=-14$ mag at $z\simeq3.65$, $z\simeq4.9$ and $z\simeq5.9$ in the same manner as in Section \ref{subsubsec:dndz_calc} (see Figure \ref{fig_dndz}). The incidence rate of LLSs, $dn/dz({\rm LLS})$, as well as $dn/dz({\rm DLA})$ and $dn/dz({\rm sub-DLA})$, were obtained from the formulae in \citet{Songaila2010} and \citet{Zafar2013}, respectively. The results of the abundance matching for DLAs evolve with redshift. At $z\simeq3.7$, $dn/dz({\rm DLA})$ can be matched with $dn/dz({\rm LAH})$ with $M_{1500} \leq -18.9^{+0.1}_{-0.1}$. Meanwhile, it requires $M_{1500} \leq -17.4^{+0.2}_{-0.1}$ and $M_{1500} \leq -16.4^{+0.1}_{-0.1}$ at $z\simeq4.9$ and $z\simeq5.9$, respectively. The same trends with redshift are found for $dn/dz({\rm sub-DLA})$ and $dn/dz({\rm LLS})$. We need to go to fainter $M_{1500}$ limits at higher redshifts to provide the same level of $dn/dz$ as that for absorbers. 

The $z$ evolution seen in this exercise can be explained by the combination of two redshift evolutions. First, we have the $dn/dz$ evolution of absorbers, which increases with increasing redshifts, implying the presence of denser and more neutral hydrogen gas at higher redshifts. This evolution is expected to be maintained by increasing rates of cold accretion and the higher mean density of the Universe at higher redshifts \citep[][]{Rahmati2015}. Second, the UV LFs evolve. The number density of such relatively bright galaxies decreases with increasing redshifts. Therefore, with the assumption of no dependence of $A_{\rm LAH}$ on $z$ and $M_{1500}$, fainter $M_{1500}$ sources are required to be accounted to reproduce $dn/dz$ of absorbers at higher redshifts.

Despite the remarkable progress with the UV-selected, as opposed to $\lya$-selected, sample, we caution against overinterpreting the results. The examination here is a very simple abundance matching with assumptions. The dependence of $A_{\rm LAH}$ on $z$, $M_{1500}$, or other galaxy's properties were not considered. Since UV brighter galaxies and lower-$z$ galaxies tend to have a larger UV size than their counterparts \citep[e.g.,][]{Shibuya2015a}, UV fainter and higher-$z$ galaxies may have a smaller $A_{\rm LAH}$. Would it be true, $dn/dz({\rm LAH},\,M_{1500} \leq M_{1500,0})$ could be overestimated with $M_{1500,0}\geq-18$ or at $z\simeq5$ and $\simeq6$. However, the correlation between $M_{1500}$ and halo size is under debate \citep[e.g.,][]{Steidel2011,Momose2016,Xue2017,Leclercq2017, Wu2020,Claeyssens2022arXiv}. No clear redshift evolution of the $\lya$ halo size has been confirmed \citep[e.g.,][]{Momose2014,Leclercq2017}. Last but not least, our modeling is based on a simple abundance matching approach in which one $\lya$ halo corresponds to one absorber. As shown in \citet{Rahmati2015} and \citet{Stern2021arXiv2}, for example, many absorbers with various $N(\HI)$ plausibly make up the CGM of individual high-z galaxies. Simulations of the CGM and absorbers greatly depend on the strength and the implementation of feedback models \citep{Faucher-Giguere2015,Rahmati2015,Suresh2015}. Since $\lya$ emission is the best tracer of the CGM \HI\ gas at high redshifts, which can provide spatial information, our results have great potential to constrain feedback models. It will be challenging as demonstrated in \citet{Mitchell2021}, but could also provide information on the mass, extent, dynamics, and physical state of the CGM gas. 

\subsection{Implications from non-LAHs}\label{subsec:imp_nonLAH}
It is also interesting that four of our galaxies do not have a significant $\lya$ halo, though they could be hidden in the noise as discussed in Section \ref{subsubsec:testhalo_nolahs}. Even aside from RID=22230, which shows potential extended $\lya$ emission with an anisotropic profile, the remaining three objects may have different reasons for the suppression of extended $\lya$ emission. RID=5479 has very bright $\lya$ emission on the scale of the galaxy's UV component as shown in Figure \ref{fig_sample_cat2}, but it does not show extended $\lya$. It may imply a small amount of \HI\ gas in the CGM, which allows $\lya$ photons to escape directly from the galaxy to the IGM. Such a galaxy may differ in halo mass from other galaxies or may be in a different evolutionary phase. A low \HI\ content of the CGM may be due to a phase with weak stellar feedback which does not push the gas out from the ISM to the CGM \citep[see Figure 11 in][and Appendix B in \citealt{Faucher-Giguere2015}]{Rahmati2015}. It could also be caused by a phase with very strong stellar and AGN feedback which either ionizes the gas or significantly disrupts or ejects gas from the CGM \citep[see Section 5.2 in][]{Trebitsch2017}. A low fraction of such objects would imply a short duty cycle of the lacking or disrupted phase. RID=23135 and RID=54891 have faint $\lya$ emission inside the galaxy's UV-component scale, suggesting that $\lya$ photons produced by the star formation could be killed by the dust in the ISM or the CGM and that the in-situ halo mechanism and satellite scenario do not work for these galaxies. In fact, a wide variety of \HI\ covering fractions are predicted in simulations: around 25\%-45\% and around 55\%-75\% as a 15-85 percentile for $\log(M_{\rm h}\ [M_\odot])\simeq11.1$ at $z\simeq3$ and $z\simeq4$, respectively \citep{Rahmati2015}. It is interesting to explore causes of a poor gas content for rare galaxies in simulations. In particular, galaxies which can allow ionizing photons to escape directly from the ISM to the IGM thanks to the low CGM \HI\ column density would play an important role in cosmic reionization.

\section{Conclusions}\label{sec:conclusions}
Thanks to the more than $100$-hour integration with MUSE AO in the MXDF \citep{Bacon2021}, we were able to examine the existence of $\lya$ haloes around UV-selected star-forming galaxies at $z\simeq2.9$--$4.4$. With the MXDF data, we constructed a sample with a  $F775W\leq27.5$ mag cut, with spectroscopic redshift constraints.  We confirmed that all of 26 sources with $z_{\rm p}\simeq2.9$--$4.40$ have a close spec-$z$ estimation to their photo-$z$, which implies high completeness values of our spec-$z$ assignments. We used 21 galaxies at $z_{\rm s}=2.9$--$4.4$, which include 17 isolated sources in HST images. The $M_{1500}$ range of our sample is -20 to -18 mag, enabling us to construct a UV-bright sample over the redshift range (11 sources, $-20.0\leq M_{1500}\leq-18.7$).  Our major results are summarized as follows. 

\begin{enumerate}
\item Among 21 galaxies, 17 were confirmed to have significant extended $\lya$ emission. We report the first individual detections of extended $\lya$ emission around non-LAEs with negative net equivalent widths of $\lya$ (for instance, RID=4587 and 4764). We measured the $\lya$ halo fraction for the sample of all the sources of $81.0^{+7.1}_{-11.2}$\%, for the 17 isolated sources of $76.4^{+8.6}_{-13.5}$\%, and for the 12 $-20.0\leq M_{1500}\leq-18.7$ sources of $91.7^{+5.1}_{-13.1}$\%. The $X_{\rm LAH}$ increases to $100.0^{+0.0}_{-20.0}$\%, for nonisolated galaxies, though it is consistent within the $1\sigma$ error bars. The high fractions are similar to that for MUSE LAEs, about 80\%, in \citet{Leclercq2017}, though the methods are different.
\item The high fractions of $\lya$ haloes imply that UV-selected galaxies generally have a significant amount of cool/warm gas in the CGM. Our study shows for the first time significant extended $\lya$ emission around most individual high-$z$ star-forming galaxies in a spec-$z$ complete sample. 
\item The mean SB profile for 17 LAHs was derived with cosmic dimming corrections for the midpoint redshift of $z\simeq3.65$. The mean $\lya$ radius above the typical $1\sigma$ uncertainty of the SB is 5.4 arcsec (40 kpc).  
\item Assuming that extended $\lya$ emission traces the same cool/warm gas as absorbing systems, we used the abundance matching technique for incidence rates in order to investigate the correspondence between gas detected in absorption and emission. Our $dn/dz({\rm LAH})$ calculated from a UVLF, the typical LAH size, and the measured $X_{\rm LAH}$ is $0.76^{+0.09}_{-0.09}$ for $M_{1500}\leq-18$ mag at $z\simeq3.65$. This is in between $dn/dz({\rm DLA})$ and $dn/dz({\rm DLA})$+$dn/dz({\rm sub-DLA})$ at the same redshift. It suggests that $\lya$ haloes trace the same gas as DLAs and sub-DLAs.
\item At higher redshifts, we need to go to a fainter $M_{1500}$ limit to reach the same $dn/dz$ as that for absorbers (DLA, sub-DLA, and LLSs). At $z\simeq3.7$, $dn/dz({\rm DLA})$ can be matched with $dn/dz({\rm LAH})$ with $M_{1500} \leq -18.9^{+0.1}_{-0.1}$, but it requires $M_{1500} \leq -17.4^{+0.2}_{-0.1}$ and $M_{1500} \leq -16.4^{+0.1}_{-0.1}$ at $z\simeq4.9$ and $z\simeq5.9$, respectively. The $z$ evolution is due to the increase of $dn/dz$ for absorbers with $z$ and the decrease of number densities of UV LFs with $z$. 
\item We found four non-LAHs, though one of them shows potential extended $\lya$ emission with an anisotropic profile (with $M_{1500}=-18.6$). Another one has very bright $\lya$ emission within the scale of the galaxy's UV component but does not show extended $\lya$ (with $M_{1500}=-19.3$), implying a small amount of the HI gas in the CGM, which may allow $\lya$ photons to escape directly from the galaxy to the IGM. The remaining sources have faint $\lya$ emission inside the galaxy's UV-component scale (with $M_{1500}=-18.0$ and $-18.4$), suggesting that $\lya$ photons produced by the star formation could be destroyed by dust in the ISM or the CGM and that the in-situ halo mechanism and satellite scenario are not at play for these galaxies. 
\end{enumerate}

We plan to extend this work to photometric-redshift samples in the near future using the MUSE HUDF survey data (9 arcmin$^2$) and upcoming MUSCATEL data (MUSE Cosmic Assembly survey Targetting Extragalactic Legacy fields; in total 36 arcmin$^2$), which will give us a more general view of $X_{\rm LAH}$. The larger cosmic volume of those surveys compared to that of the MXDF (0.84 arcmin$^2$ in this study) will be useful to extend the dynamic range to brighter galaxies. In order to constrain the physics of $\lya$ haloes and the trends of galaxies with and without haloes, we will investigate the LAH properties and the properties of the host galaxies with a larger sample. Finally, with MXDF data, our sample will allow spatially resolved spectral analysis of $\lya$ emission as those in the literature \citep[][see also \citealt{Patricio2016} and \citealt{Matthee2020barXiv2}, as well as \citealt{Claeyssens2022arXiv} for the Lensed Lyman-Alpha MUSE Arc Sample, LLAMAS]{Claeyssens2019,Leclercq2020}.

\begin{acknowledgements}
We thank the anonymous referee for constructive comments and suggestions. We would like to express our gratitude to Edmund Christian Herenz, Leindert Boogard, Miroslava Dessauges, Moupiya Maji, Valentin Mauerhofer, Charlotte Paola Simmonds Wagemann, Masami Ouchi, Kazuhiro Shimasaku, Akio Inoue, and Rieko Momose for giving insightful comments and suggestions. HK is grateful to Liam McCarney for useful suggestions on English writing through the UniGE's Tandems linguistiques. HK acknowledges support from Swiss Government Excellence Scholarships and Japan Society for the Promotion of Science (JSPS) Overseas Research Fellowship. HK, FL, and AV are supported by the SNF grant PP00P2 176808. AV and TG are supported by the ERC Starting Grant 757258"TRIPLE". This work was supported by the Programme National Cosmology et Galaxies (PNCG) of CNRS/INSU with INP and IN2P3, co-funded by CEA and CNES. This work is based on observations taken by VLT, which is operated by European Southern Observatory. This research made use of Astropy\footnote{\url{http://www.astropy.org}}, which is a community-developed core Python package for Astronomy \citep{TheAstropyCollaboration2013,TheAstropyCollaboration2018}, and other software and packages:\textsf{MARZ}, \textsf{MPDAF} \citep{Piqueras2019}, \textsf{PHOTUTILS}, \textsf{Numpy} \citep{Harris2020}, \textsf{Scipy} \citep{Virtanen2020}, and \textsf{matplotlib} \citep{Hunter2007}.

\end{acknowledgements}


\begin{appendix}
\section{Details of the sample}\label{ap:sample}
\subsection{MXDF catalog construction}\label{ap:catalog}
Here we give a short summary of the MXDF catalog construction and the full explanation is given in Bacon et al. (in prep.). The MUSE MXDF catalog was constructed in two ways and was merged into one catalog: a blind source detection with the 3D matched filtering software, \textsf{ORIGIN} \citep[][]{Mary2020}, and source extraction and deblending based on HST prior information with the \textsf{ODHIN} software \citep[][]{Bacher2017}, in a similar manner to those used in the MUSE-HUDF catalog \citep[][Bacon et al. in prep.]{Inami2017}. \textsf{ORIGIN} is optimized for the detection of compact sources with faint spatial-spectral emission signatures in MUSE datacubes. It subtracts the continuum from the MUSE cube first, and the detection is based on the local maxima of generalized likelihood ratio test statistics obtained for a set of spatial-spectral profiles of emission line emitters. It provides an estimation of the purity. The threshold value of the purity in the MXDF catalog is 0.8 (i.e., a false detection rate of 0.2).  The \textsf{ODHIN} software performs deblending of sources in MUSE datacubes, based on the prior information from HST images with a higher spatial resolution \citep[][]{Rafelski2015}. The selection cut for the prior sources is a S/N of 0.8 per spectral pixel. We extracted 1D spectra at the position of the \textsf{ORIGIN}-detected sources and that of the \textsf{ODHIN} HST-prior sources from the 3D data cube for the redshift determination and visual inspection. To assign redshifts, we used an updated version of the redshift fitting software, \textsf{MARZ}, which is based on a cross-correlation algorithm \citep[e.g.,][see also \citealt{Inami2017}]{Hilton2012,Baldry2014} for both \textsf{ORIGIN} and \textsf{ODHIN} sources. The redshift solutions provided by \textsf{MARZ} were visually inspected by the MUSE team members independently with the source inspector software (Bacon et al. in prep.). We assigned HST counterparts from the \citet{Rafelski2015} catalog when available, and merged an \textsf{ORIGIN} and an \textsf{ODHIN} source into one when they overlapped by choosing the detection method. After reconciliating conclusions between the different inspectors, we summarized sources into the final catalog.

Following \citet{Inami2017} with some improvements, we have three levels of confidence for spec-$z$ (ZCONF). ZCONF=1 is for a possible redshift with low confidence. It can be assigned due to low S/N values of lines (e.g., $S/N\lesssim3$), poor line fittings, existence of valid alternative redshift solutions, noisy narrow bands, or additional lines which present in the spectrum with reasonable S/Ns but can not be explained by the proposed redshift solutions. ZCONF=2 is for a good redshift. Regarding non-$\lya$ emitters (e.g., $z\lesssim2.9$), it is assigned for sources with multiple lines detected with good S/N values of $S/N\gtrsim5$. For instance, a resolved [O\,{\sc II}]$\lambda\lambda$3726,3729 doublet with a clear narrow band would be sufficient for ZCONF=2. Regarding $\lya$ emitters, it is required to have a $\lya$ line with a good S/N of $S/N\gtrsim5$ and a width and an asymmetry compatible with $\lya$ line shapes. The highest level, ZCONF=3, indicates a secure redshift. For non-$\lya$ emitters, the criteria are similar to those of ZCONF=2, but it is required to have more lines and higher S/N, together with high S/N narrow band images. For $\lya$ emitters, if there is no additional line except for $\lya$, it is required to have a $\lya$ line with a high S/N value of $S/N\gtrsim7$ and a $\lya$-like line profile: a pronounced red asymmetrical line profile\footnote{$\lya$ lines are fit with a skewed Gaussian with an asymmetry parameter $\gamma$: $F(\lambda)=F_{\rm max}\left[1+ {\rm erf} \left( \gamma\frac{\lambda-\lambda_0}{\sqrt{2}\sigma}\right) \right] \left[ {\rm exp} \left( - \frac{(\lambda-\lambda_0)^2}{2\sigma^2}\right) \right]$, where $F_{\rm max}$, $\sigma$, and $\lambda_0$ represent the amplitude, the width, and the peak wavelength, respectively, and the erf is error function (see Bacon et al. in prep. for more details). } (highly asymmetric of $\gamma>2$) and/or a blue bump, or a double peaked line profile. The S/N thresholds are not strictly defined because other information was taken into account. In the case of an ORIGIN detection, and if the source can be matched to an HST counterpart, it adds confidence to the detection. In addition, if a photo-$z$ is reliable and a good match to the MUSE spec-$z$, it adds confidence to the redshift assignment. This could lead to a higher ZCONF than in the case of an ORIGIN source without HST counterpart. Meanwhile, faint HST sources often do not have a reliable photo-$z$, which can be in disagreement with a MUSE spec-$z$. It is not taken as a strong negative constraint (see Bacon et al. in prep.). In addition to the three ZCONF levels, ZCONF=0 is defined in the MXDF catalog (DR2 v0.8) for a source with an ORIGIN detection for which no spectroscopic redshift could be identified. However, in this paper, we independently defined ZCONF=0 as no spec-$z$ in the MXDF catalog with a tentative (best-estimate) assignment of a redshift or a constraint of a redshift.

\subsection{Spec-$z$ and photo-$z$ distribution}\label{ap:zpzs}

\begin{figure}
   \centering
   \includegraphics[width=\hsize]{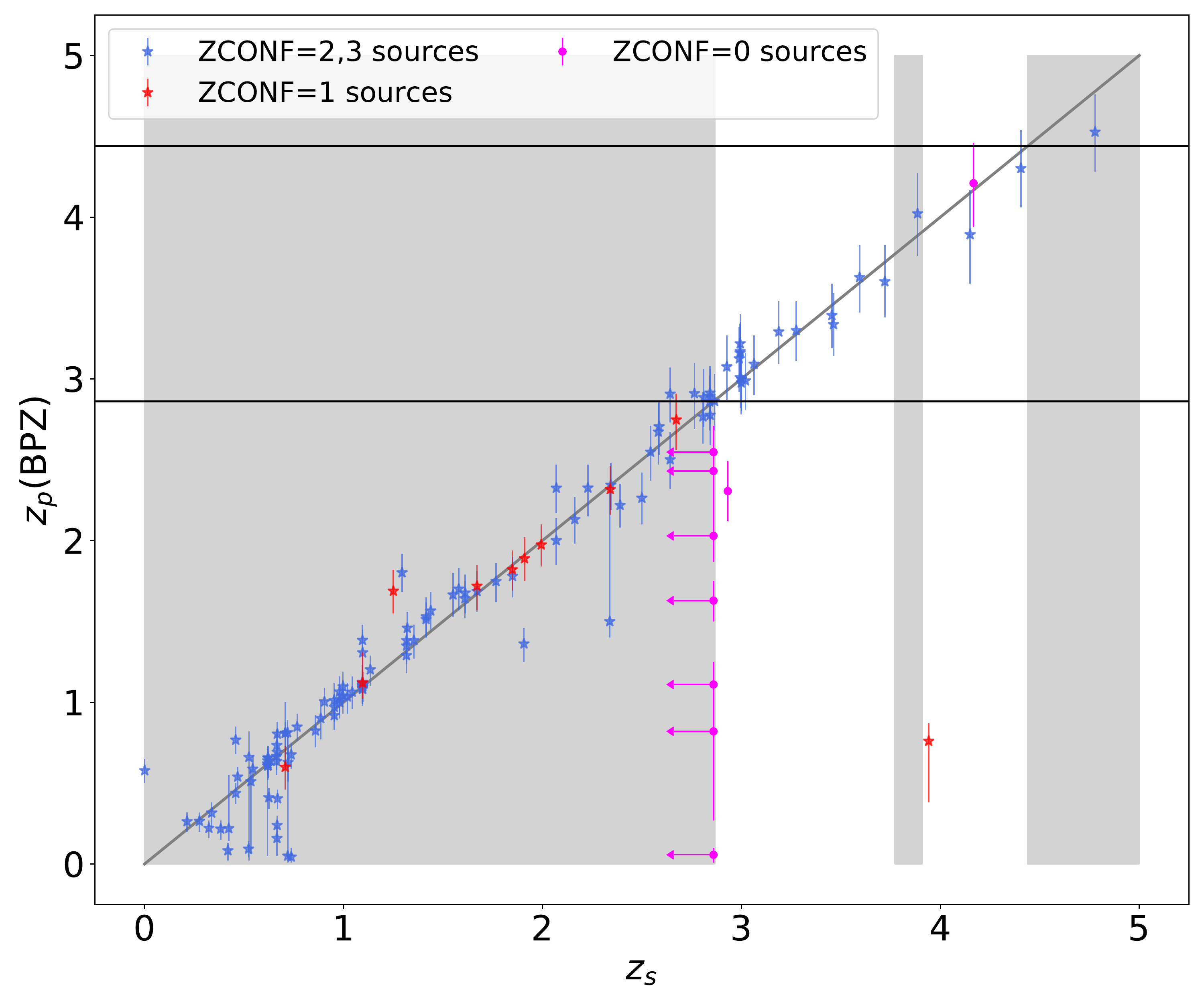}
      \caption{Photo-$z$ vs. spec-$z$ for 142 $F775W<27.5$ mag sources. The blue stars, the red stars, and the magenta dots represent galaxies with ZCONF=2 or 3, 1, and 0, respectively. The gray shaded areas indicate the spec-$z$ outside the targeted redshift (z=2.86-4.44) and the AO gap. The gray diagonal line and the black lines show the $z_{\rm p}$ = $z_{\rm s}$ relation, and $z_{\rm p} =2.86$ and $4.44$ lines, respectively. The photometric redshifts were estimated in \citet{Rafelski2015} with the Bayesian Photometric Redshift (BPZ) algorithm \citep{Benitez2000,Benitez2004,Coe2006}. The error bars of $z_{\rm p}$ show 95\% upper and lower limits. }
         \label{fig_zp_zs}
\end{figure}

In order to check a potential bias in our sample selection, we plotted the distribution of photo-$z$ and spec-$z$ for 142 $F775W<27.5$ mag sources from \citet{Rafelski2015} in the MXDF with more than 100-hour integration (see Figure \ref{fig_zp_zs}). As described in Section \ref{subsec:sample}, 123 sources have ZCONF=2 or 3, 10 sources have ZCONF=1, and 9 sources are not included in the MXDF catalog (ZCONF=0). Most of the ZCONF= 1 to 3 sources have a spec-$z$, which is consistent with their photo-$z$ (blue and red stars). RID=23135 is the single ZCONF=1 source in our sample (red star with $z_{\rm s}=3.94$ and $z_{\rm p}=0.76^{+0.11}_{-0.38}$). Although the photo-$z$ and the spec-$z$ do not match, the source has a strong \textsf{ORIGIN}-detected $\lya$ emission with $S/N=6.4$. Regarding ZCONF=0 sources (magenta circles), 7 sources were suggested to be at $z<2.86$ from clear continua at the blue edge of the MUSE spectra, which are consistent with their photo-$z$. The remaining two sources, RID=54891 and 6693, were included in our sample to minimize a selection bias despite ZCONF=0. The tentative spec-$z$ for RID=54891 was determined from a low-$S/N$ $\lya$ line at $z_{\rm s}=2.94$, while it was based on a $\lya$ break and stacks of low-$S/N$ UV absorption lines for RID=6693 (see Section \ref{subsec:sample} for the process of spec-$z$ assignment). Although not all the sources have a secure or reliable spec-$z$, our spec-$z$ estimations are almost complete with $F775W<27.5$ mag.

Generally, at $z\simeq3$--$6$, the $\lya$ line is the most common $z_{\rm s}$ indicator in follow-up observations, which could introduce a certain sample bias in favor of $\lya$ emission. We also checked spec-$z$ estimates for $z_{\rm p}\gtrsim3$ sources with $F775W<27.5$ mag. Among 142 sources, 26 sources have a best-fit photo-$z$ in $2.86\leq z_{\rm p} \leq 4.44$ (between the black lines). All of them except for RID=6693 have a spec-$z$ with ZCONF=2 or 3, most of which are consistent with their photo-$z$. As LAE fractions are not high \citep[e.g.,][]{Kusakabe2020}, it means that our sample construction is not significantly biased  a $\lya$ selection despite the redshift range. This result reinforces the high spec-$z$ completeness in our sample construction.

\subsection{$\lya$ equivalent width and LAE fraction}\label{ap:ew_xlae}
\begin{figure*}
   \centering
   \includegraphics[width=\hsize]{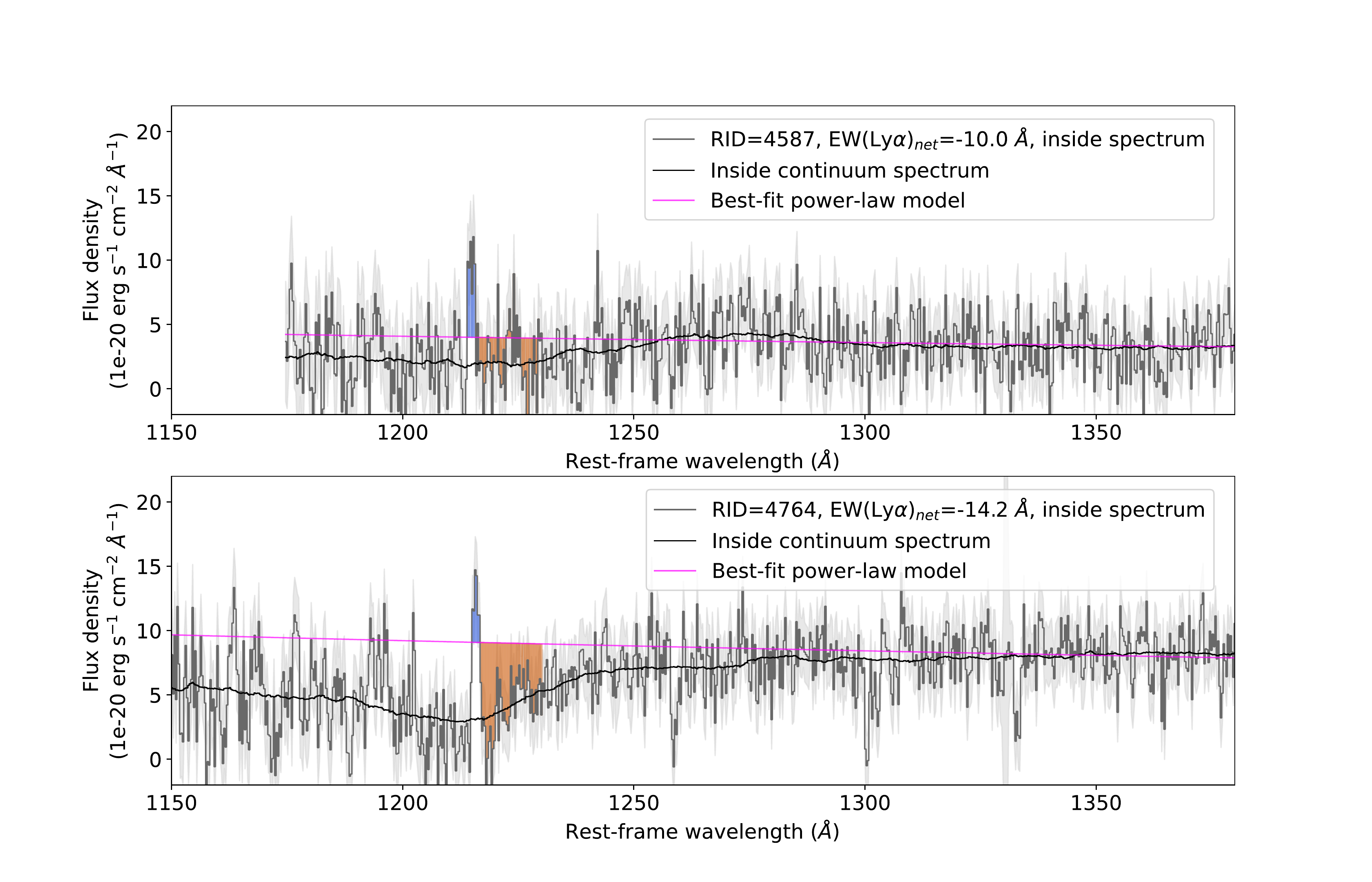}
      \caption{ 1D spectra of the two non-LAEs with LAHs, RID=4587 and 4764 discussed in Section \ref{subsubsec:testhalo_results}, as examples of our inside spectra. The dark gray lines, the faint gray shaded areas, the black lines, and the magenta lines show the inner spectra, its 1$\sigma$ uncertainties, the inner continuum spectra, and the best-fit power-law models for the continua, respectively. The blue and orange shaded areas indicate fluxes integrated for the $\lya$ emission and $\lya$ absorption measurements (the spectral windows, see Appendix \ref{ap:ew_xlae} for more details), respectively. We note that the $\lya$ absorption flux (more precisely, the upper limit of the absorption flux) was obtained by multiplying two to the integrated flux in the orange shaded spectral window. }
         \label{fig_ew}
\end{figure*}

We used rest-frame $EW(\lya)$ measurements to confirm the presence of LAHs around non-LAEs in Section \ref{subsubsec:testhalo_results} and to compare the $X_{\rm LAE}$ with a given $EW(\lya)$ cut for that in \citet{Kusakabe2020} in Section \ref{subsec:sample} in order to validate that our sample is not biased s the LAE selection. Therefore, the $EW(\lya)$ values in this paper were measured in the galaxy's stellar-component scale following \citet{Kusakabe2020}. Here we give a brief summary of the methods for the $EW(\lya)$ measurements and the $X_{\rm LAE}$ calculations. We note that investigating $\lya$ line properties including $EW(\lya)$ for our individual sources is beyond the scope of this paper. We plan to improve the method for measuring $EW(\lya)$ in the MUSE 3D cube and discuss them in a future project. 

 We extracted the 1D spectra inside the target's continuum-component mask (hereafter, inside spectra). Some fraction of our sources shows a wide $\lya$ absorption feature that could extend up to $\lambda\simeq1260$ \AA\ or even longer. Indeed, $\lya$ absorption troughs have already been identified for individual LBGs at $z\simeq3$ \citep[][see also \citealt{Chen2021}]{Kornei2010}, though the feature has not frequently been found for spectra of individual high-$z$ galaxies perhaps because of limited S/N values. Some of our sources also show a $\lya$ emission line on top of the wide absorption feature \citep[c.f., about 10\% of the sample in ][]{Kornei2010}, which are similar to local Green Pea galaxies \citep[e.g.,][]{McKinney2019,Jaskot2019} as well as stacked MUSE LAEs at $z\sim2.9$--$4.6$ \citep[e.g.,][]{Feltre2020}. Thanks to the very deep MUSE data, we need to go beyond the standard method for $EW(\lya)$ measurements for high-$z$ galaxies. We measured the $EW(\lya)$ for $\lya$ emission and $\lya$ absorption ($EW(\lya)_{\rm emi.}$ and $EW(\lya)_{\rm abs.}$, respectively) on the inside spectra with a method based on a similar idea in, for instance, \citet{Kornei2010} and calculated the net $EW(\lya)$, $EW(\lya)_{\rm net.}$, by summing $EW(\lya)_{\rm emi.}$ and $EW(\lya)_{\rm abs.}$ \citep[similar idea to that in][]{McKinney2019,Jaskot2019}. Figure \ref{fig_ew} shows the 1D spectra of the two non-LAEs with LAHs discussed in Section \ref{subsubsec:testhalo_results} as examples of our inside spectra. First of all, we estimated the continuum spectrum around $\lya$ by fitting the inside spectrum at rest-frame $\lambda\geq1270$ \AA\ with power-law models \citep[see Section 4.3 in][for the choice of the wavelength limit]{Matthee2021}. To obtain a $\lya$ emission flux, we integrated $\lya$ fluxes above the best-fit power-law continuum within a spectral window, whose fluxes are consecutively above the best-fit power-law around the $\lya$ peak (the blue shaded area). The $EW(\lya)_{\rm emi.}$ was calculated by dividing the emission flux with the continuum at the $\lya$ wavelength estimated from the power-law model and converted to that in the rest frame. Regarding the $\lya$ absorption, we estimated the absorption flux only from redder wavelengths than the $\lya$ wavelength, as the bluer wavelengths are generally affected by the IGM absorption. We defined the spectral window to measure the absorption fluxes (the orange shaded area). The blue edge of the absorption window is the next wavelength slice of the red edge of the emission window. The red edge of the absorption window was defined with the spectrum extracted in the same target's continuum-component mask from the continuum minicube (hereafter inside continuum spectrum; Section \ref{subsec:consub}). The red edge corresponds to the longest wavelength pixel among the consecutive increasing pixels at the redder side of the blue edge, where the flux on the inside continuum spectrum (the black line) is fainter than that of the best-fit continuum (the magenta line). Considering the possible contamination of N{\sc v}$\lambda1243$ P-Cygni profile and the limited S/N of the inside spectra for all the sources, we gave a conservative upper limit of the red edge as 1230 \AA. We integrated relative fluxes to the power-law continuum within the spectral window. Then, we obtained the $\lya$ absorption flux by multiplying two to the integrated flux. We calculated $EW(\lya)_{\rm abs.}$ by dividing the negative absorption flux with the continuum at the $\lya$ wavelength and converted it to that in the rest-frame. These $EW(\lya)_{\rm abs.}$ values would be actually the upper limits of the true $EW(\lya)$ for absorption because of the limited spectral windows (the orange shades). In this work, non-LAEs were defined with $EW(\lya)_{\rm net.}$= $EW(\lya)_{\rm emi.}$ + $EW(\lya)_{\rm abs.}$ $\leq0$ \AA. The two non-LAEs with LAHs introduced in Section \ref{subsubsec:testhalo_results}, RID=4587 and 4764, have $EW(\lya)_{\rm net.}\leq-10.0\pm2.9$ \AA\ and $-14.2\pm1.0$ \AA, respectively. Moreover, their $EW(\lya)_{\rm emi.}$ are as small as $2.8\pm0.6$ \AA\ and $0.7\pm0.3$ \AA, respectively. Even when we ignore the $\lya$ absorption ($EW(\lya)_{\rm abs.}\leq-12.7\pm2.8$ \AA\ and $-14.9\pm1.0$ \AA, respectively), their $EW(\lya)_{\rm emi.}$ values are too low to be categorized as LAEs with the typical 20 \AA\ selection criteria. Their $EW(\lya)_{\rm emi.}$ are also significantly smaller than the minimum $EW(\lya)$ of the MUSE LAEs with LAHs in \citet{Leclercq2017}, $\sim20$ \AA\, though $\lya$ halo fluxes are included in their $EW(\lya)$. These results validate our first individual detections of LAHs around high-$z$ non-LAEs. 

The LAE fractions for our sample were calculated with $EW(\lya)_{\rm emi.}$ in a similar method in that of \citet{Kusakabe2020} using the MUSE-HUDF data, which cannot take $\lya$ absorption into account as their continua are not detected in the MUSE spectra (i.e., nondetection of absorption) for most of the sources. $\lya$ fluxes from extended emission were not included in the $EW(\lya)$ for both samples. We did not correct the incompleteness of the detection of $\lya$ emission for our sample as the completeness is expected to be high, and we estimated the uncertainties on the $X_{\rm LAE}$ from the binomial proportion confidence interval. The $X_{\rm LAE}$ with the typical threshold for LAE selections, $EW(\lya)\geq20$ \AA, for out entire sample with $-20\leq M_{1500}\leq-18.0$ at $z=2.9$--$4.4$ is $0.33^{+0.11}_{-0.09}$. For a fair comparison, we also calculated the $X_{\rm LAE}$ with $EW(\lya)\geq65$ \AA\ for our entire sample as $0.14^{+0.09}_{-0.06}$. As discussed in Section \ref{subsec:sample}, this value is similar to those in \citet{Kusakabe2020}, suggesting that our sample is not biased. 

\subsection{nonisolated sources}\label{ap:nonisolated}

\begin{figure}
   \centering
   \includegraphics[width=\hsize]{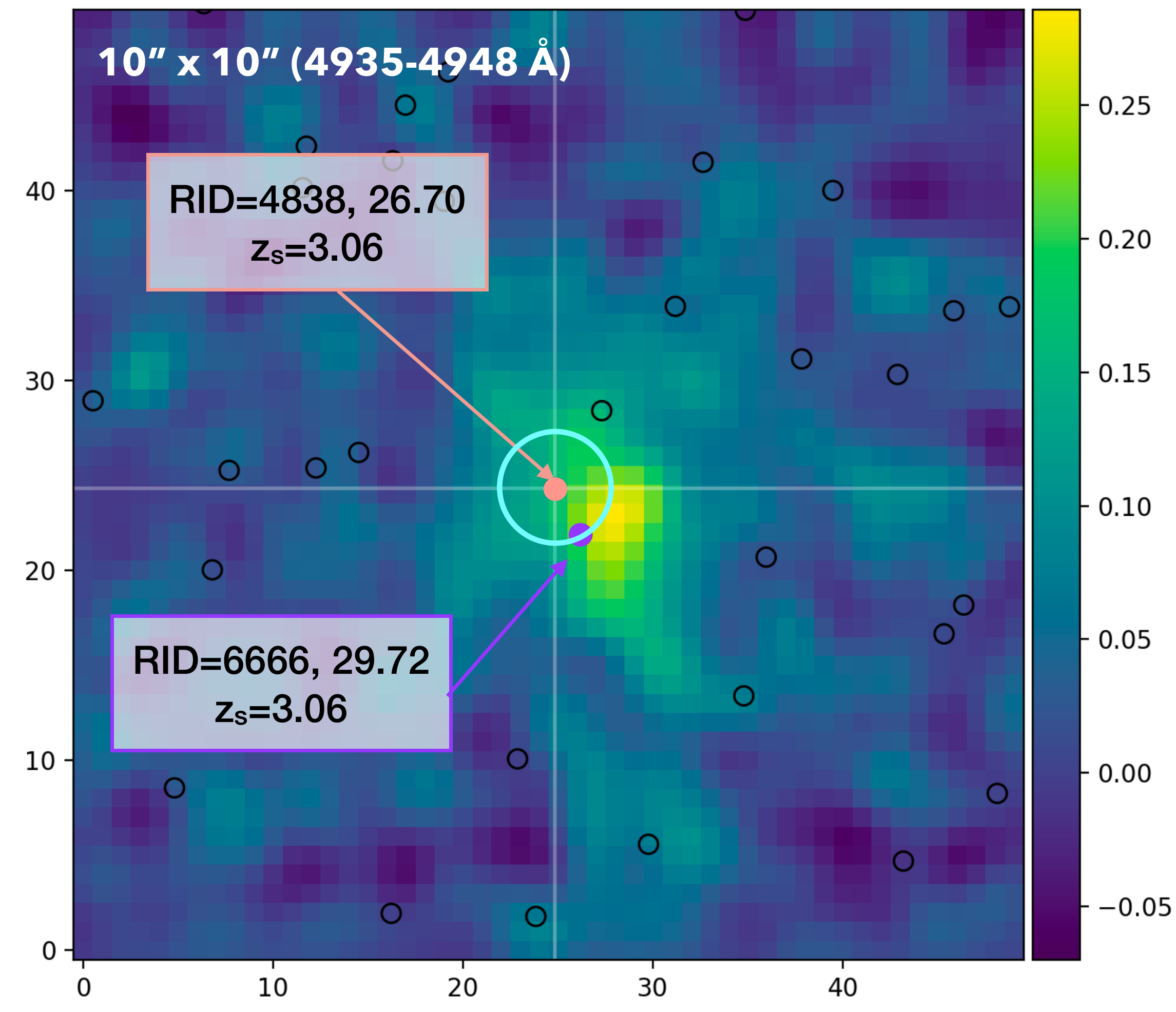}
      \caption{$\lya$ NB centered at RID=4838. The $\lya$ NB was extracted from the MXDF data cube with a wavelength range of 4935--4948 \AA\ and a size of 10"$\times$10" by \textsf{source inspector}, which was used to visually inspect the MXDF sources. The color bar shows the flux with an arbitrary normalization. The filled pink and filled purple circles indicate a nonisolated galaxy in our sample and its neighbor, which is not included in our sample with a spec-z measurement, respectively. The black open circles show sources in the catalog of \citet{Rafelski2015}. The cyan circle indicates a $0\farcs6$-radius circle centered at the position of RID=4838. The ID, $F775W$ magnitude, and $z_{\rm s}$ are shown for each object.   
       }
         \label{fig_group1_NB}
\end{figure}

\begin{figure}
   \centering
   \includegraphics[width=\hsize]{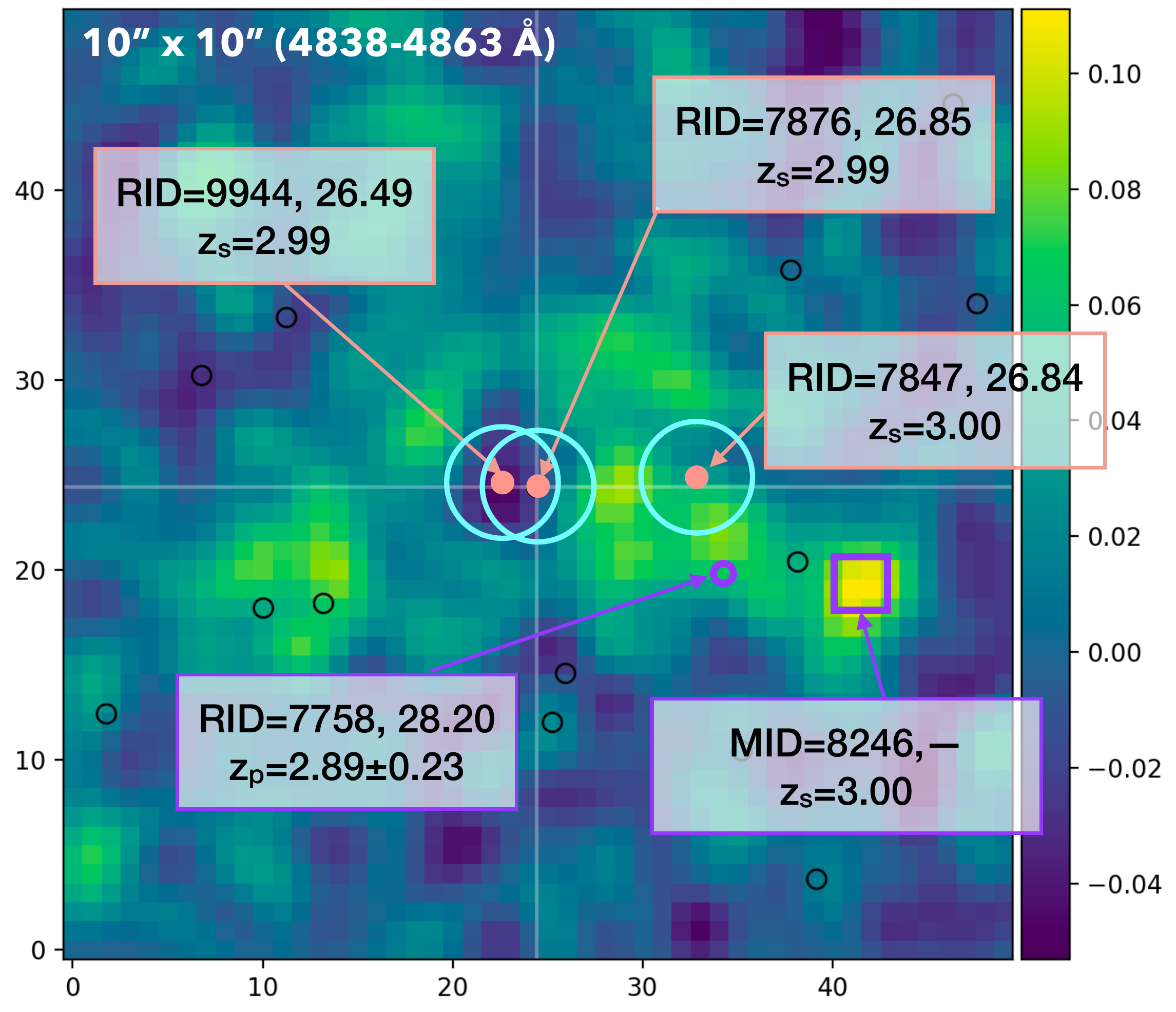}
      \caption{Same as Figure \ref{fig_group1_NB}, but for RIDs=7847, 7876, and 9944. The $\lya$ NB is centered at RID=7876 with a wavelength range of 4838--4863 \AA\ and a size of 10"$\times$10". The filled pink circles indicate three nonisolated galaxies in our sample. The purple open square shows a close MUSE-detected $\lya$ emitter without an HST counterpart. The purple open circle presents a potential neighbor, which does not have a spec-z but has a close photo-z. The black open circles show sources in the catalog of \citet{Rafelski2015}. The cyan circles indicate $0\farcs6$-radius circles centered at the position of RID=7847, 7876, and 9944. The ID, $F775W$ magnitude (if available), and $z_{\rm s}$ or $z_{\rm p}$ are shown for each object.   
      }
         \label{fig_group2_NB}
\end{figure}

As described in Section \ref{subsec:sample}, our sample includes four nonisolated galaxies, which have a neighboring galaxy with an HST detection within $0\farcs6$ \citep{Inami2017} with close spec-$z$ ($|\Delta V| \lesssim 500$ km s$^{-1}$). Figure \ref{fig_group1_NB} shows a NB image of a nonisolated galaxy. RID=4838 at $z_{\rm s}=3.06$ ($F775W$=26.70, ZCONF=2) has a UV-faint neighbor RID=6666 at $z_{\rm s}=3.06$ ($F775W$=29.72, ZCONF=2) with \textsf{ORIGIN}-detected $\lya$ emission, which is not included in our sample. Interestingly, these sources are a part of a cosmic web filament found with MUSE in \citet{Bacon2021}, called "G02", whose structure extends to a total length of 1.1 pMpc with a width of 47 kpc (see their Figure 12). Therefore, RID=4838 was categorized as a nonisolated galaxy.

As shown in Figure \ref{fig_group2_NB}, RID=7876 ($F775W$=26.85) and RID=9944 ($F775W$=26.49) are located within $0\farcs6$ each other. Because of the limited spatial resolution of MUSE, a MUSE source (MID=103, $z_{\rm s}=2.99$, ZCONF=3), which shows $\lya$ emission and absorption depending on wavelength and position, conservatively has two HST counterparts of RID=7876 and RID=9944. The $\lya$ absorption feature gets stronger at the position of RID=9944 than that of RID=7876\footnote{RID=7876 and RID=9944 are treated as a single HST source in the Cosmic Assembly Near-IR Deep Extragalactic Legacy Survey \citep[CANDELS, ID=110794,][]{Whitaker2019} and 3D-HST \citep[ID=28521,][]{Skelton2014}. Following the other MUSE GTO papers, we used the catalog of \citet{Rafelski2015} as a HST prior list (see also Figures \ref{fig_hstmask1} and \ref{fig_hstmask2})}. Moreover, MID=103 is also within $0\farcs6$ from RID=7847 at $z_{\rm s}=3.00$ ($F775W$=26.84, ZCONF=3). Near them, we can see an HST non-detected source at $z_{\rm s}=3.00$ (ZCONF=2, MID=8246) and a potential neighbor of RID=7758 at $z_{\rm p}=2.89$ ($F775W$=28.20). These sources are also included in an overdensity of MUSE-detected LAEs, "G01" \citep{Bacon2021}. As a conservative choice, we categorized RID=7847 as a nonisolated galaxy, in addition to RID=7876 and 9944. 

In summary, we conclude that the environments around four nonisolated galaxies in our sample would be different from those of the rest of the galaxies in our sample, which may affect halo properties. We calculated $\lya$ halo fractions separately for the isolated sources and the nonisolated sources in Section \ref{subsec:flah} for a conservative discussion.

\section{Continuum subtraction}\label{ap:consub}
\begin{figure}
   \centering
   \includegraphics[width=\hsize]{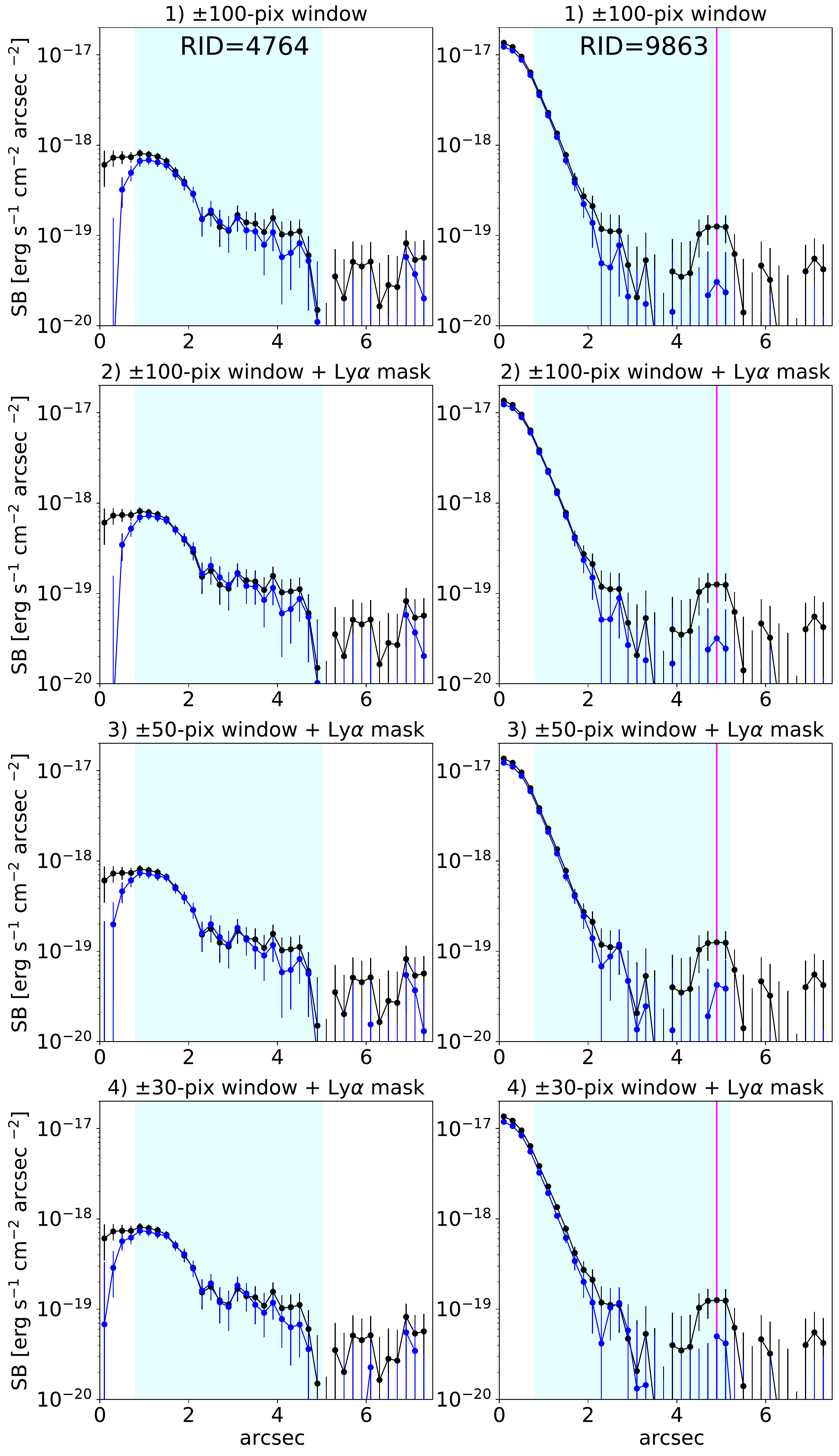}
      \caption{ Examples of the difference in the radial SB profiles for four different settings of the continuum subtractions. The black and blue lines indicate the SB profiles for the original minicubes and the continuum-subtracted minicubes, respectively. The neighboring masks were applied for both. The cyan shaded areas indicate the radii outside the target's continuum-component masks and inside the $r_{\rm CoG}$. {\it Left column:} RID=4764, which has an artificial absorption trough with the general setting of a 200-pix ($\pm100$-pix) window. {\it Right column:} RID=9863, which has a neighbor (at the radius shown by the magenta solid line) that cannot be eliminated only by the neighboring object mask. From top to bottom, each panel represents the settings: 1) 200-pix ($\pm100$-pix) window, 2) 200-pix ($\pm100$-pix) window + Ly$\alpha$ mask ($\pm400$ km s$^{-1}$ around Ly$\alpha$), 3) 100-pix ($\pm50$-pix) window + Ly$\alpha$ mask, and 4) 60-pix ($\pm30$-pix) window + Ly$\alpha$ mask, respectively. 
       }
         \label{fig_consub}
\end{figure}

Figure \ref{fig_consub} shows two examples of the difference in the radial SB profiles for the four different settings of the continuum subtractions. One of the objects, which has an artificial absorption trough on the SB profile with the general setting of a 200-pix ($\pm100$-pix) window, is RID=4764. The artificial absorption becomes less significant with the smaller spectral windows. Although the profile inside the target's continuum-component mask changes depending on the settings of the continuum subtraction, the profiles at $r=r_{\rm in}$--$r_{\rm CoG}$ are independent of the settings. Another example is RID=9863, which has a neighbor that cannot be eliminated only by the neighboring object mask. The right panels emphasize the fact that the continuum subtraction is essential to remove neighboring continuum-detected objects from the minicube. The neighboring object masks can only cover continuum-bright objects and prevent uncertainties due to continuum subtractions. Therefore, we needed both of the continuum subtractions and the neighboring object masks. The second example also shows that the SB profiles at $r=r_{\rm in}$--$r_{\rm CoG}$ are independent of the settings except around neighbors, implying that our tests for the existence of a $\lya$ halo at $r=r_{\rm in}$--$r_{\rm CoG}$ are robust and stable.

\section{Masks}\label{ap:hstmask}
\begin{figure}
   \centering
   \includegraphics[width=0.85\hsize]{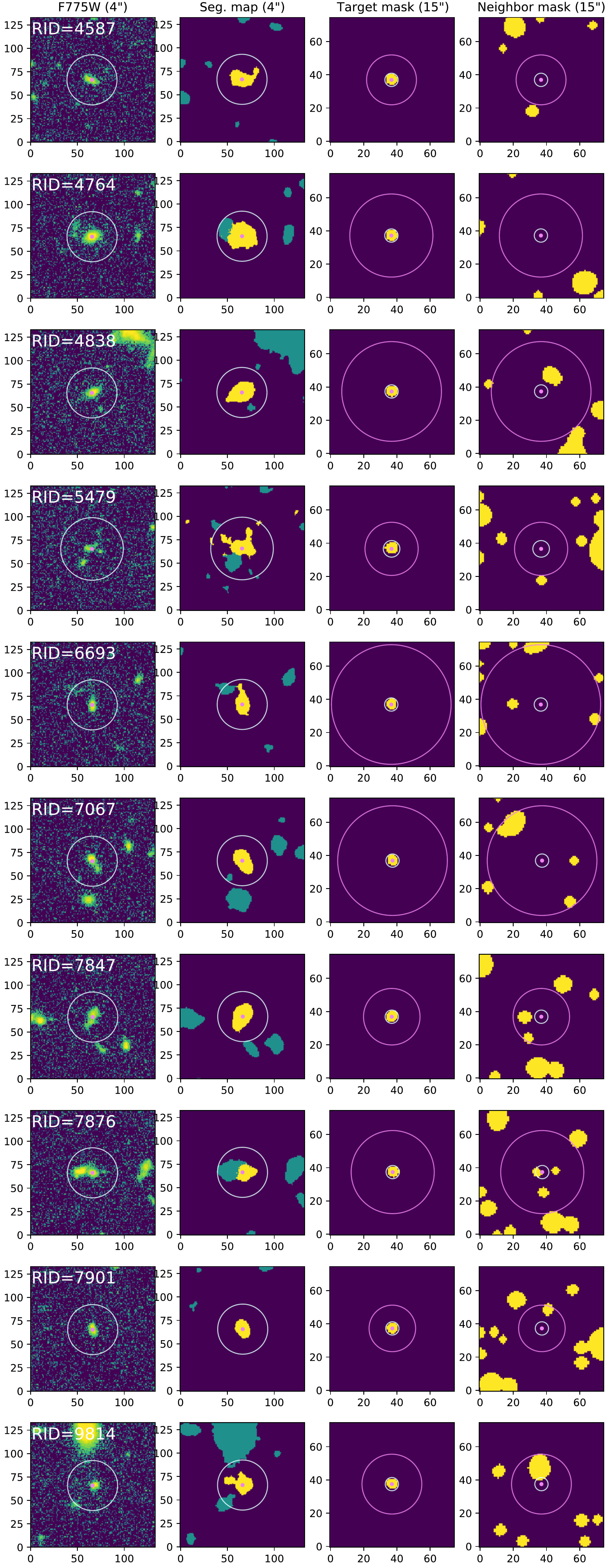}
      \caption{HST cutouts, segmentation maps, and masks for the first 10 sources in order of RID. {\it First column:} 4$''$ $\times$ 4$''$ HST $F775W$ cutout images. The purple dots and circles represent the positions of the UV-selected galaxy and r$_{\rm in}$, respectively.  {\it Second column:} 4$''$ $\times$ 4$''$ cutouts of the HST segmentation map. The yellow and green regions indicate the areas in which the target and neighboring galaxies are detected on the HST images, respectively. The segmentation map gives the boundaries of the objects. The dark purple regions indicate the sky. {\it Third column:} target's continuum-component masks (in the spatial resolution of MUSE, 15$''$ $\times$ 15$''$). The yellow and dark purple regions present the mask (the area of the main part of the galaxy) and the sky, respectively. The violet circles show $r_{\rm CoG}$. {\it Fourth column:} Neighboring object masks (neighbor mask; in the spatial resolution of MUSE, 15$''$ $\times$ 15$''$). The yellow and dark purple regions present the mask (the area of the main part of neighboring galaxies that have a bright continuum) and the sky, respectively.
              }
         \label{fig_hstmask1}
\end{figure}

\begin{figure}
   \centering
   \includegraphics[width=0.85\hsize]{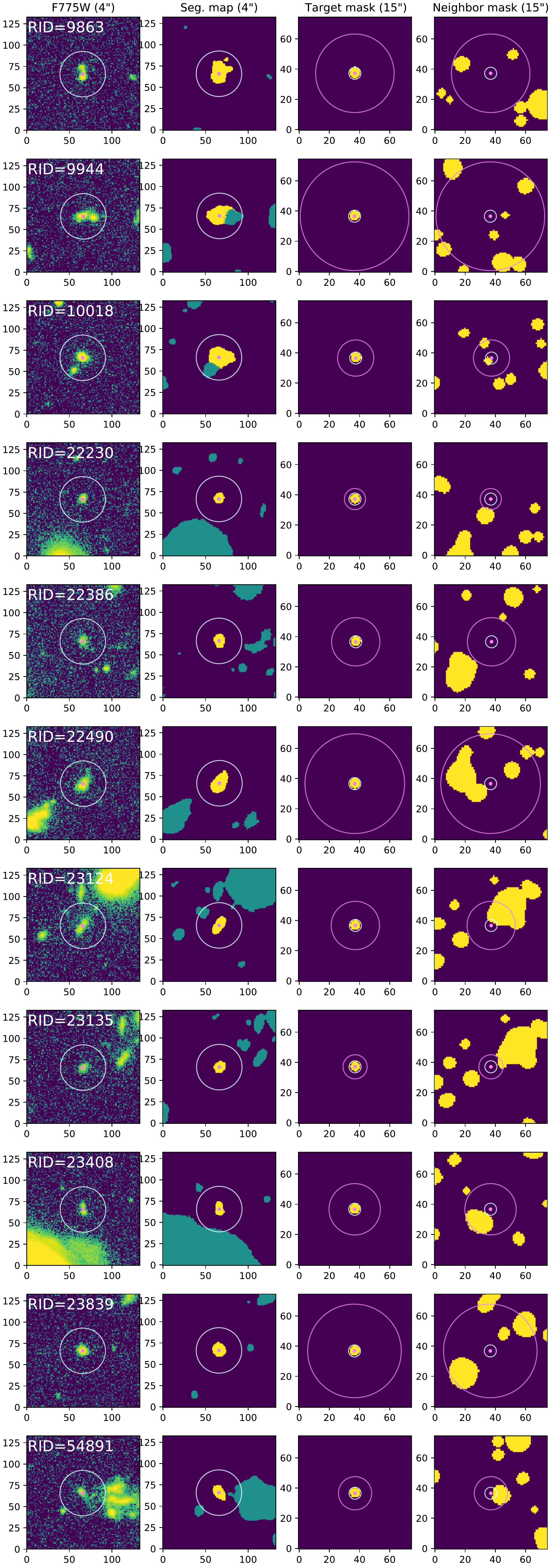}
      \caption{Same as Figure \ref{fig_hstmask1}, but for the last 11 sources in order of RID. 
              }
         \label{fig_hstmask2}
\end{figure}

Figures \ref{fig_hstmask1} and \ref{fig_hstmask2} show 4$''$ $\times$ 4 $''$ HST $F775W$ cutout images, 4$''$ $\times$ 4$''$ cutouts of the HST segmentation map, the target's continuum-component masks (in the spatial resolution of MUSE, 15$''$ $\times$ 15$''$), and the neighboring object masks (in the spatial resolution of MUSE, 15$''$ $\times$ 15 $''$), respectively. We used the HST segmentation map in \citet{Rafelski2015} following \citet{Inami2017}, which indicates areas in which galaxies are detected and defines the boundaries of the objects. The segmentation map was created with Source Extractor \citep{Bertin1996} with multiple thresholds for detection and deblending, in order to deblend sources and optimize detections and photometries, as no single set of such thresholds perfectly detects bright, faint, large, and small sources simultaneously (see Section 3.5 and Table 2 in \citealt{Rafelski2015}). They had four iterations. The parameter set for the first iteration is called ``Deep'' with \textsf{detect\_thresh=1.1$\sigma$} and \textsf{detect\_minarea=9}, with which source definitions near bright targets were poorly defined. The second ``Shallow'' run has \textsf{detect\_thresh=3.5$\sigma$} and gave better detections for bright sources. They merged the two catalogs into one, but both runs have difficulty in deblending sources. Then, they had two more iterations called ``Deep Deblend'' and ``Shallow Deblend'', which are with normal deblending parameters and lower deblending thresholds (see their Table 2). The resulting catalogs are merged into a single catalog with a single segmentation map \citep{Rafelski2015}.

\section{Completeness simulations and non-LAHs}\label{ap:comp}
\begin{figure*}
   \centering
   \includegraphics[width=1\hsize]{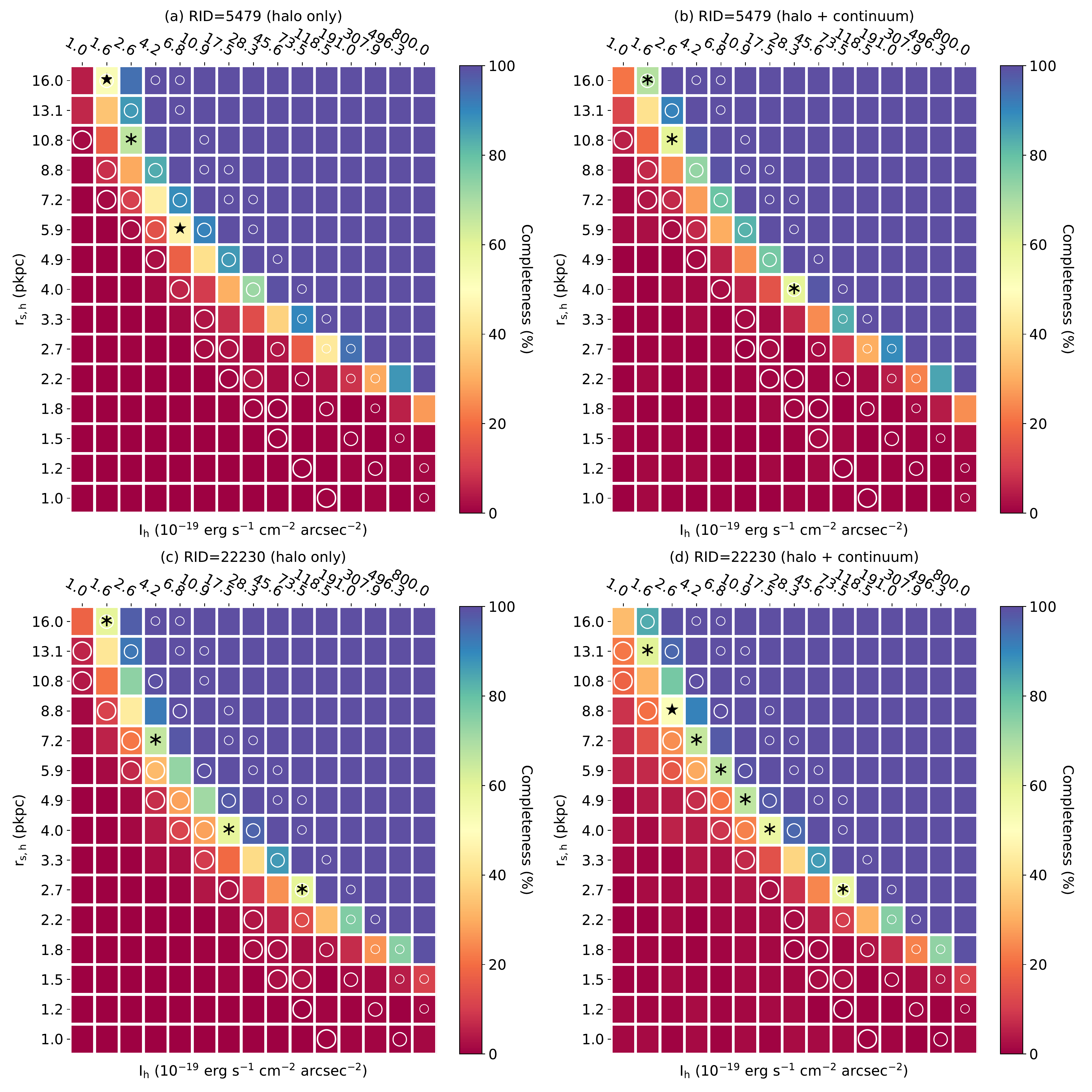}
      \caption{Heat map of the simulated completeness on the diagram of $I_{\rm h}$ and $r_{\rm s, h}$ of halo models for the first two non-LAH sources (RID=5479 and 22230). Left and right panels show completeness maps for halo only models and those for halo and continuum-like components models. 
      Black stars and asterisks on the map indicate the halo parameter sets that have completeness values from 45\% to 54\% and those from 54\% to 70\%, respectively, while white large, middle-size, and small circles represent the parameter sets whose $\lya$ fluxes of the halo models are from $1\times10^{-18}$ erg s$^{-1}$ cm$^{-2}$ to $2\times10^{-18}$ erg s$^{-1}$ cm$^{-2}$, those from $4\times10^{-18}$ erg s$^{-1}$ cm$^{-2}$ to $6\times10^{-18}$ erg s$^{-1}$ cm$^{-2}$, and those from $1\times10^{-17}$ erg s$^{-1}$ cm$^{-2}$ to $2\times10^{-17}$ erg s$^{-1}$ cm$^{-2}$, respectively. {\it (a) and (b)}: RID=5479. The flux of the continuum-like component is $f(\lya)=1.5\times10^{-17}$ erg s$^{-1}$ cm$^{-2}$. 
 }
         \label{fig_comp_4a}
\end{figure*}

\begin{figure*}
   \centering
   \includegraphics[width=1\hsize]{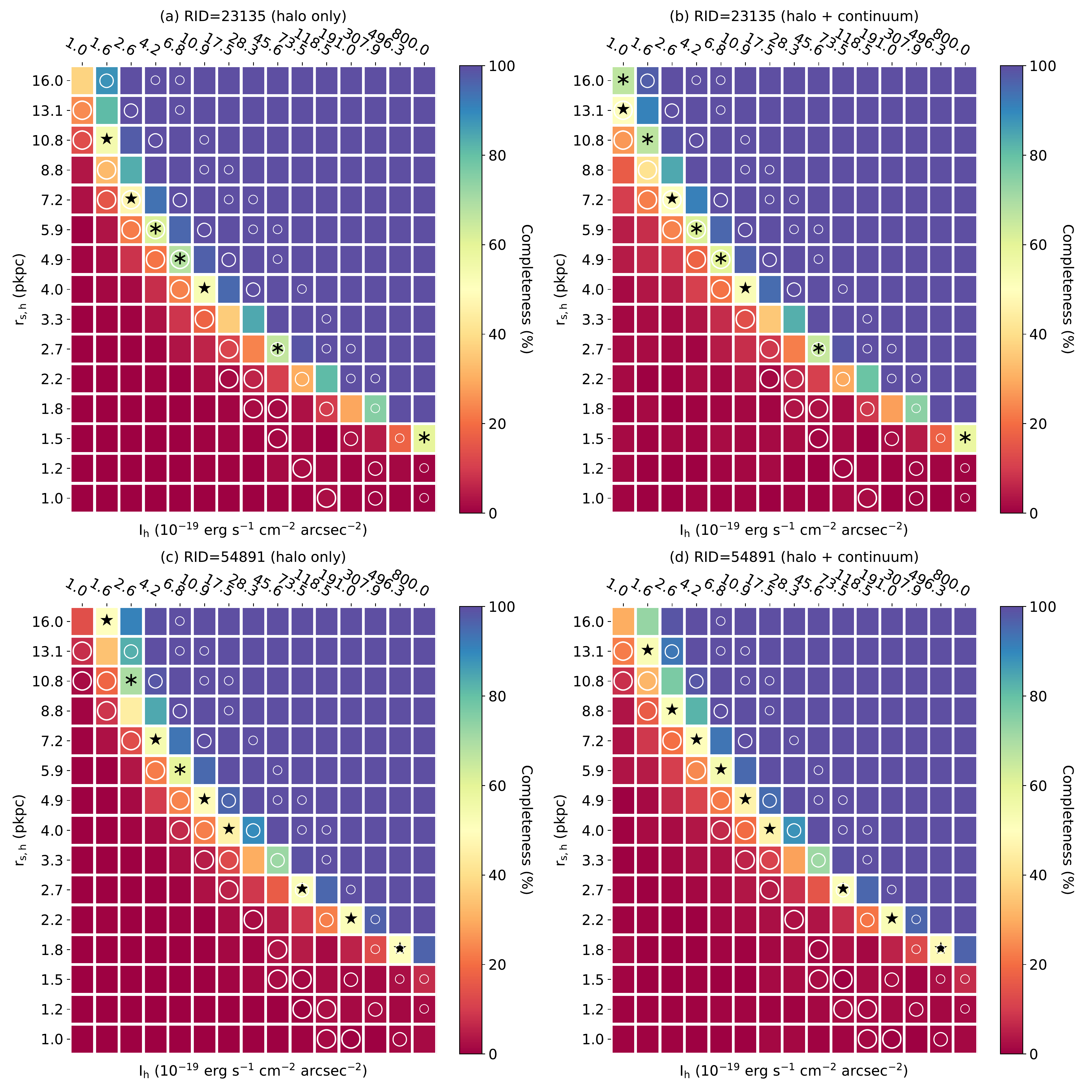}
      \caption{Same as Figure \ref{fig_comp_4a}, but for the last two non-LAH sources (RID=23135 and 54891) {\it (a) and (b)}: RID=23135 with $f(\lya)=3.7\times10^{-19}$ erg s$^{-1}$ cm$^{-2}$. {\it (c) and (d) }:RID=54891 with $f(\lya)=1.6\times10^{-19}$ erg s$^{-1}$ cm$^{-2}$.
 }
         \label{fig_comp_4b}
\end{figure*}

\begin{figure*}
   \centering
   \includegraphics[width=1\hsize]{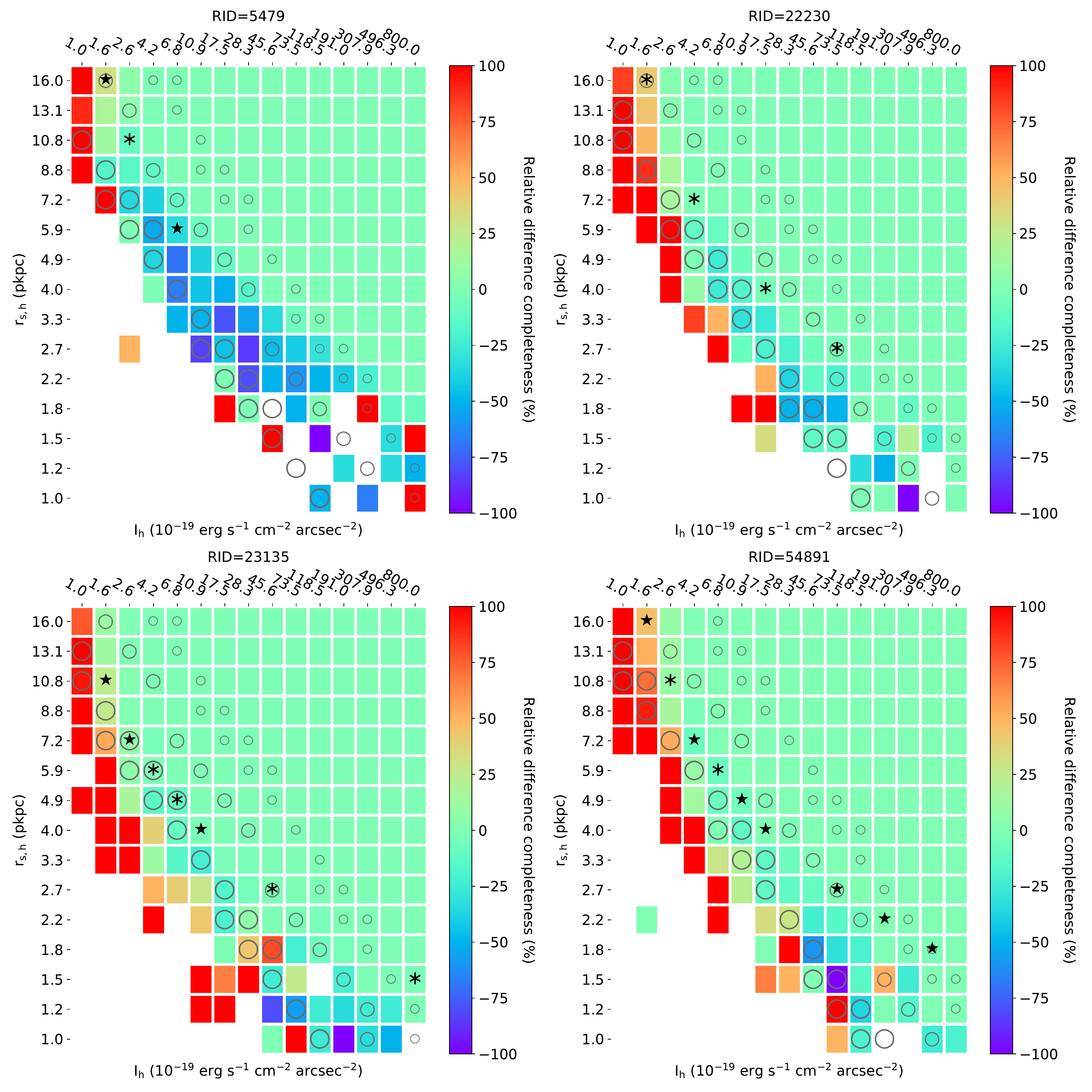}
      \caption{Heat map of the relative difference in the completeness for four non-LAHs on the diagram of $I_{\rm h}$ and $r_{\rm s, h}$ of halo models. The relative difference in completeness is defined as ($f_{\rm comp}^{h+c}$-$f_{\rm comp}^{h}$)/$f_{\rm comp}^{h}$, where $f_{\rm comp}^{h+c}$ and $f_{\rm comp}^{h}$ are the completeness for halo and continuum-like model, and that for halo only model, respectively. The heat map gets white when the relative difference is not available ($f_{\rm comp}^{h}$=0) or $f_{\rm comp}^{h}\sim0.0033$. Black stars and asterisks on the map indicate the halo parameter sets that have completeness values for halo models from 45\% to 54\% and those from 54\% to 70\%, respectively, while gray large, middle-size, and small circles represent the parameter sets whose $\lya$ fluxes of the halo models are from $1\times10^{-18}$ erg s$^{-1}$ cm$^{-2}$ to $2\times10^{-18}$ erg s$^{-1}$ cm$^{-2}$, those from $4\times10^{-18}$ erg s$^{-1}$ cm$^{-2}$ to $6\times10^{-18}$ erg s$^{-1}$ cm$^{-2}$, and those from $1\times10^{-17}$ erg s$^{-1}$ cm$^{-2}$ to $2\times10^{-17}$ erg s$^{-1}$ cm$^{-2}$, respectively. }
         \label{fig_comp_rel}
\end{figure*}

\subsection{Completeness simulations}\label{ap:comp_simu}

We simulated the completeness of the LAH test (Section \ref{subsec:testhalo}) for our four individual non-LAHs. We created mock NB images with different halo parameters and repeated the procedures for the halo test as follows.

First, we created 300 NB images with sky background noises by producing random values for pixels, which follow the Gaussian distribution. The Gaussian width $\sigma$ was assumed to be the median of the square root of the variance of the actual optimized NB for each source.

Second, we created various 2D $\lya$ halo models assuming an exponential profile as SB profile, $SB(r)$, used in the literature \citep[e.g.,][]{Wisotzki2016,Leclercq2017,Zhang2020}: 
\begin{equation}
SB(r)=I_{\rm h}\exp{\left(-\frac{r}{r_{\rm s, h}}\right)}, 
\end{equation}
where $I_{\rm h}$ and $r_{\rm s, h}$ are the central surface brightness and the scale length for haloes. The $I_{\rm h}$ is varied from $1\times10^{-19}$ erg s$^{-1}$ cm$^{-2}$ arcsec$^{-2}$ to $8\times10^{-17}$ erg s$^{-1}$ cm$^{-2}$ arcsec$^{-2}$ with 15 steps in the log scale, while the $r_{\rm s, h}$ is from 1.0 kpc to 16 kpc with 15 steps in the log scale. The 2D halo models were convolved with the MUSE PSF at the $\lya$ wavelength for each source. In total, we had 225 halo models for each source. We added each halo model to 300 mock images with background noises (in total, 67500 images). Then, we measured $r_{\rm CoG}$ (Section \ref{subsec:3dcog}) and did the same SB test for the existence of the $\lya$ halo (Section \ref{subsec:testhalo}). We calculated a completeness value for each halo parameter set for each source by dividing the number of confirmed haloes with 300. We also checked the median $r_{\rm CoG}$ for each halo parameter set.

We also tested for the case when a source has a $\lya$ component with a continuum-like profile. We fixed the model profile of the continuum-like $\lya$ component to that of the target's continuum-component image, which was obtained from the MUSE PSF convolution to the HST counterpart (Section \ref{subsec:mask}). We set the total flux of the target's continuum-component profile to the $\lya$ flux, f(\lya), which is measured from the flux-maximized NB and the target's continuum-component mask with an aperture correction. The aperture correction factor was derived for each source from the target's continuum-component image and the target's continuum-component mask. We added the continuum-like $\lya$ component to 300 mock images for each halo parameter set and then repeated the other procedures. The results of the completeness for non-LAHs are shown in Figures \ref{fig_comp_4a} and \ref{fig_comp_4b} for RID=5479 and 22230, and 23135 and 54891, respectively. 

Generally, there is a sharp transition between the parameter space with high completeness values in blue, and the parameter space with low completeness values in red, depending on $I_{\rm h}$ and $r_{\rm s, h}$ (see the drastic change of colors near black stars in Figures \ref{fig_comp_4a} and \ref{fig_comp_4b}). The halo tests for the non-LAH sources are complete when the halo fluxes are equal to or brighter than $\simeq 4\times10^{-18}$ erg s$^{-1}$ cm$^{-2}$ to $6\times10^{-18}$ erg s$^{-1}$ cm$^{-2}$ in most of the parameter ranges. At a fixed halo flux range (see white circles with a given size), completeness decreases both when $r_{\rm s, h}$ is too small and too large. Compact haloes with $r_{\rm s, h}\lesssim1$--$3$ kpc mostly cannot be detected due to the MUSE PSF and the method in which we only used $r=r_{\rm in}$--$r_{\rm CoG}$. We note that \citet{Leclercq2017} found compact haloes with $r_{\rm s, h}=2$--$3$ kpc with shallower MUSE data without AO, because of the difference in the methods. We also would like to note that the $r_{\rm e}$ in UV is $\simeq0.5$--$1.7$ kpc, which is relatively close to our limit. The halo only models and halo and continuum-like component models have similar distribution of completeness and median $r_{\rm CoG}$ in most of the case. However, the completeness of the two cases are slightly different from each other when the values are low as seen in Figure \ref{fig_comp_rel} (lower than 70\%; black asterisks). In particular, if a galaxy has a bright continuum-like $\lya$ component like RID=5479, its extended emission can be hidden by the continuum-like component for some cases. The non-LAH, RID=5479, has a $\simeq$0.25 to 1 times lower completeness for the two-component model than that of the halo only model, when the only halo completeness is $\simeq$50\%--70\% with relatively bright $I_{\rm h}$ (i.e., most of the cases with halo fluxes lower than $6\times10^{-18}$ erg s$^{-1}$ cm$^{-2}$). However, a bright-continuum component can also help the halo detection by increasing the S/N required for the test, when $I_{\rm h}$ is low. The parameter space in which the completeness for both cases is different is narrow for RID=5479, and the trend holds for both directions. We conclude that it does not cause a serious bias to our tests. With regard to $r_{\rm CoG}$, it is natural that the median value gets slightly higher for halo and continuum-like components models than halo only counterparts, since the continuum-like components can enhance the S/N on the outskirts. With the completeness obtained here, we give an overview of non-LAHs in Section \ref{subsubsec:testhalo_nolahs}, focusing on an object included in the UV-bright sample (RID=5479), and closely look at the other sources below. 

\subsection{RID=22230}\label{ap:22230}

This object has a high confidence level of ZCONF=2 for $z_{\rm s}=3.46$ in the MXDF catalog and $M_{1500}=-18.60$. It shows a clear $\lya$ emission line on the 1D spectrum for $r\leq r_{\rm in}=0\farcs8$ as shown in Figure \ref{fig_sample_cat3} and has S/N($r_{\rm in}$--$r_{\rm CoG}$)=6.72 with $r_{\rm CoG}=1\farcs4$. However, the SB profile at $r\geq r_{\rm in}$ is consistent with that of the MUSE PSF-convolved HST profile (continuum-like component), with $p_{0}=0.273$, which is higher than the threshold of 0.05. The completeness simulations for RID=22230 is shown in Figure \ref{fig_comp_4a}. For instance, the completeness decreases to $\simeq50$\% at $r_{\rm s, h}$=4.5 kpc and $I_{\rm h}=8.8\times10^{-19}$ erg s$^{-1}$ cm$^{-2}$ arcsec$^{-2}$.

On the radial SB profile shown in Figure \ref{fig_SBtest}, we can see an absorption-like dip at the top of the profile with surrounding emission. The dip could be created by the combination of $\lya$ absorption and spatially offset and anisotropic distribution of $\lya$, which can be confirmed on the NB image by eye (Figure \ref{fig_sample_cat3}). The S/N of the extended flux (at $r\geq r_{\rm in}$) would be higher if we only use angle-limited spatial pixels with a direction of the extended $\lya$ emission. If we could consider the SB profile at $r\leq r_{\rm in}$, which is found to be different from the shape of the MUSE PSF-convolved HST profile by eye, we might be able to confirm a potential halo. Deeper and higher-spatial resolution data set than MXDF will help improve the test, but with the current data set and method, this object is classified as a non-LAH. This object is not included in the UV-bright sample and does not influence on our main conclusions, or could even support our main conclusions if it was classified as a LAH.

\subsection{RID=23135}\label{ap:23135}

This object has a confidence level of ZCONF=1 for $z_{\rm s}=3.94$ in the MXDF catalog and $M_{1500}=-18.39$. As discussed in Section \ref{subsec:sample}, it has a faint but ORIGIN-detected $\lya$ emission line, which can be seen on the 1D spectrum for $r\leq r_{\rm in}=0\farcs8$ in Figure \ref{fig_sample_cat4}. However, the object does not have a sufficient number of spectral slices with S/N>1.5 at $r= r_{\rm in}$--$r_{\rm CoG}$ that can be used to create the optimized NBs (see Section \ref{subsec:3dcog}). As a consequence, it has a low S/N($r_{\rm in}$--$r_{\rm CoG}$) (<0) on the NB with $r_{\rm CoG}$=$1\farcs6$. Therefore, it is clear that this object does not have a significant $\lya$ halo above the observational limit. The completeness simulations for RID=23135 are shown in Figure \ref{fig_comp_4b}. For instance, the completeness decreases to $\simeq50$\% at $r_{\rm s, h}$=4.9 kpc and $I_{\rm h}=5.5\times10^{-19}$ erg s$^{-1}$ cm$^{-2}$ arcsec$^{-2}$. 

We would like to note that the nondetection of a halo should not be a direct result of a wrong redshift (i.e., low ZCONF value). On the contrary, if it had a clear extended $\lya$ emission line, it could be a redshift indicator and raise the ZCONF value in the MXDF catalog. Since the UV magnitude is faint, this object is not included in the UV-bright sample. The $\lya$ halo fraction for the sample including all sources could be even higher, if we excluded this object.

\subsection{RID=54891}\label{ap:54891}
This object has a confidence level of ZCONF=0 for $z_{\rm s}=2.94$ and $M_{1500}=-18.02$. As discussed in Section \ref{subsec:sample}, it does not have ORIGIN-detected lines but has a faint $\lya$ line (see the NB and SB profile in Figure \ref{fig_sample_cat4}). Similar to RID=23135, the object does not have a sufficient number of spectral slices at $r= r_{\rm in}$--$r_{\rm CoG}$ for an optimized NB (see Section \ref{subsec:3dcog}; $r_{\rm in}=0\farcs8$, $r_{\rm CoG}=2\farcs2$), the S/N($r_{\rm in}$--$r_{\rm CoG}$) on the NB is only 1.29. Therefore, it is clear that this object does not have a significant $\lya$ halo. The completeness simulations for RID=54891 is shown in Figure \ref{fig_comp_4b}. For instance, the completeness decreases to $\simeq50$\% at $r_{\rm s, h}$=6.3 kpc and $I_{\rm h}=5.7\times10^{-19}$ erg s$^{-1}$ cm$^{-2}$ arcsec$^{-2}$. Since the UV magnitude is faint, this object is not included in the UV-bright sample.

\end{appendix}

\end{document}